\documentclass{article}

\usepackage{algorithm,algpseudocode,amsmath,setspace,amssymb,bm,longtable,lscape,caption,graphicx,tikz,subcaption,multicol,natbib,placeins,rotating,pdfpages,multirow,amsthm,wrapfig,enumerate}
\usetikzlibrary{shapes,backgrounds}

\begin{document}
\title{Joint Structure Learning of Multiple Non-Exchangeable Networks}
\author{Chris. J. Oates \\ Department of Statistics, University of Warwick, UK. \\ Sach Mukherjee \\ Department of Biochemistry, Netherlands Cancer Institute, NL.}
\maketitle

\begin{abstract}
Several methods have recently been developed for joint structure learning of 
multiple (related) graphical models or networks.
These methods treat individual networks as exchangeable, such that each pair of networks are equally encouraged to have similar structures. However, in many practical applications, exchangeability in this sense may not hold, as some pairs of networks may be more closely related than others, for example due to group and sub-group structure in the data. 
Here we present a novel Bayesian formulation that generalises joint structure learning beyond the exchangeable case.
In addition to a general framework for joint learning, we (i) provide a novel default prior over the joint structure space that requires no user input; (ii) allow for latent networks; (iii) give an efficient, exact algorithm for the  case of time series data and dynamic Bayesian networks. 
We present empirical results on non-exchangeable populations, including a real data example from biology, where cell-line-specific networks are related according to genomic features.
\end{abstract}

\section{Introduction}

Structure learning remains an important and challenging problem. 
Often we seek to learn multiple graphs or {\it networks} $\{G _i \}_{i \in \mathcal{I}}$ that are expected to be related but that may be non-identical.  
For example in biomedical applications, multivariate data $\{\bm{y}_i\}_{i \in \mathcal{I}}$ pertaining to the same biological process (e.g. gene regulation or protein signaling) may be obtained from multiple, related samples $i \in \mathcal{I}$ (e.g. patients or laboratory models) that are expected to be non-identical with respect to conditional independence structure \citep{Penfold,Danaher,Oates4}.
In such situations, it is natural to consider joint learning that allows for information sharing between the inference problems indexed by $i \in \mathcal{I}$. Several techniques have been proposed for such joint structure learning, including Bayesian techniques for graphical models \citep{Werhli,Penfold,Oates4} and penalised likelihood estimators for Gaussian graphical models \citep[GMMs;][]{Chiquet,Guo,Yang,Danaher,Mohan}. These methods have been shown to improve estimation of individual graphs (or networks, we use both terms interchangeably)  $G_i$, especially in the regime where local sample sizes $n_i$ are not large.

Existing joint structure learning methods operate by shrinking estimated networks towards each other under an exchangeability assumption (i.e. the $\{G_i\}_{i \in \mathcal{I}}$ are treated as exchangeable random variables). 
However, in practice, relationships between datasets $\{\bm{y}_i\}_{i \in \mathcal{I}}$ (and their underlying networks) may be complex, e.g. hierarchical, with group and sub-group structure.
For example, in biology, datasets from multiple species may be related according to a complex evolutionary history \citep{Baumbach}, while cells within a tumour are related according to their lineage within the tumour \citep{Gerlinger}.
Similarly, in a data mining application, networks with nodes corresponding to products in an inventory \citep{Taylor} may be arranged into groups and sub-groups based on market structure or region.

This paper introduces a richer class of Bayesian joint estimators known as structure learning trees (SLTs) that subsume previous exchangeable formulations whilst permitting more complex, non-exchangeable relationships between networks.
An SLT is a rooted tree $T$ whose vertices are themselves networks and whose edges describe relationships between the networks (e.g. group and sub-group membership).
The tree $T$ encodes possibly non-exchangeable relationships between networks that are exploited during joint structure learning.
In this paper we restrict attention to the case where the tree $T$ can be reasonably specified {\it a priori}.
For example in biology, depending on the setting, established taxonomies such as phylogeny, tissue type, disease (sub)-type etc. can be used to specify  $T$.
Such taxonomies are often supported by a wealth of experimental evidence, 
and it is therefore natural to leverage them for improved structure learning. 
In the case where $T$ may be uncertain, we provide empirical results that investigate the extent to which SLT estimators are robust to $T$ misspecification. 
 
This paper is organised as follows:
In Section \ref{defs} we introduce SLTs, generalising existing work on joint structure learning to the non-exchangeable setting. 
Prior specification for SLTs is achieved by appealing to the intuitive notion that model constraints should be inherited along the edges of the tree. 
This heuristic allows specification of a default structural prior over all networks jointly that has essentially no user-set hyperparameters.
Section \ref{inference} provides an exact belief propagation algorithm for inference of both data-generating and latent network structures, while Section \ref{likelihood} focuses on time series data and dynamic Bayesian networks.
Empirical results on simulated data in Section \ref{results 1} assess the performance of SLTs, including cases where the joint structural prior is misspecified. 
Section \ref{results 2} shows results on proteomic time series data from multiple cancer cell lines that illustrate the use of SLTs in a topical application. 
Finally we close with a discussion in Section \ref{discussion}.

\section{Methods} 

\subsection{A Bayesian hierarchical model} \label{defs}

\paragraph{Structure learning trees} 
We consider joint structure learning of multiple {\it networks} $G_i = (V,E_i)$, $i \in \mathcal{I}$, that share the same {\it vertex set} $V = \{ 1, \dots, P \}$ but may differ with respect to their {\it edge sets} $E_i \subseteq V \times V$. 
Let $\mathcal{G}$ denote the space of all networks over vertex set $V$, up to restrictions associated with any particular model class (e.g. acyclicity, undirected edges etc.).
We define a \emph{structure learning tree} $T = (\mathcal{I},E_{T})$ as a rooted tree whose vertices are used\footnote{It will be convenient to interchange between an index $i \in \mathcal{I}$ and its corresponding network $G_i$.} to index individual networks $G_i$, with all edges $e \in E_T$ directed away from the root. 
Examples of SLTs are displayed in Figs. \ref{ex1} and \ref{panel}.
The root network is denoted by $G_1$.
Existing methods for joint estimation (see Introduction) can be regarded as a special case of the general SLT where $T$ has a star topology with centre $G_1$.

 \paragraph{Latent networks.}
In classical structure learning, network structure is latent in the sense that it not directly observed. 
SLTs allow for further latency; specifically we consider the situation in which data $\{ \mathbf{y}_i\}_{i \in \iota}$ are available conditional upon only a subset $\iota \subseteq \mathcal{I}$ of the networks of interest. 
The remaining nodes $\mathcal{I} \setminus \iota$ are {\it doubly latent} in the sense that neither they, nor data directly conditional upon them, are observed.
Latent nodes may be used to describe hidden (e.g. group level) structure (as in our biological example in Sec. \ref{results 2}).
Learning in an SLT exploits relationships between networks as encoded in $T$ to allow joint estimation of \emph{all} networks $\{G_i\}_{i \in \mathcal{I}}$, whilst respecting non-exchangeable relationships between these networks.

\paragraph{A default, subset prior.} \label{sec prior}
To formulate a joint statistical model we begin by placing a prior $p(G_1|G^0)$ on the root network $\overline{G_1}$.
Henceforth $\overline{G_i}$ represents the true (unknown) value of the network corresponding to $i \in \mathcal{I}$ whilst $G_i$ will be used to denote a possible structure for $\overline{G_i}$, and $G^0$ is a fixed ``prior network'' (see below). 
Then we define a joint structural prior over all networks $\{\overline{G_i}\}_{i \in \mathcal{I}}$ that factorises along the edges of $T$:
\begin{eqnarray}
p(\{G_i\}_{i \in \mathcal{I}}|G^0,E_T) \label{factors in SLT} 
= p(G_1|G^0) \prod_{(i,j) \in E_T} p(G_j|G_i) 
\end{eqnarray}
Previously proposed structural priors \citep[e.g.][]{Mukherjee2,Werhli} could in principle be used to specify the conditional density $p(G'|G)$. 
Recent work on the joint estimation of multiple exchangeable networks has focused on Boltzmann priors $p(G'|G) \propto \exp(-\lambda d(G,G'))$ for some measure of distance $d : \mathcal{G} \times \mathcal{G} \rightarrow [0,\infty)$ \citep{Werhli,Penfold,Oates4} and analogous penalised likelihoods \citep{Chiquet,Guo,Yang,Danaher,Mohan}.
However, such priors can be difficult to specify in the exchangeable case \citep{Werhli,Penfold,Oates4} and generalise poorly to the non-exchangeable case since each edge $e \in E_T$ in principle introduces an associated hyperparameter $\lambda_e \in [0,\infty)$.

To control complexity of prior specification,  we make use of the simple heuristic that network structure must be a subset of the structure of all network ancestors according to $T$:
\begin{eqnarray}
p(G_j|G_i) \propto \mathbb{I}\{ E_j \subseteq E_i \} \eta(G_j)
\label{model eq}
\end{eqnarray}
Here $\mathbb{I}$ is the indicator function and $\eta$ provides multiplicity correction for varying $G_j \in \mathcal{G}$ (see below). 
Under Eqn. \ref{model eq} the prior $p(G_1|G^0)$ encodes prior certainty that particular edges cannot exist, in \emph{any} network in $\{\overline{G_i}\}_{i \in \mathcal{I}}$.
If the networks are interpreted as causal graphical models then $G^0$ describes a set of conditional independence assumptions.
Thus in our formulation, inferred causation is explicitly conditional on prior causal hypotheses $G^0$ \citep{Pearl}.

\paragraph{Multiplicity correction.}

Multiplicity correction plays an important role in Bayesian structure learning beyond the penalty on model complexity provided by the marginal likelihood. 
This is clearly illustrated in the context of  variable selection, where a uniform prior over variable subsets has the undesirable property that the prior mass on all models with exactly one predictor goes to zero as the number of predictors grows large; such a prior cannot make sense in settings where one expects that a single predictor should have some reasonable prior mass.
Following \cite{Scott} we employ a binomial multiplicity correction
\begin{eqnarray}
\eta(G) = \prod_{p \in V} \binom{P}{d_p(G)}^{-1} \mathbb{I}\{d_p(G) \leq d_{\max}\} \label{multiplicity}
\end{eqnarray}
where $d_p(G)$ is the in-degree of vertex $p$ in $G$.
Here $d_{\max}$ represents a constraint on in-degree; such constraints are widely used to facilitate inference in graphical models \citep[e.g.][]{Hill}.

\subsection{Exact inference}

\paragraph{Marginal belief propagation.} \label{inference}

In this Section we describe how marginalisation and belief propagation combine to facilitate efficient, exact inference in SLT models. 
Taken together with a ``local'' likelihood $p(\bm{y}_i|\bm{\theta}_i,G_i)$, Eqn. \ref{factors in SLT} defines a Bayesian network on both discrete ($\overline{G_i}$) and possibly continuous ($\bm{\theta}_i$) variables (SFig. \ref{model}).
Efficient inference will require marginalisation of continuous variables; for data $\bm{y}_i$ we require that the ``marginal likelihoods'' $p(\bm{y}_i|G_i) = \int p(\bm{y}_i|\bm{\theta}_i,G_i) p(\bm{\theta}_i|G_i) d\bm{\theta}_i$ are pre-computed and cached for all $i \in \iota$.
Here $p(\bm{y}_i|G_i)$ is a convenient shorthand for $p(\bm{y}_i|E_i)$, the evidence for a particular topology $\overline{E_i} = E_i$, and $\bm{\theta}_i$ are parameters required to specify the local data-generating model.
For many models of interest, including dynamic Bayesian networks (see Sec. \ref{likelihood}), marginal likelihood  may be computed in closed form by exploiting conjugate prior specifications.
Otherwise, Monte Carlo and related numerical techniques may be used to approximate marginal likelihood in more complex models (e.g. \cite{Calderhead}).

The marginalised SLT (SFig. \ref{marginal model}) is then a discrete Bayesian network with respect to $T$. 
A factor graph representation of the marginal SLT model is shown in SFig. \ref{factor}.
Exact inference over factor graphs can be achieved efficiently using belief propagation \citep{Pearl3}, provided the factor graph is acyclic.
By restricting attention to tree structures $T$ in Sec. \ref{defs} we have guaranteed that the factor graph is acyclic. 
Belief propagation therefore yields posterior distributions $p_i(G_i | \bm{y})$ over structure for each $i \in \mathcal{I}$.
Pseudocode for our approach is provided in Supp. Sec. \ref{prop alg}.

\paragraph{Model averaging.}

Evidence in favour of an edge $(k,l)$ in a network $\overline{G_i}$ is summarised by the posterior marginal inclusion probability obtained by averaging over all possible structures $G_i$ for $\overline{G_i}$:
\begin{eqnarray}
p((k,l) \in \overline{E_i} | \bm{y}) = \sum_{G_i \in \mathcal{G}} \mathbb{I}\{(k,l) \in \overline{E_i}\} p_i(G_i|\bm{y}). \label{mave}
\end{eqnarray}
Here $\bm{y} = \{\bm{y}_i\}_{i \in \iota}$ contains all data.
The subset constraints of Eqn. \ref{model eq} manifest in the posterior as $p((k,l) \in \overline{E_i}|\bm{y}) \geq p((k,l) \in \overline{E_j}|\bm{y})$ whenever $j$ is a descendant of $i$ in $T$.

\subsection{Explicit formulae for time series} \label{likelihood}

\paragraph{FFDBN models.} For graphical models and time series data we provide explicit formulae:
We follow previous work by \cite{Murphy,Hill}, adopting a ``feed-forward'' dynamic Bayesian network (FFDBN) model for time series data.
For clarity of notation we consider a specific fixed network $G$, suppressing dependence upon $i \in \mathcal{I}$.
FFDBNs prohibit contemporaneous edges; this confers computational advantages (see \cite{Hill} for full details). Key features of FFDBNs include; (i) feedback can be explicitly modelled through time, (ii) the likelihood factorises over variables $p \in V$, reducing computational complexity (see below), (iii) conjugate priors and closed form expressions for marginal likelihood are available, and (iv) experimental designs involving interventions may be integrated in line with a causal calculus \citep{Spencer}.

In a FFDBN the value $Y_p(t)$ of variable $p$ at (discrete) time $t$ is dependent upon covariates $\bm{Y}(t-1) = \left[ Y_1(t-1), \dots , Y_P(t-1) \right]$.
When multiple time series are available, the vector $\bm{Y}_p = \left[ Y_p^1(1), \dots , Y_p^1(n) \; , \; Y_p^2(1), \dots , Y_p^2(n) \dots \right]$ denotes the concatenated time series, with the subscript indexing a specific variable $p \in V$.
We write $\text{pa}_G(p)$ for the parents of vertex $p$ in the network $G$.
In this paper we restrict attention to linear models that, for variable $p$, may be expressed as $\bm{Y}_p = \bm{X}_0\bm{\alpha} + \bm{X}_{\text{pa}_G(p)}\bm{\beta} + \bm{\epsilon}$ where $\bm{\epsilon} \sim N(\bm{0}_{n\times 1},\sigma^2\bm{I}_{n \times n})$.
The matrix $\bm{X}_0 = [\bm{1}_{\{t=1\}} \; \bm{1}_{\{t>1\}}]_{n \times 2}$ contains a term for the initial time point in each series. 
The elements of $\bm{X}_{\text{pa}_G(p)}$ corresponding to initial observations $(\bm{Y}_p)_{\{t=1\}}$ are simply set to zero.
Parameters $\bm{\theta} = \{\bm{\alpha},\bm{\beta},\sigma\}$ are specific to variable $p$ and network $G$.
In the linear case the model-specific component $\bm{X}_{\text{pa}_G(p)}$ of the design matrix consists of the predictors $\bm{Y}_{\text{pa}_G(p)}(t-1)$, where $\bm{Y}_A$ denotes the elements of the vector $\bm{Y}$ belonging to the set $A$.

\paragraph{Intervention.} In Sec. \ref{results} we consider experimental designs that involve targeted intervention on vertices in the data-generating networks. 
We followed the approach described in \cite{Spencer} to integrate interventional data in line with a causal calculus. Specifically, for the type of intervention in the experimental data (drug inhibition of kinases), using a ``perfect out fixed effects'' (POFE) approach (we direct the interested reader to the reference for full details).
This changes the network structure to model the intervention in line with the {\it do}-calculus \citep{Pearl} and also 
includes a fixed effect $\bm{X}_1\bm{\gamma}$ in the regression model for $\bm{Y}_p$, such that $\bm{X}_1$ indicates whether or not intervention(s) were used for each data-point.


\paragraph{Prior specification.}
We used a standard conjugate formulation for the linear model. Specifically, we employed a Jeffreys prior $p(\bm{\alpha},\sigma|\text{pa}_G(p)) \propto 1/\sigma$ for $\sigma>0$ over the common parameters. 
Prior to inference, the non-interventional components of the design matrix were orthogonalised \citep[following][]{Deltell} using the transformation 
\begin{eqnarray}
(\bm{X}_{\text{pa}_G(p)})_{ak} \mapsto \sum_{l=1}^n (\bm{I}_{n \times n}-\bm{P}_0)_{al} (\bm{X}_{\text{pa}_G(p)})_{lk}, 
\end{eqnarray}
where $\bm{P}_0 = \bm{X}_0(\bm{X}_0^T\bm{X}_0)^{-1}\bm{X}_0^T$.
We then assumed a unit-information $g$-prior for regression coefficients \citep{Zellner}, given by 
\begin{eqnarray}
\bm{\beta}|\bm{\alpha},\sigma,\text{pa}_G(p) \sim N(\bm{0}_{b \times 1},n\sigma^2(\bm{X}_{\text{pa}_G(p)}^T\bm{X}_{\text{pa}_G(p)})^{-1})
\end{eqnarray}
where $b = \dim(\bm{\beta})$.
(When interventional designs are used, the pair $(\bm{\beta},\bm{\gamma})|\bm{\alpha},\sigma,\text{pa}_G(p)$ are jointly assigned a $g$-prior.)

\paragraph{Marginal likelihood.} With the above specification, the evidence in favour of $\text{pa}_G(p)$ can be obtained in closed-form:
\begin{align}
p(\bm{y}_p|\text{pa}_G(p)) \propto \frac{1}{(n+1)^{b/2}}  \times \left(\bm{y}_p^T\left(\bm{I}_{n\times n}-\bm{P}_0 -\frac{n}{n+1}\bm{P}_{\text{pa}_G(p)}\right)\bm{y}_p\right)^{-\frac{n-a}{2}} \label{ML}
\end{align}
where $\bm{P}_{\text{pa}_G(p)} = \bm{X}_{\text{pa}_G(p)}(\bm{X}_{\text{pa}_G(p)}^T\bm{X}_{\text{pa}_G(p)})^{-1}\bm{X}_{\text{pa}_G(p)}^T$, $a = \dim(\bm{\alpha})$ and $b = \dim(\bm{\beta})$.
Note that the left hand side of Eqn. \ref{ML} is an abuse of notation since dependence on covariates $\bm{Y}_{\text{pa}_G(p)}(t-1)$ is suppressed (a formal treatment is presented in \cite{Oates4}).

\paragraph{Computation.} From the factorisation property of FFDBNs, the total marginal likelihood is simply given by the product 
\begin{eqnarray}
p(\bm{y}|G) = \prod_{p \in V} p(\bm{y}_p|\text{pa}_G(p)).
\label{factorised}
\end{eqnarray}
For FFDBNs the parent sets $\text{pa}_G(p)$ ($1 \leq p \leq P$) are Fisher-orthogonal; computational complexity may therefore be significantly reduced by decomposing the SLT into $P$ independent SLTs, each targeting one parent set $\text{pa}_G(p)$. MATLAB R2013b code implementing our procedure is provided in the Supplement.

Although we have focussed on FFDBNs, our procedure applies to other classes of network models, such as Bayesian networks and Gaussian graphical models. 
The availability of explicit formulae for FFDBNs motivates their use for the computational study presented below.

\section{Results} \label{results}

\subsection{Simulated data} \label{results 1}

To probe empirical performance of SLTs, we simulated data from a known tree $T$ and assessed ability to infer the true data-generating networks $\{\overline{G_i}\}_{i \in \iota}$.
In all experiments we placed $2P$ edges uniformly at random to generate a root network $\overline{G_1}$ subject to the in-degree constraint $d_p(G_1) = 2$ for all $p \in V$.
Two child networks $\overline{G_{11}},\overline{G_{12}}$ were then generated, each containing $P$ edges drawn as described below. Finally 10 networks $\overline{G_{1ij}}$ were generated by sampling $\rho P$ edges as described below.
We use concatenated subscripts to uniquely identify nodes in $T$; for example $G_{12}$ corresponds to child 2 of network $G_1$.

Existing joint structure learning methodologies require exchangeability of networks, while SLT instead imposes a tree structure capturing non-exchangeable relationships. We considered 5 data-generating regimes designed to mimic various applied settings, including those in which the SLT assumptions are violated: 

\begin{figure*}
\centering
\includegraphics[width = \textwidth,clip,trim = 2.5cm 0cm 0cm 1.5cm]{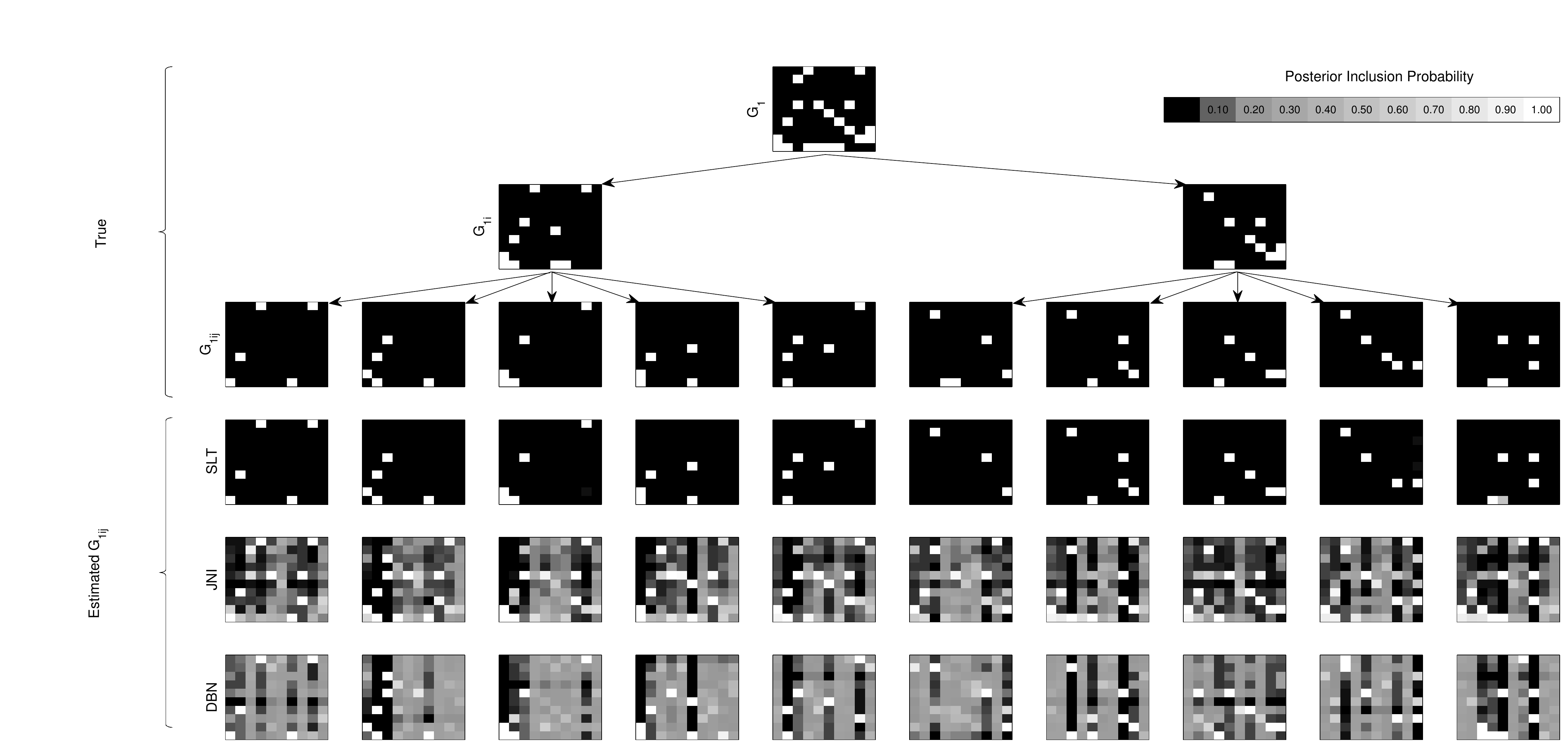}
\caption{Results on data generated from disjoint sub-groups (regime (1), see text), where ``doubly latent'' networks $\overline{G_{11}}$, $\overline{G_{12}}$ share no common edges. 
Top: Data-generating networks and associated tree structure.
Bottom: Estimates for individual network structure.
[Inference methods: ``SLT'' = structure learning trees, ``JNI'' = joint (exchangeable) network inference \citep{Oates4}, ``DBN'' = classical network inference applied to each network separately (see text for details).
Data consisted of $n = 60$ time points; see Supplement for full details of the data-generating set-up.].
}
\label{ex1}
\end{figure*}

\begin{enumerate}[(1)]
\item {\bf Disjoint sub-groups.} Edges in each (non-root) network are drawn at random from the parent in $T$, conditional upon the networks $\overline{G_{11}}$, $\overline{G_{12}}$ having disjoint edge sets.
This regime strongly violates the exchangeability assumption implicit in existing joint structure learning methodologies.
\item {\bf Weakly exchangable.} Here networks $\overline{G_{11}}$, $\overline{G_{12}}$ are generated independently, conditional upon $\overline{G_1}$, such that they are likely to share common edges. As above, all edges in each (non-root) network are drawn at random from the parent in $T$. Whilst exchangeability is violated, this regime ought to be more favourable to existing exchangeable estimators than regime (1) above.
\item {\bf Fully exchangeable.} The networks $\overline{G_{11}} = \overline{G_{12}}$ are taken equal, rendering the networks $\overline{G_{1ij}}$ fully exchangeable. In this regime SLT should lack efficiency relative to exchangeable estimators.
\item {\bf Misspecified tree.} This data-generating regime is equivalent to the disjoint sub-groups regime (1), however the SLT estimator 
is based not on the true data-generating tree, but rather on  a tree $T'$ uniformly sampled from the space of all trees. Thus while the networks are non-exchangeable, the SLT is misspecified with high probability. This mimics the scenario in which an {\it a priori} assumed tree is used that is in fact largely incorrect.
\item {\bf Subset violation.} All edges in each (non-root) network are drawn such that 20\% of edges in each child network are not edges in its parent network in $T$. In this regime, sub-groups exist among the networks, but the key assumption (Eqn. \ref{model eq}) of the parameter-free structural prior is violated with high probability.
\end{enumerate}

Time series data $\bm{y}_{1ij}$ of length $n$ were generated from each of the 10 networks $\overline{G_{1ij}}$ according to a linear autoregressive process with interventions described in Supp. Sec. \ref{sim set up}.
No data were made available on the networks $\overline{G_1},\overline{G_{11}},\overline{G_{12}}$, which are doubly latent.
For all simulation experiments we fixed $P=10$.
The entire process was repeated 10 times.
We compared SLT to: 
\begin{enumerate}[(A)]
\item Non-joint network inference (``DBN"), the default approach of carrying out structural inference using a FFDBN for each dataset $\bm{y}_{1ij}$ independently.
\item Joint network inference \citep[``JNI";][]{Oates4}. This Bayesian method is a special (exchangeable) case of our proposed SLT methodology. Hyperparameters were chosen according to the heuristics of \cite{Oates4}.
\end{enumerate}

We note that alternative exchangeable estimators to (B) include \citet{Danaher} and \citet{Penfold}, but the former has not been adapted for time series data and heavy computational demands of the latter preclude systematic empirical comparison.
To ensure fair comparison, the same in-degree restriction $d_{\text{max}} = 2$ (which includes the data-generating networks) was used for all methods.
Moreover, to prevent confounding by differing formulations of likelihood, we based each method on the same FFDBN likelihood (as described in Sec. \ref{likelihood}). Thus, all methods share the same basic time series formulation and differ only with respect to whether and how they share information between networks.
No specific prior information was given regarding network topology, except for the tree structure $T$ (in regimes 1-3,5) which was exploited by SLT.

\begin{figure*}[t!]
\centering
\begin{subfigure}{0.32\textwidth}
\includegraphics[width = \textwidth]{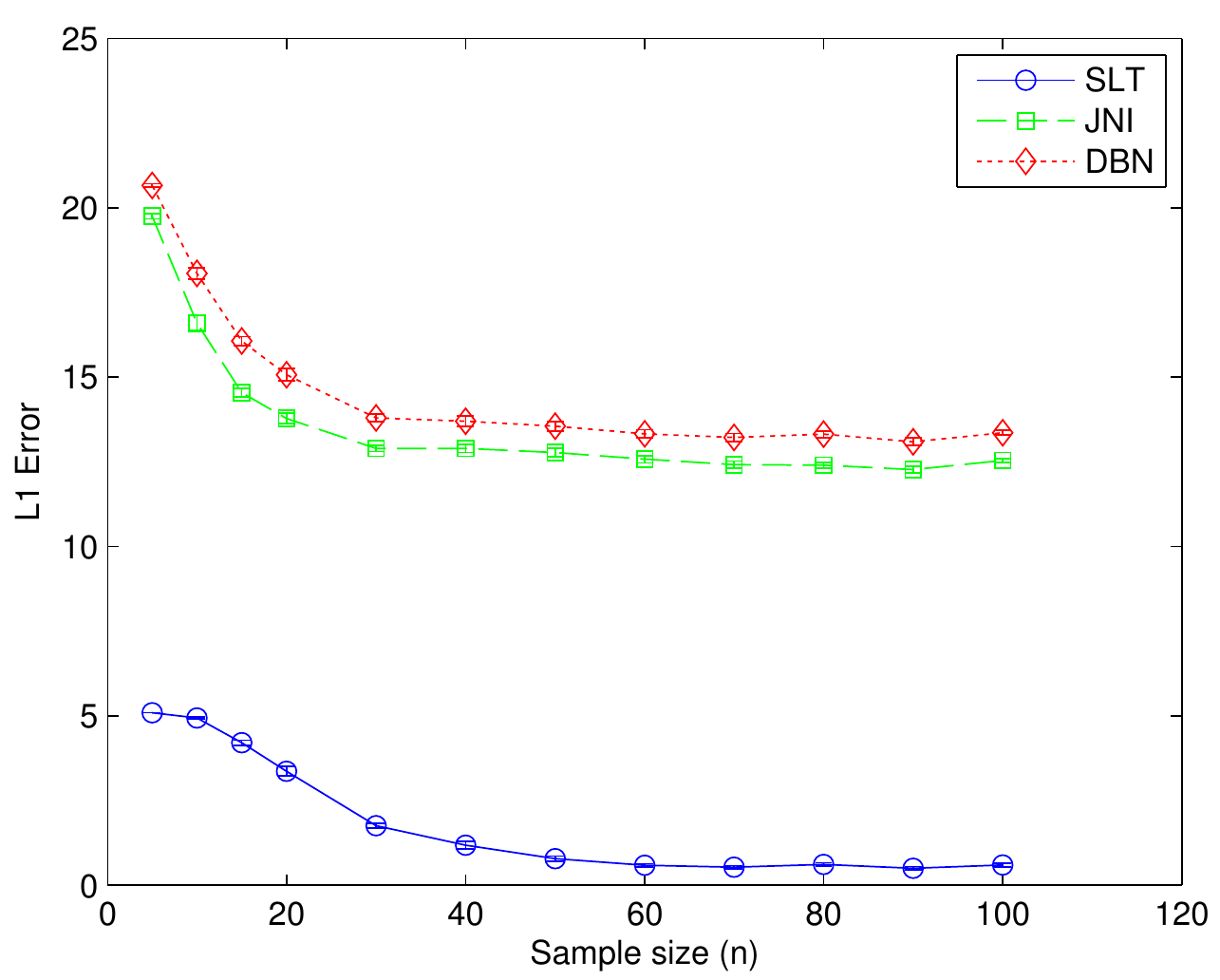}
\includegraphics[width = \textwidth]{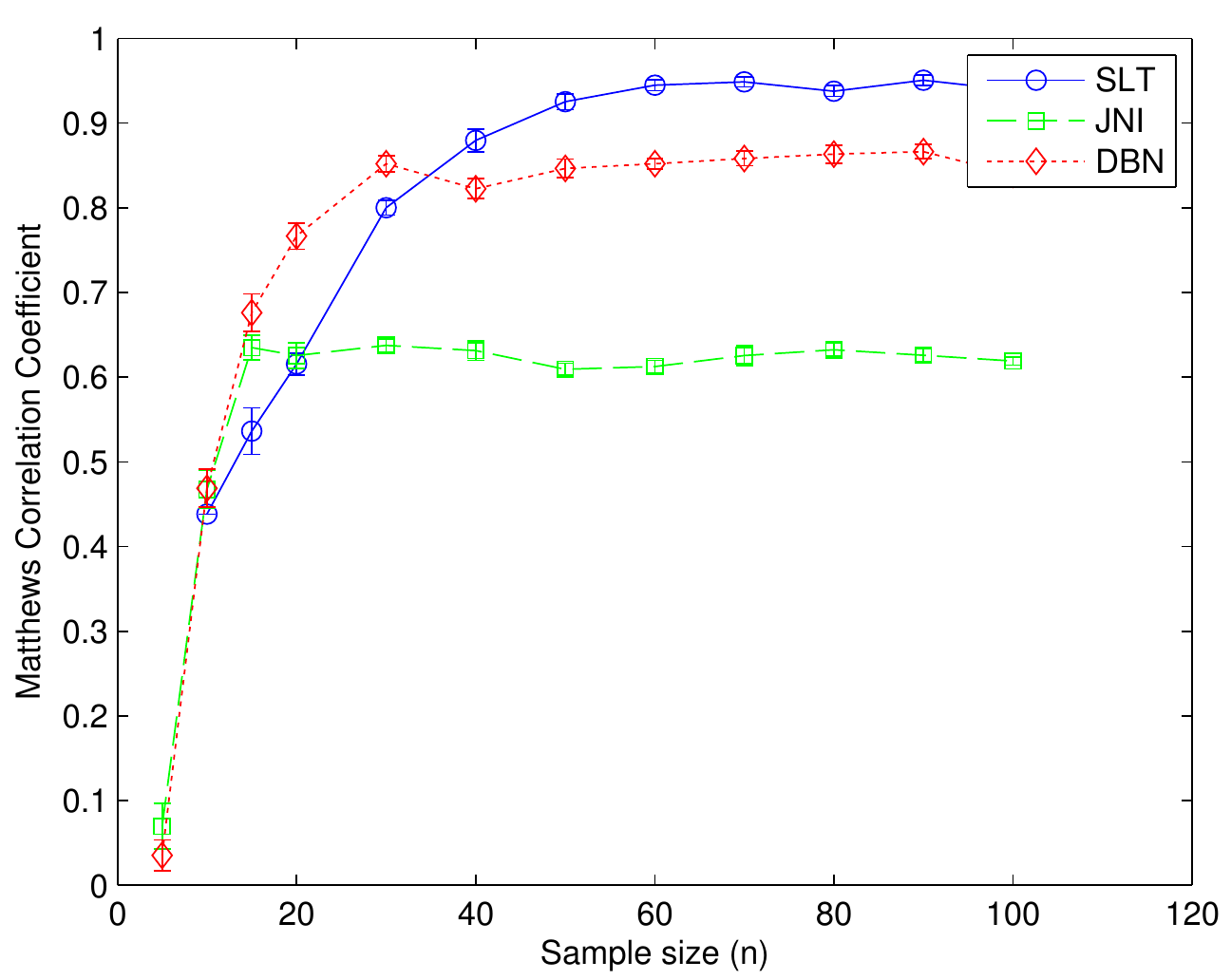}
\includegraphics[width = \textwidth]{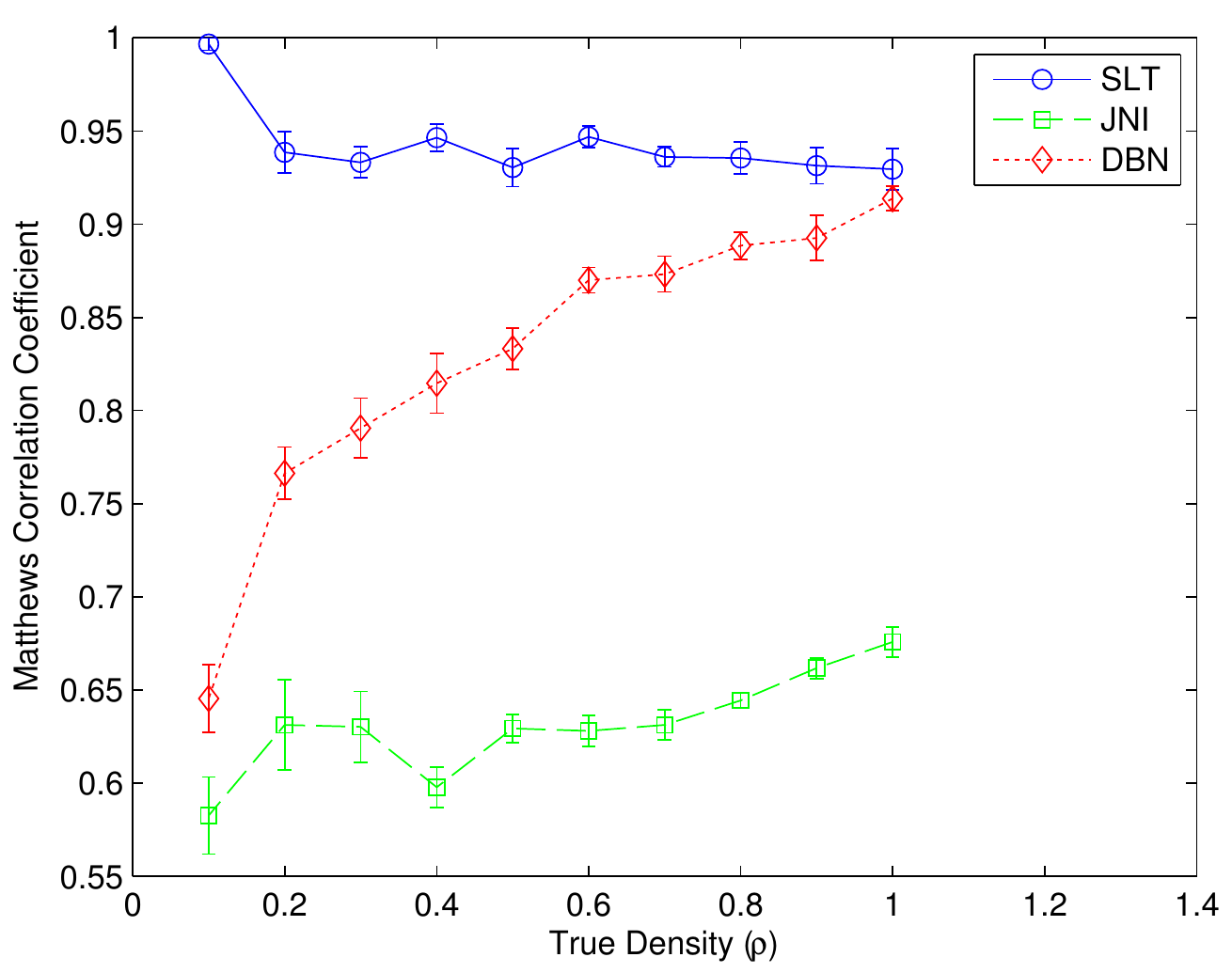}
\caption{Disjoint sub-groups}
\label{reg1}
\end{subfigure}
\begin{subfigure}{0.32\textwidth}
\includegraphics[width = \textwidth]{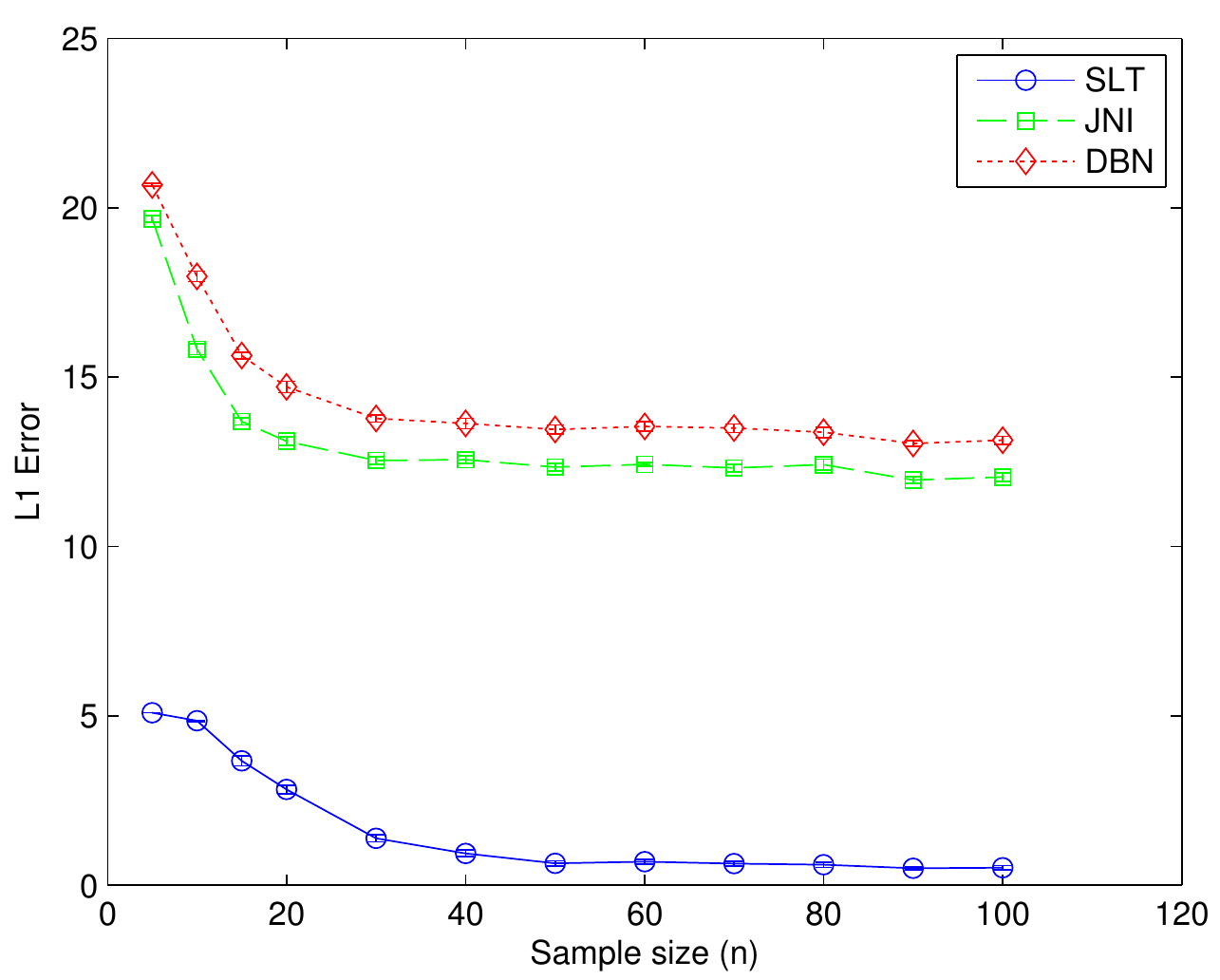}
\includegraphics[width = \textwidth]{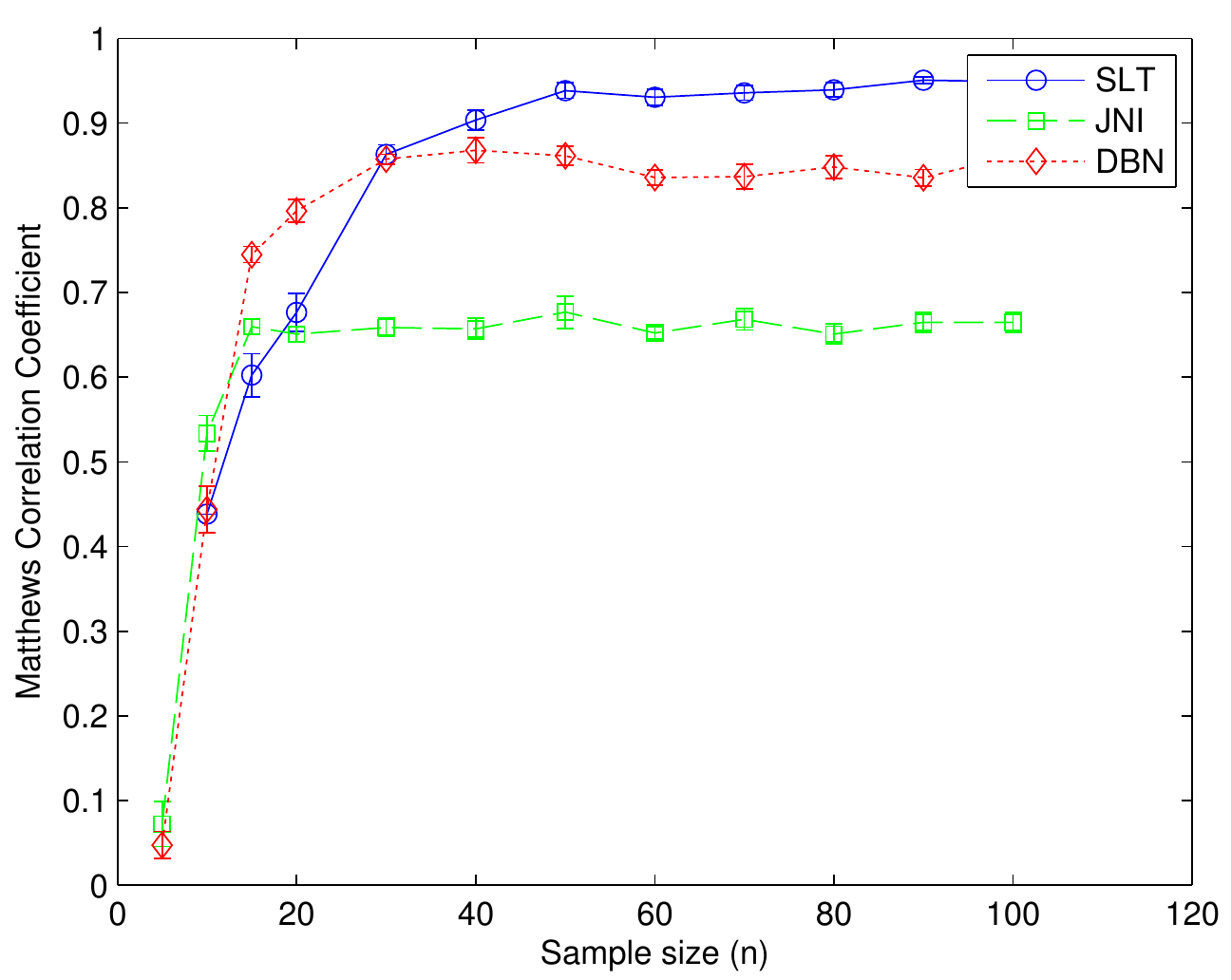}
\includegraphics[width = \textwidth]{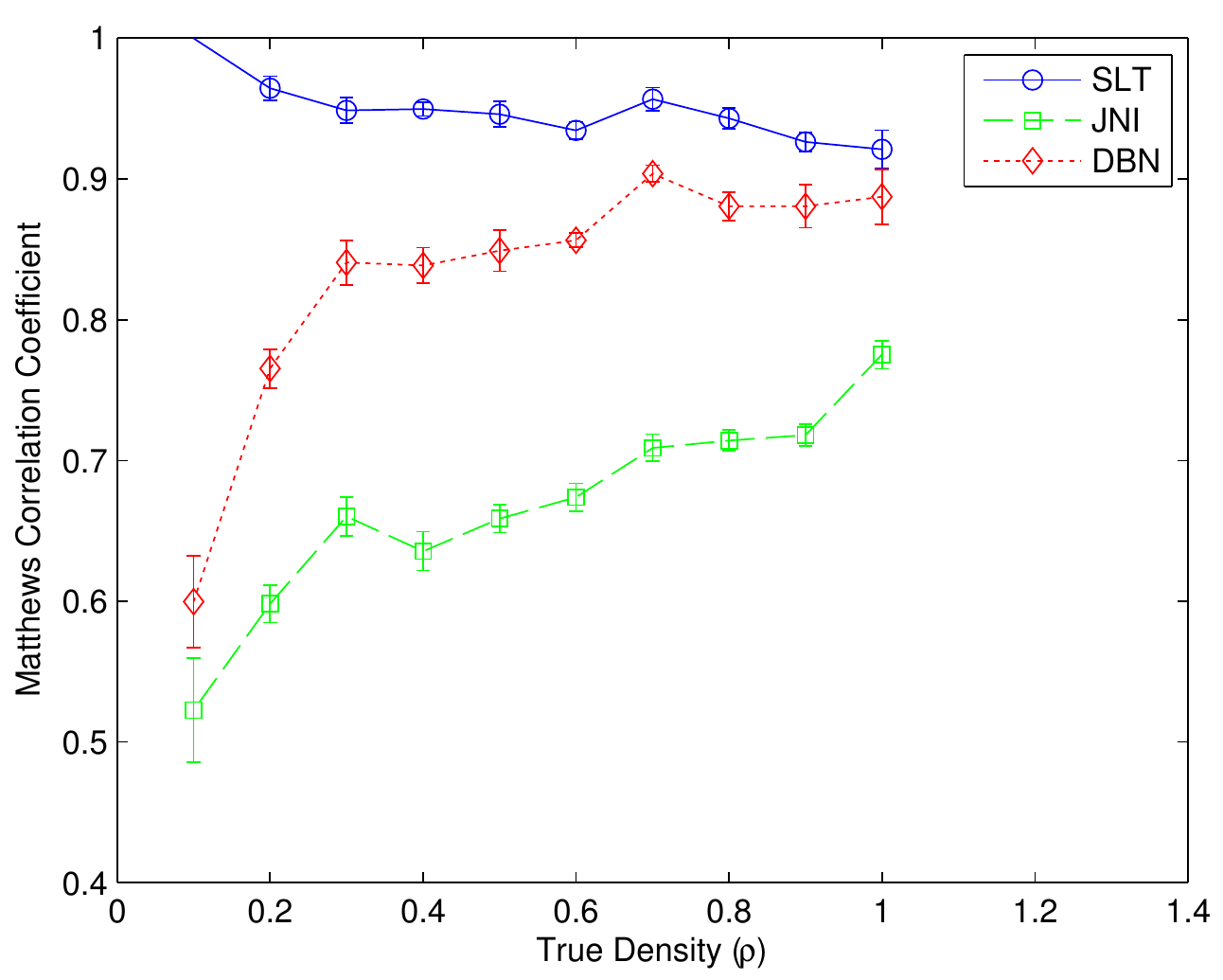}
\caption{Weakly exchangeable}
\label{reg2}
\end{subfigure}
\begin{subfigure}{0.32\textwidth}
\includegraphics[width = \textwidth]{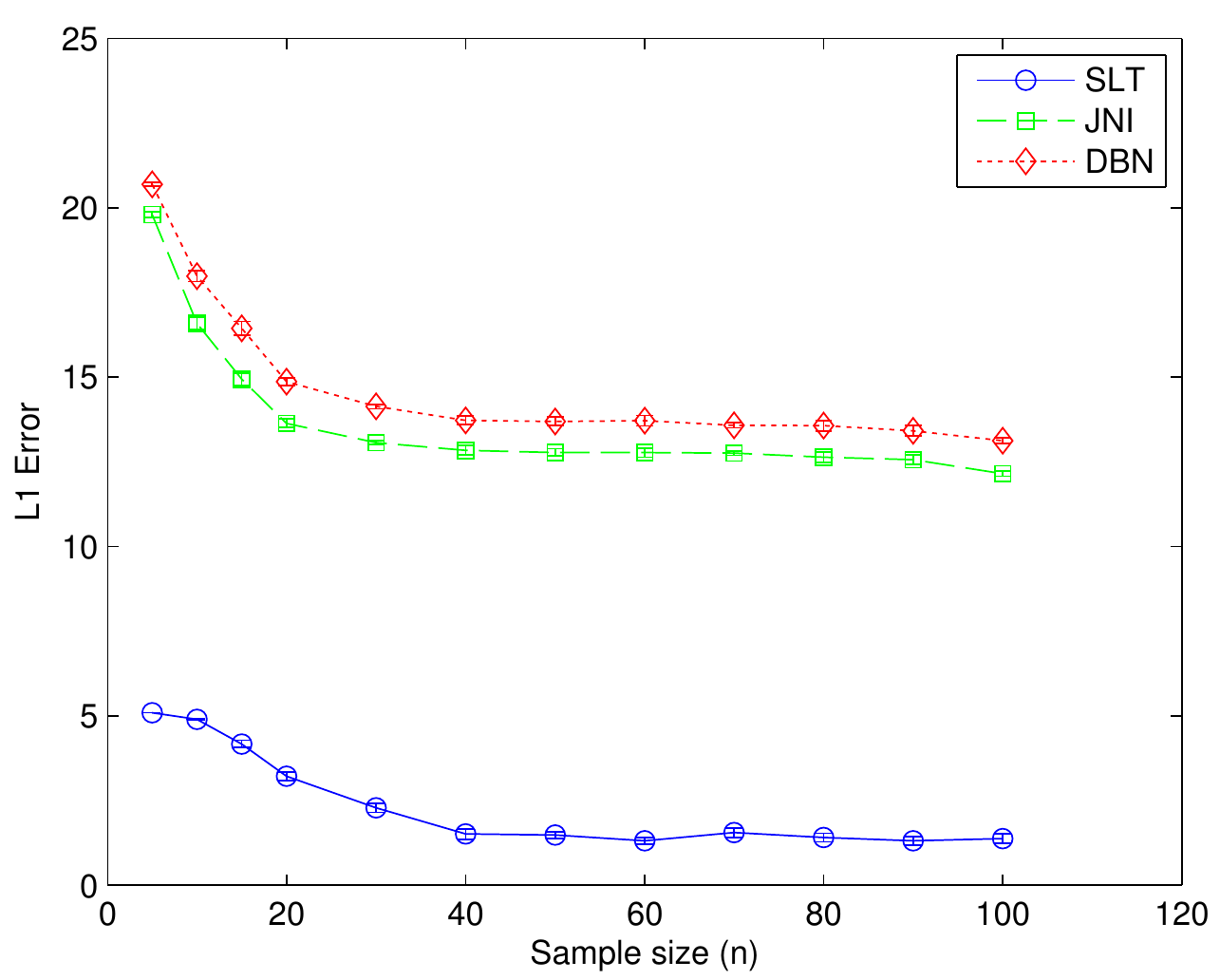}
\includegraphics[width = \textwidth]{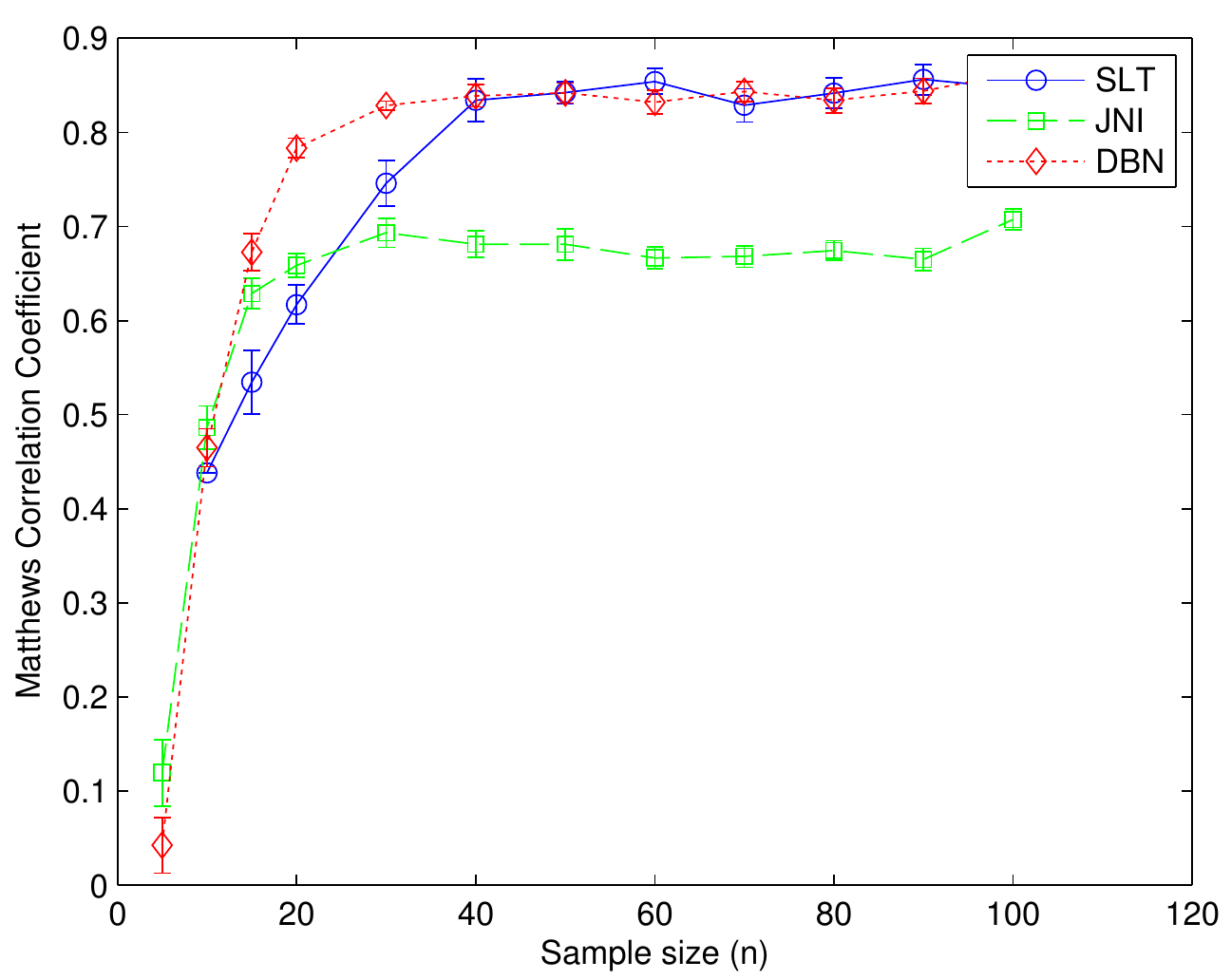}
\includegraphics[width = \textwidth]{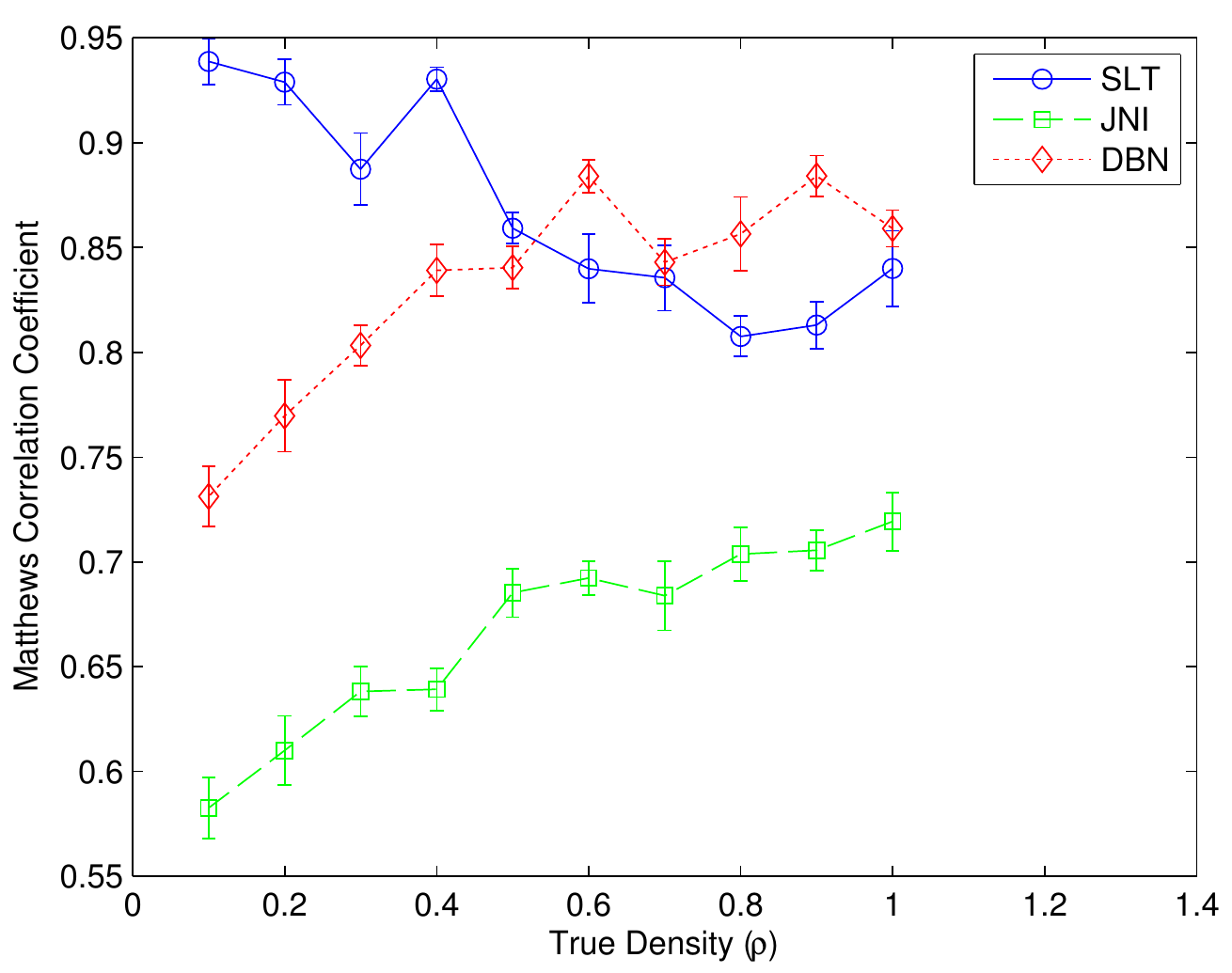}
\caption{Subset violation}
\label{reg3}
\end{subfigure}
\caption{Results on simulated data generated from SLTs in different regimes, as described in the Main Text. [Inference methods: ``SLT'' = structure learning trees, ``JNI'' = joint network inference \citep{Oates4}, ``DBN'' = classical network inference applied to each network separately. Performance metrics: ``L1 Error'' = average $\ell_1$ distance between true and inferred (weighted) adjacency matrices, ``Matthews Correlation Coefficient'' = average MCC for thresholded network estimators. Error bars display standard error computed over 10 data-generating networks and for each network 10 sampled datasets. We considered both varying $n$ for fixed $\rho = 0.5$ and varying $\rho$ for fixed $n=60$.]}
\label{reg}
\end{figure*}

We considered the thresholded network estimator, which consists of edges with marginal posterior inclusion probability (Eqn. \ref{mave}) $>0.5$.
Performance at sample size $n$ and density $\rho$ was quantified using metrics from classifier analysis (see Supp. Sec. \ref{metrics}), averaged over all data-generating networks and all datasets.
Here we focus on the Matthews correlation coefficient (MCC), which is regarded as a balanced measure, suitable for use when the underlying class distribution is skewed.
To quantify performance of the posterior inclusion probabilities themselves, we also considered the $\ell_1$ distance to the true data-generating networks.
Further details regarding performance measures (including additionally AUPR and AUROC) appear in Supp. Sec. \ref{metrics}.

\begin{figure*}[t!]
\centering
\includegraphics[width = 0.24\textwidth]{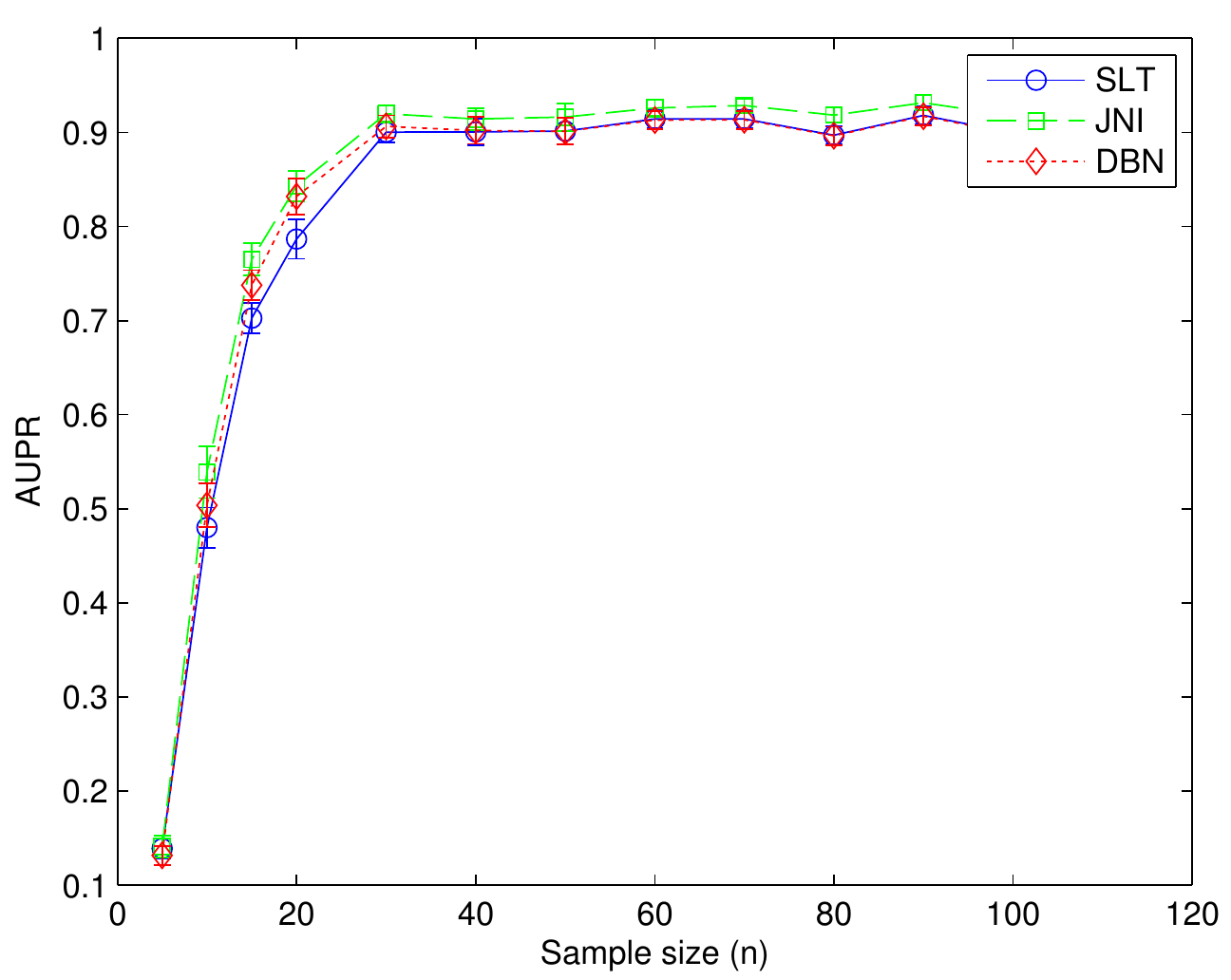}
\includegraphics[width = 0.24\textwidth]{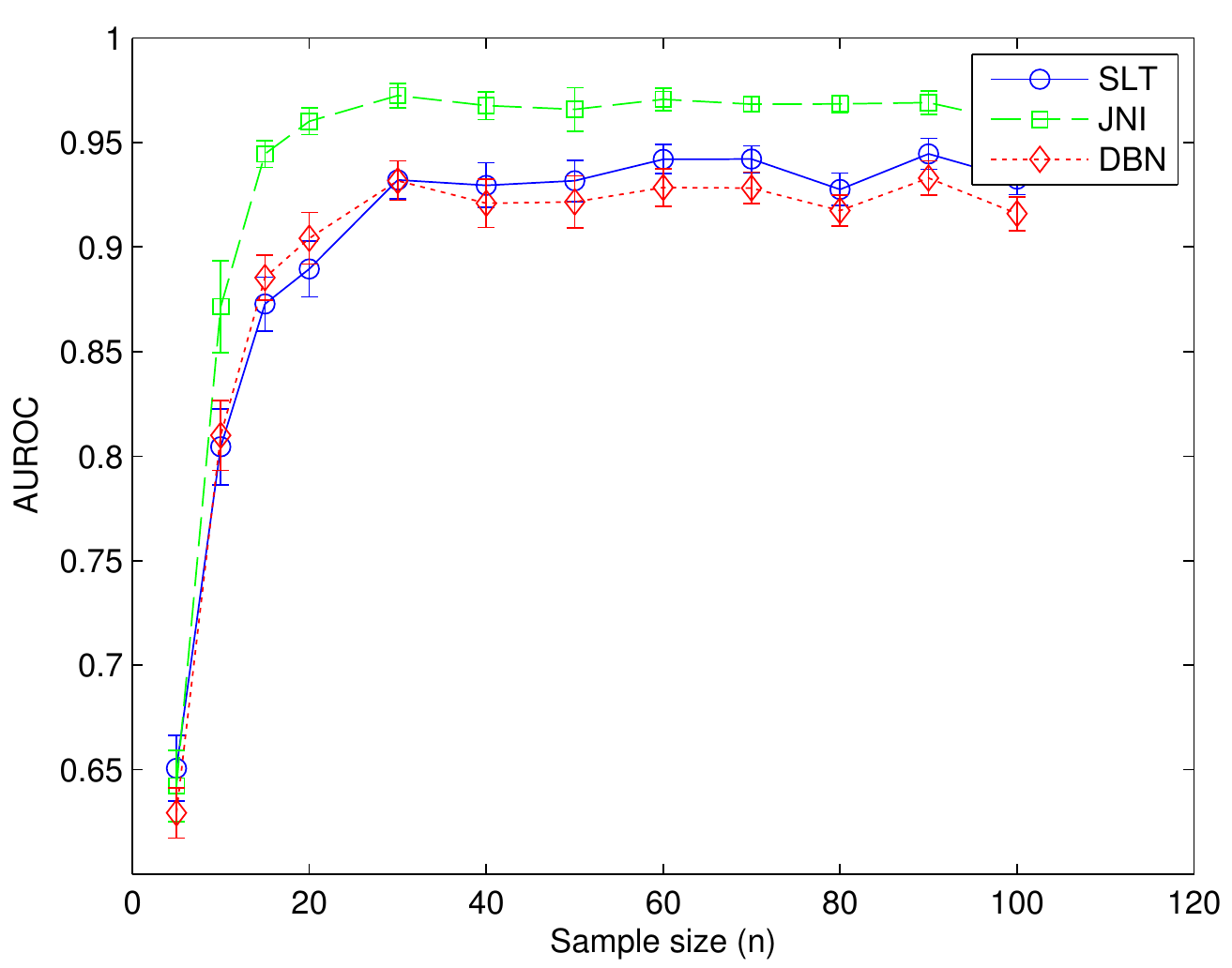}
\includegraphics[width = 0.24\textwidth]{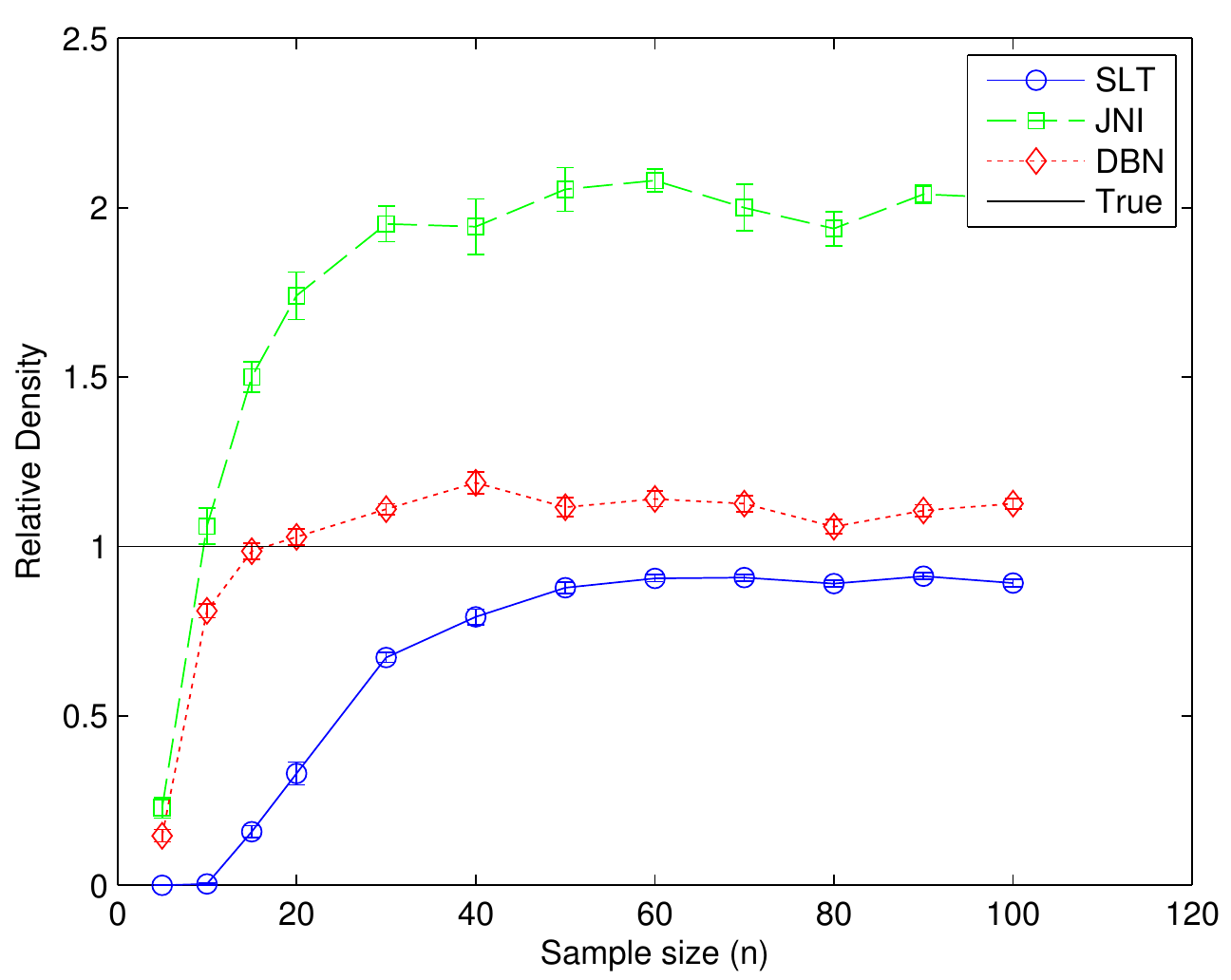}
\includegraphics[width = 0.24\textwidth]{Figures/vary_n_1_L1_Error.pdf}

\includegraphics[width = 0.24\textwidth]{Figures/vary_n_1_Matthews_Correlation_Coefficient.pdf}
\includegraphics[width = 0.24\textwidth]{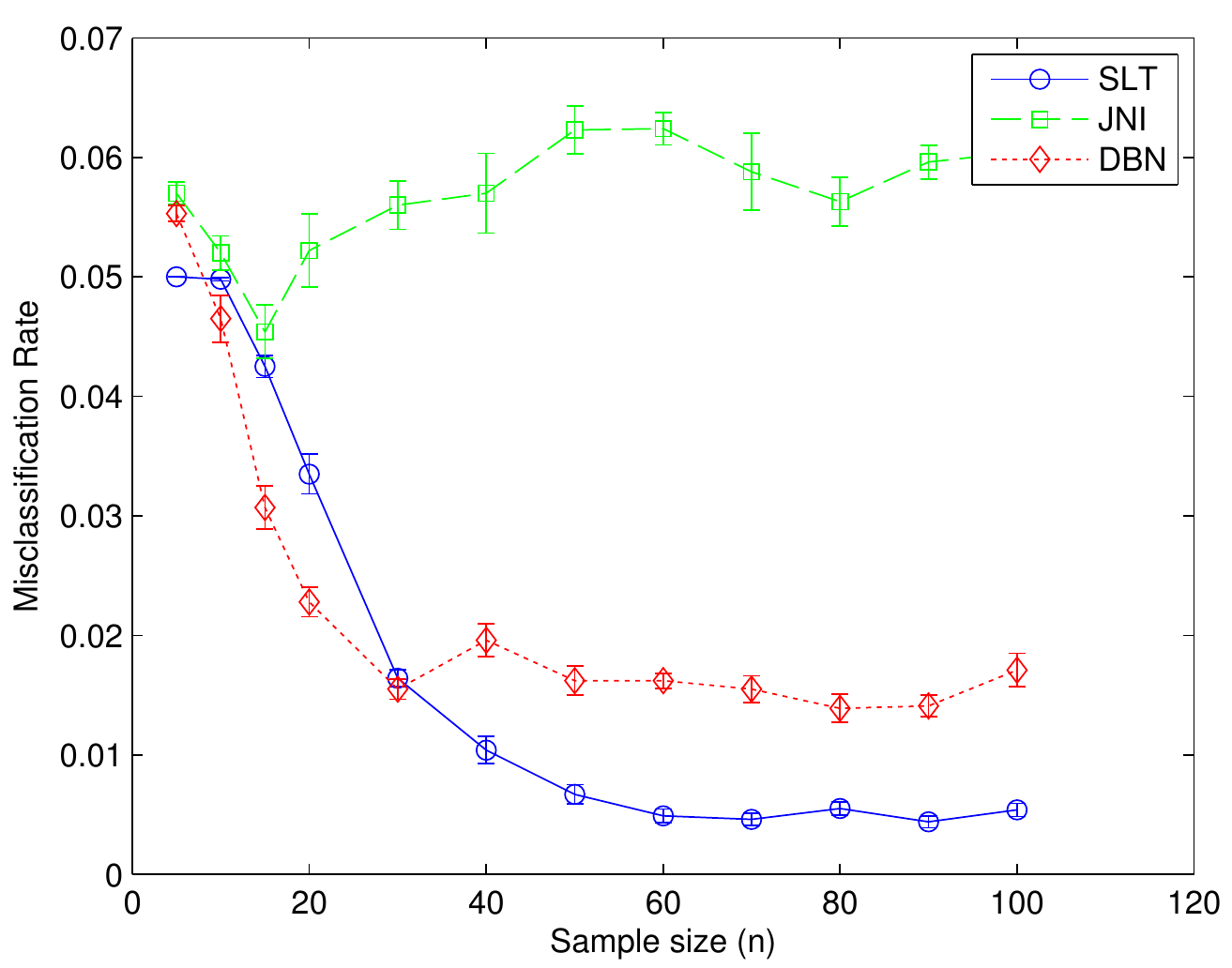}
\includegraphics[width = 0.24\textwidth]{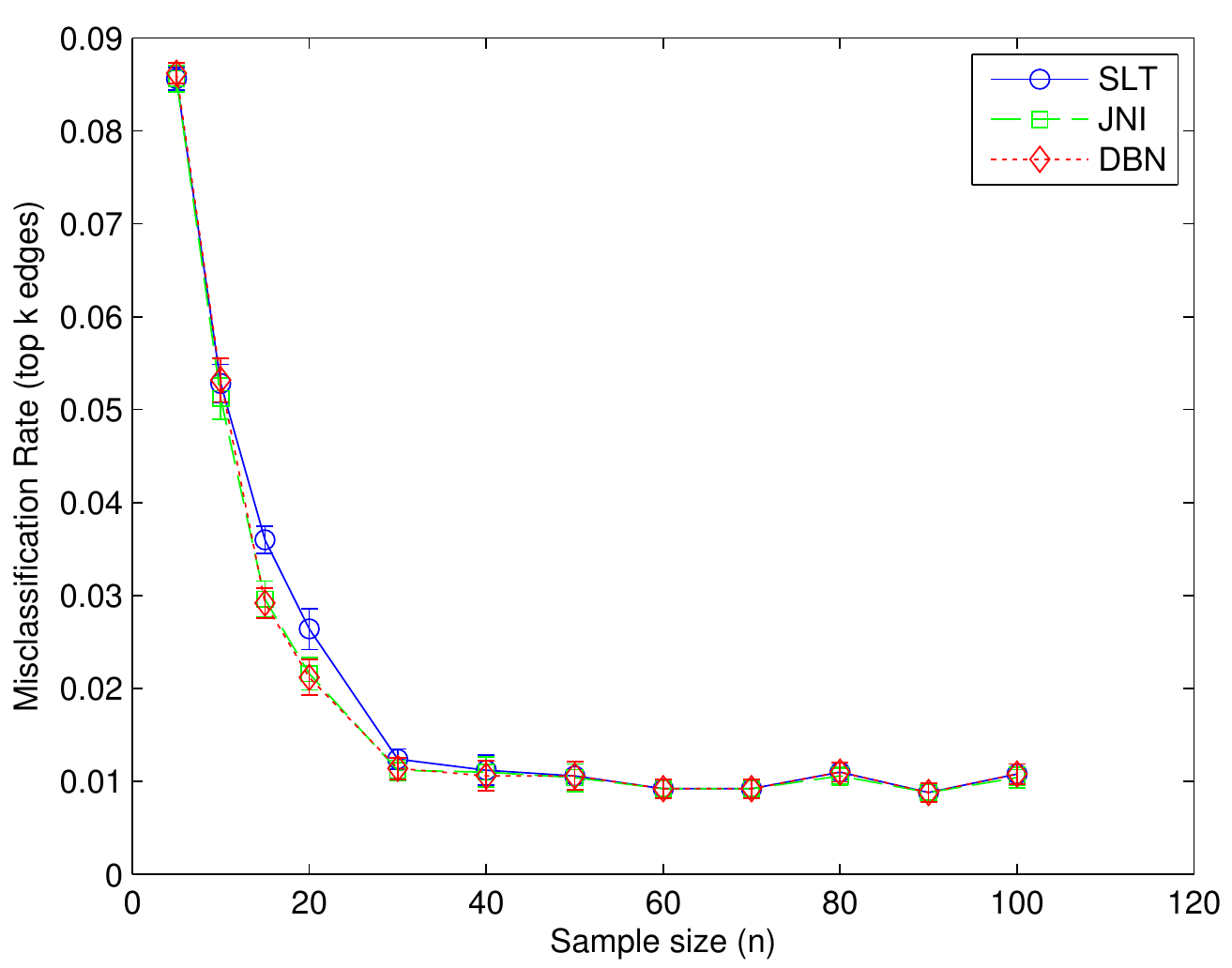}
\includegraphics[width = 0.24\textwidth]{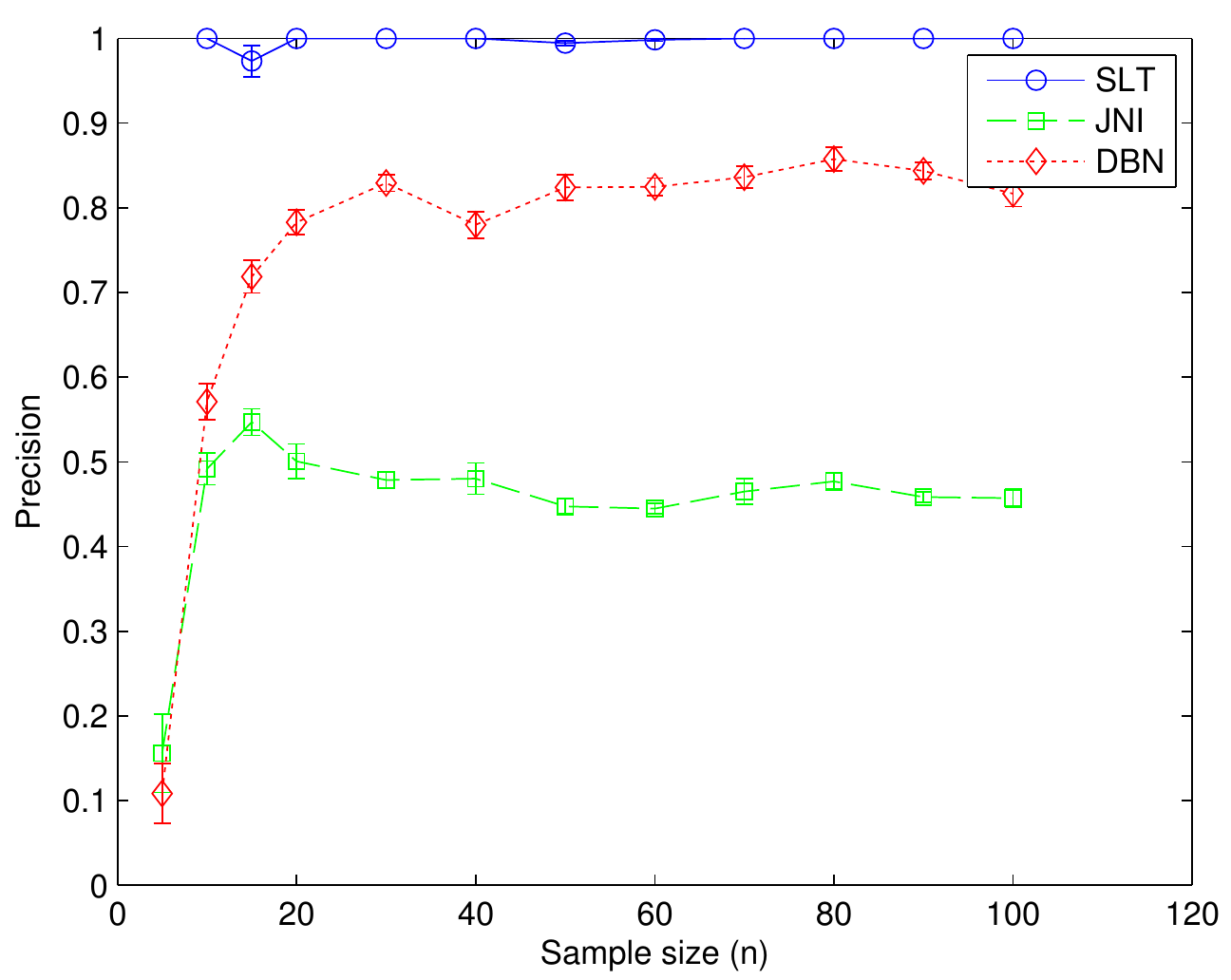}

\includegraphics[width = 0.24\textwidth]{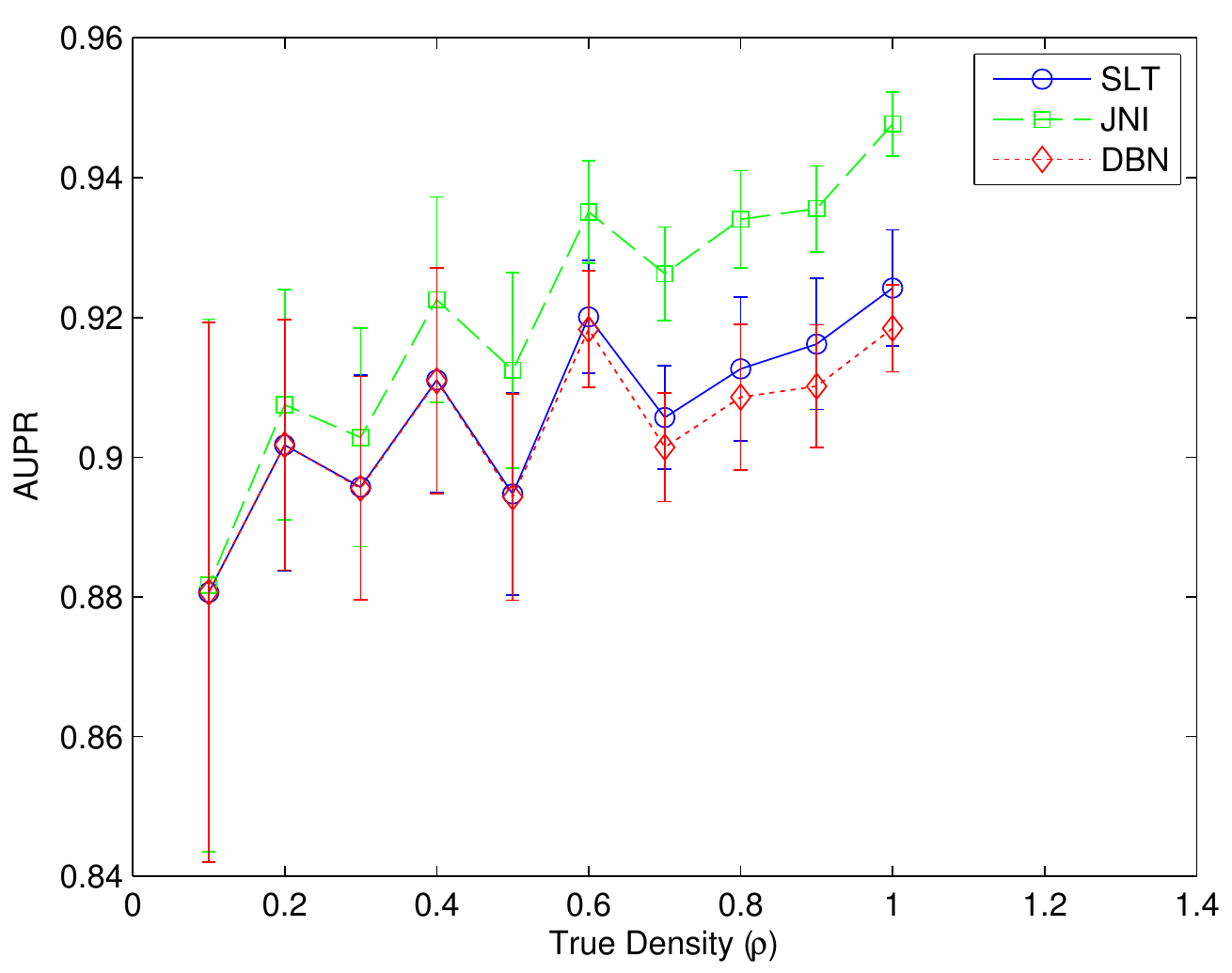}
\includegraphics[width = 0.24\textwidth]{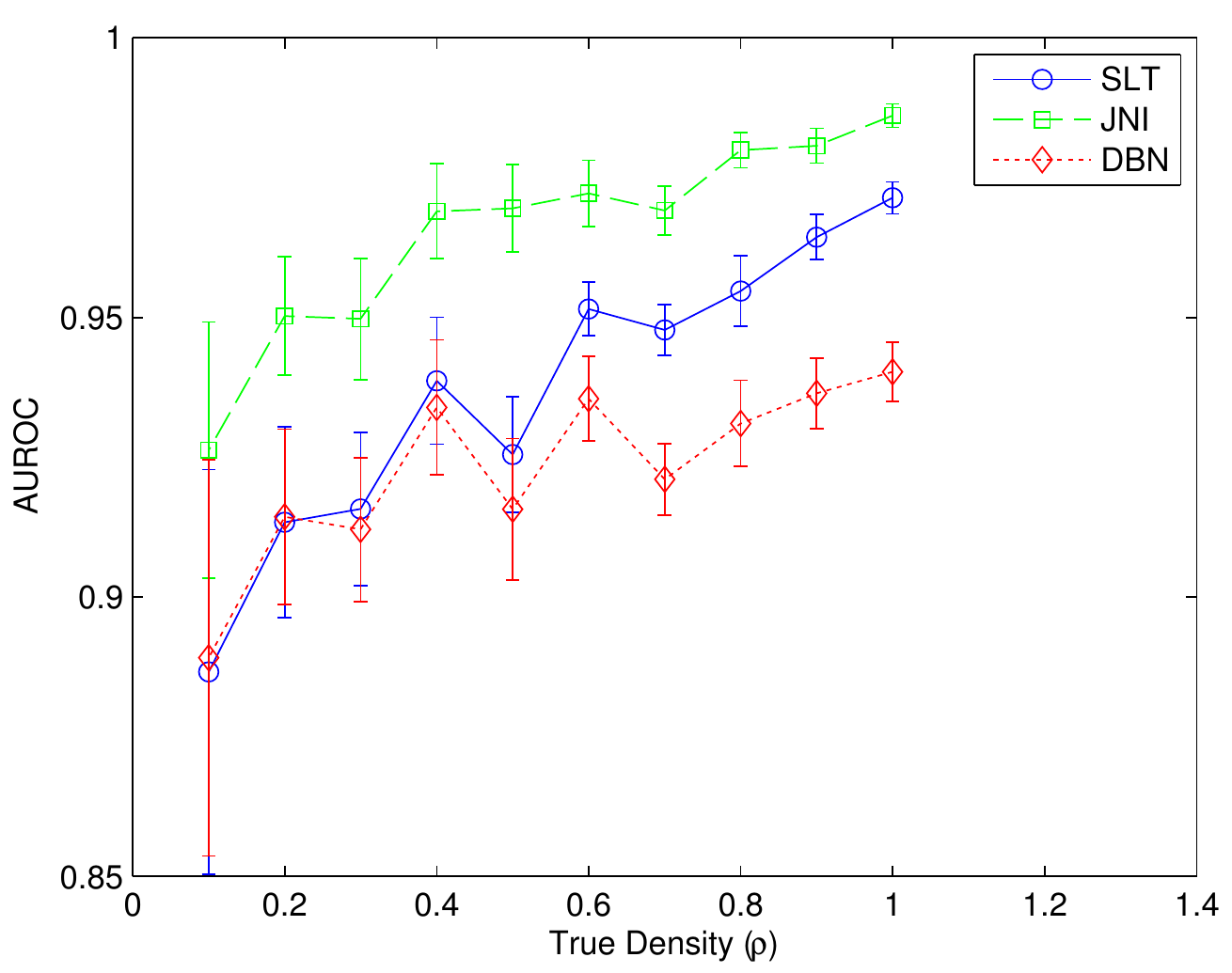}
\includegraphics[width = 0.24\textwidth]{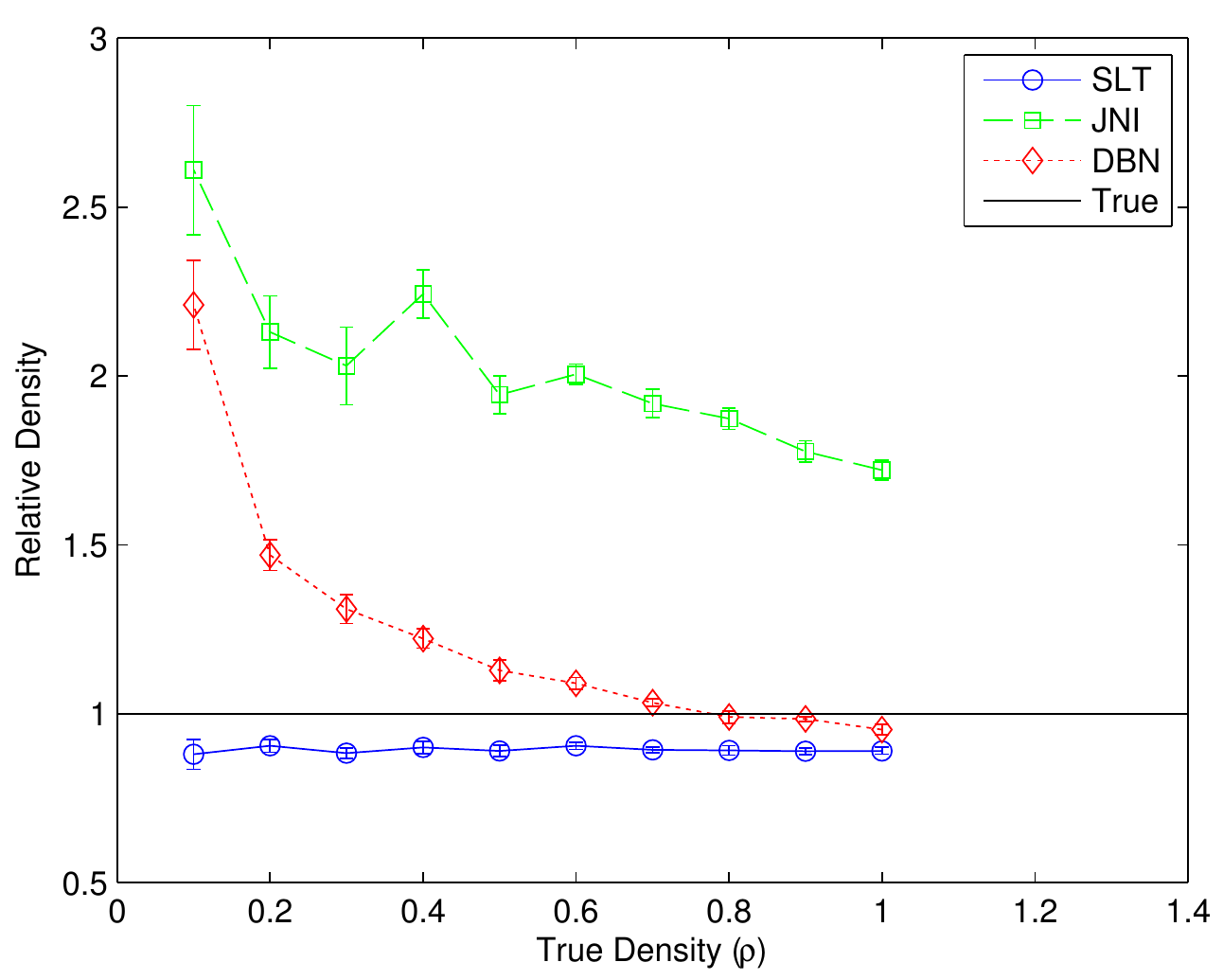}
\includegraphics[width = 0.24\textwidth]{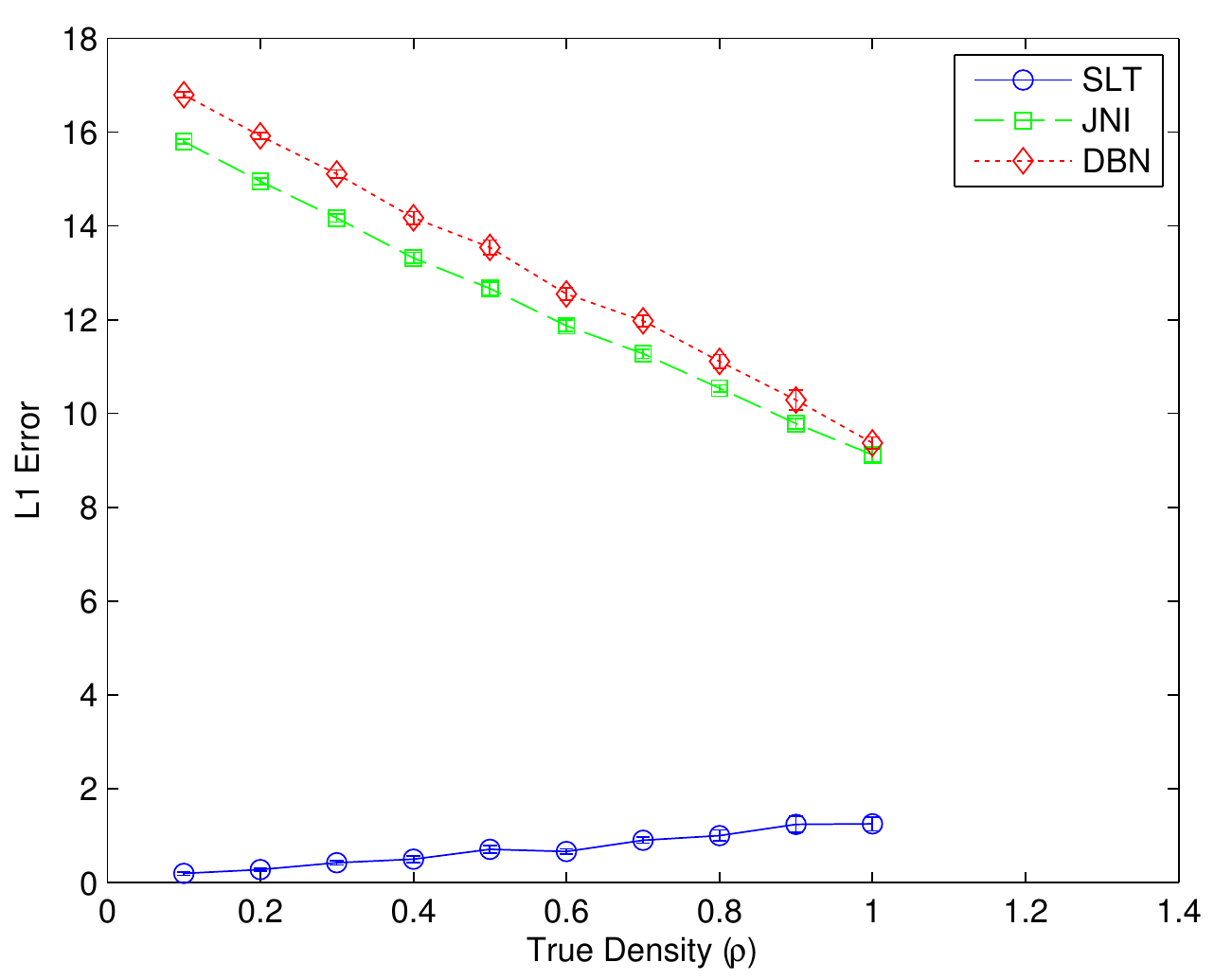}

\includegraphics[width = 0.24\textwidth]{Figures/vary_sparsity_1_Matthews_Correlation_Coefficient.pdf}
\includegraphics[width = 0.24\textwidth]{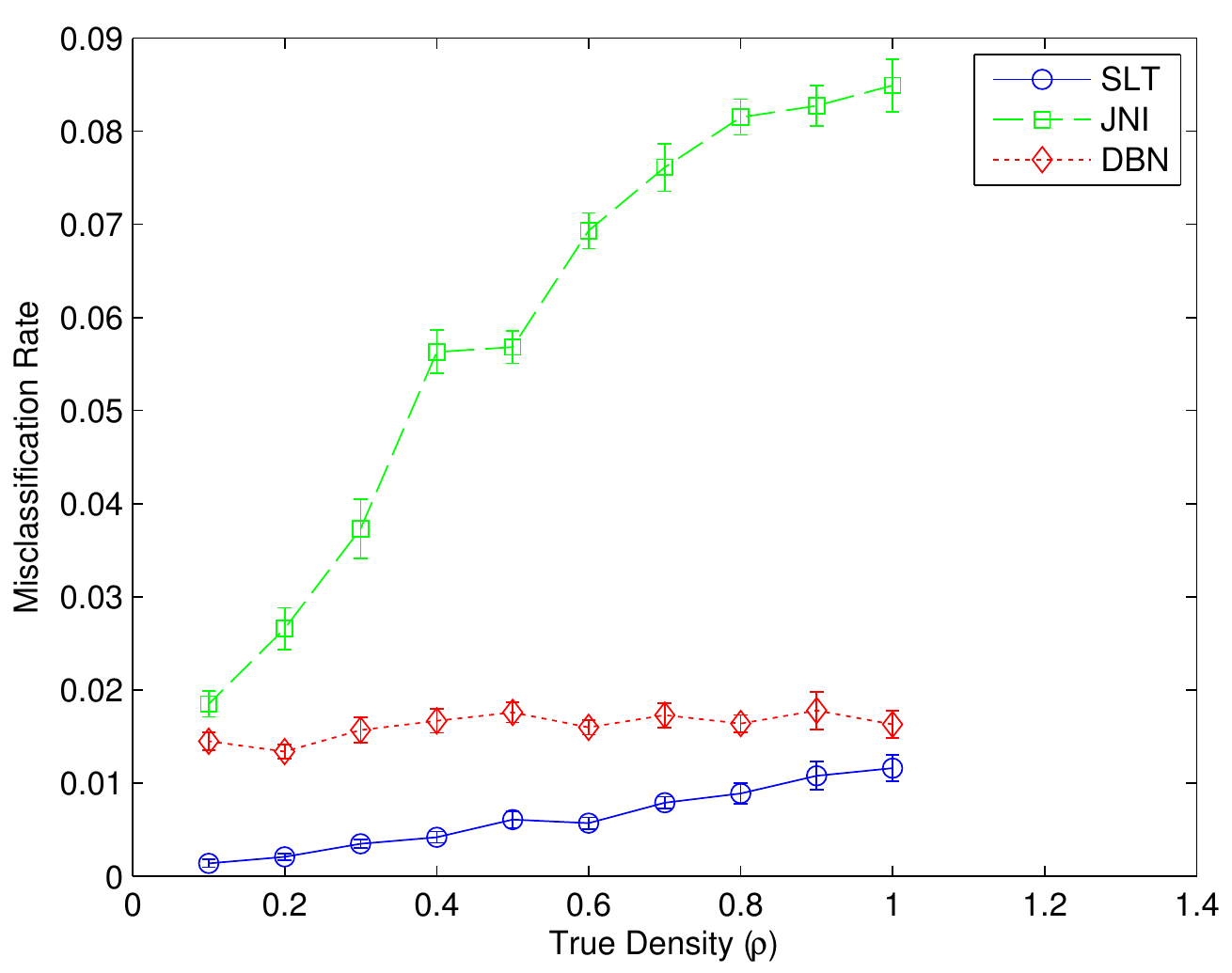}
\includegraphics[width = 0.24\textwidth]{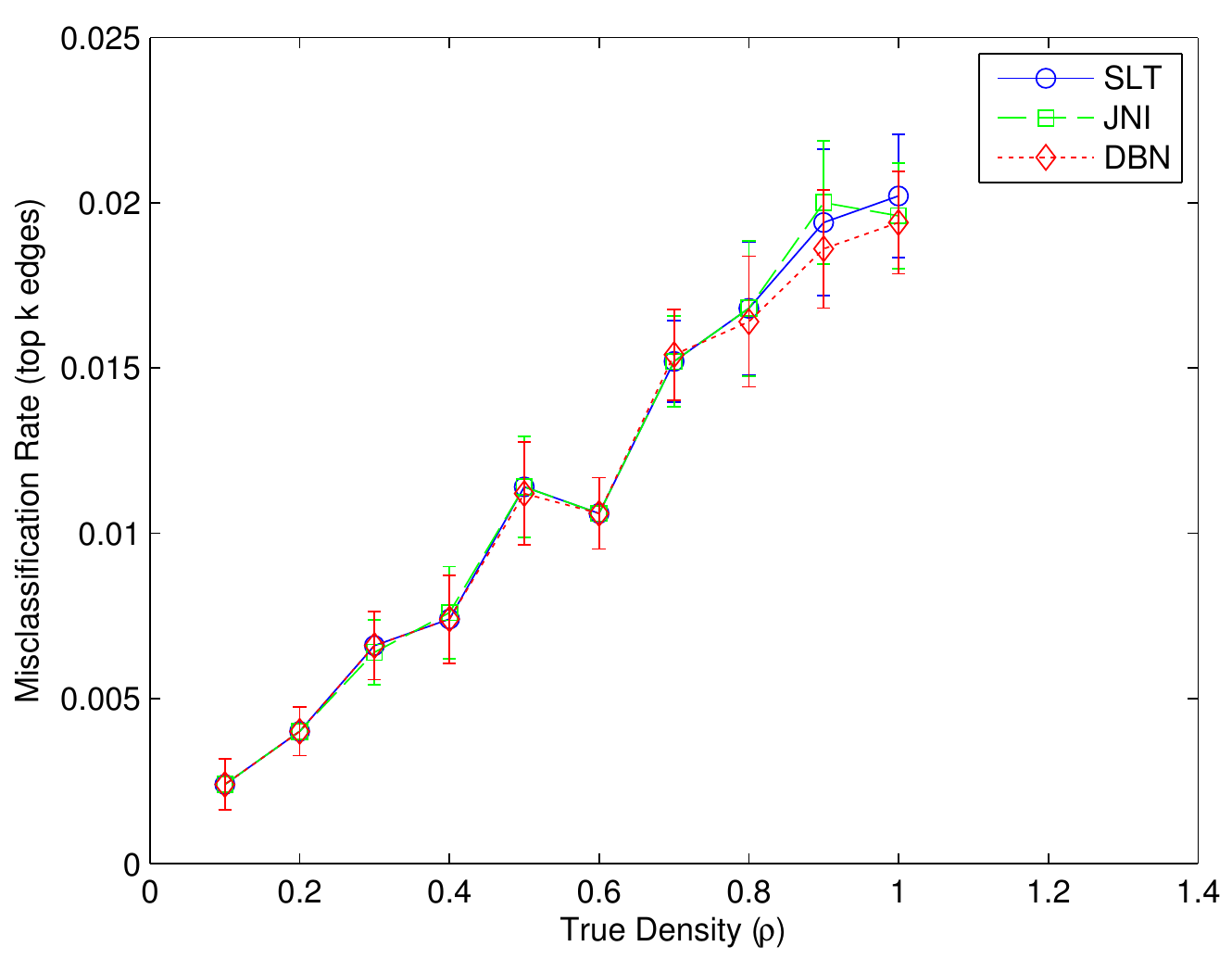}
\includegraphics[width = 0.24\textwidth]{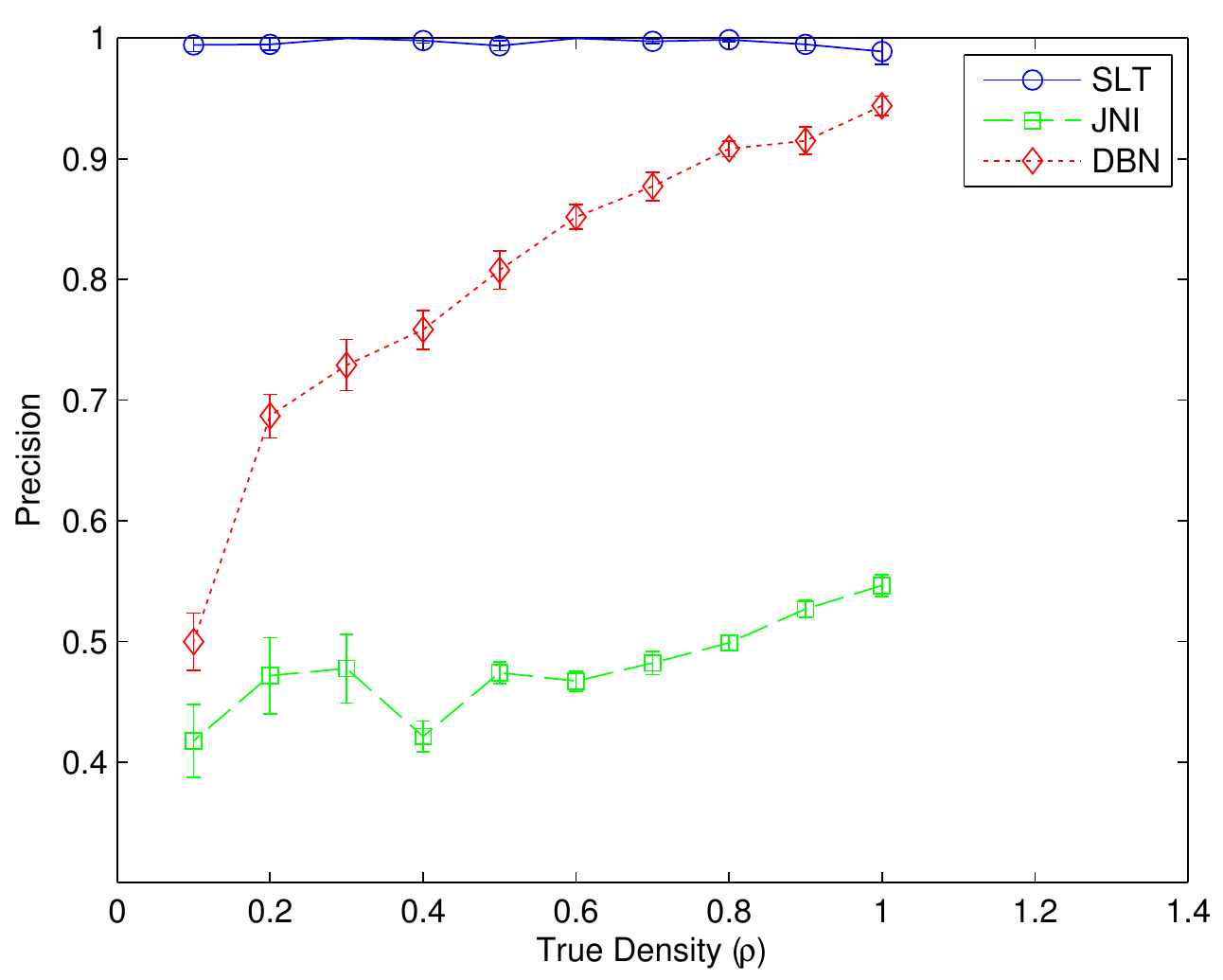}

\caption{Results on simulated data generated from a 2-tier SLT with disjoint sub-group structure. [Network estimators: ``SLT'' = structure learning trees; ``JNI'' = joint network inference; ``DBN'' = inference for each network independently. For each estimator we considered both thresholded and un-thresholded adjacency matrices. Performance scores: ``AUROC'' = area under the receiver operating characteristic curve; ``AUPR'' = area under the precision-recall curve; ``L1 Error'' = $\ell_1$ distance from the true adjacency matrices to the inferred weighted adjacency matrices; ``top k edges'' = the $\rho P$ most probable edges. Performance scores were averaged over all 10 data-generating networks and all 10 datasets; error bars denote standard errors of mean performance over datasets. We considered both varying $n$ for fixed $\rho = 0.5$ and varying $\rho$ for fixed $n=60$.]}
\label{sim res1}
\end{figure*}

Intuitively, SLT should provide an advantage over JNI when the data structure contains distinct sub-groups with respect to network topology.
Fig. \ref{ex1} displays typical inferences in the ``disjoint sub-group'' regime (1) when $n = 60$, $\rho = 1/2$; SLT is noticeably sparser than JNI and DBN whilst achieving high MCC (Fig. \ref{reg1}) and essentially perfect precision (SFig. \ref{sim res1}).
As a consequence, over all sample sizes $n$ which we considered, SLT is considerably closer than JNI and DBN to the true network structures in the $\ell_1$ norm (Fig. \ref{reg1}).
This ability to generate a clear decision boundary in the posterior is not demonstrated by JNI and DBN, which produce less sparse matrices of posterior inclusion probabilities (Fig. \ref{reg}). 
This is expected, since JNI erroneously shares information \emph{equally} among all networks, whilst DBN is statistically inefficient and therefore subject to higher variance.

\begin{figure*}[t!]
\centering
\includegraphics[width = 0.24\textwidth]{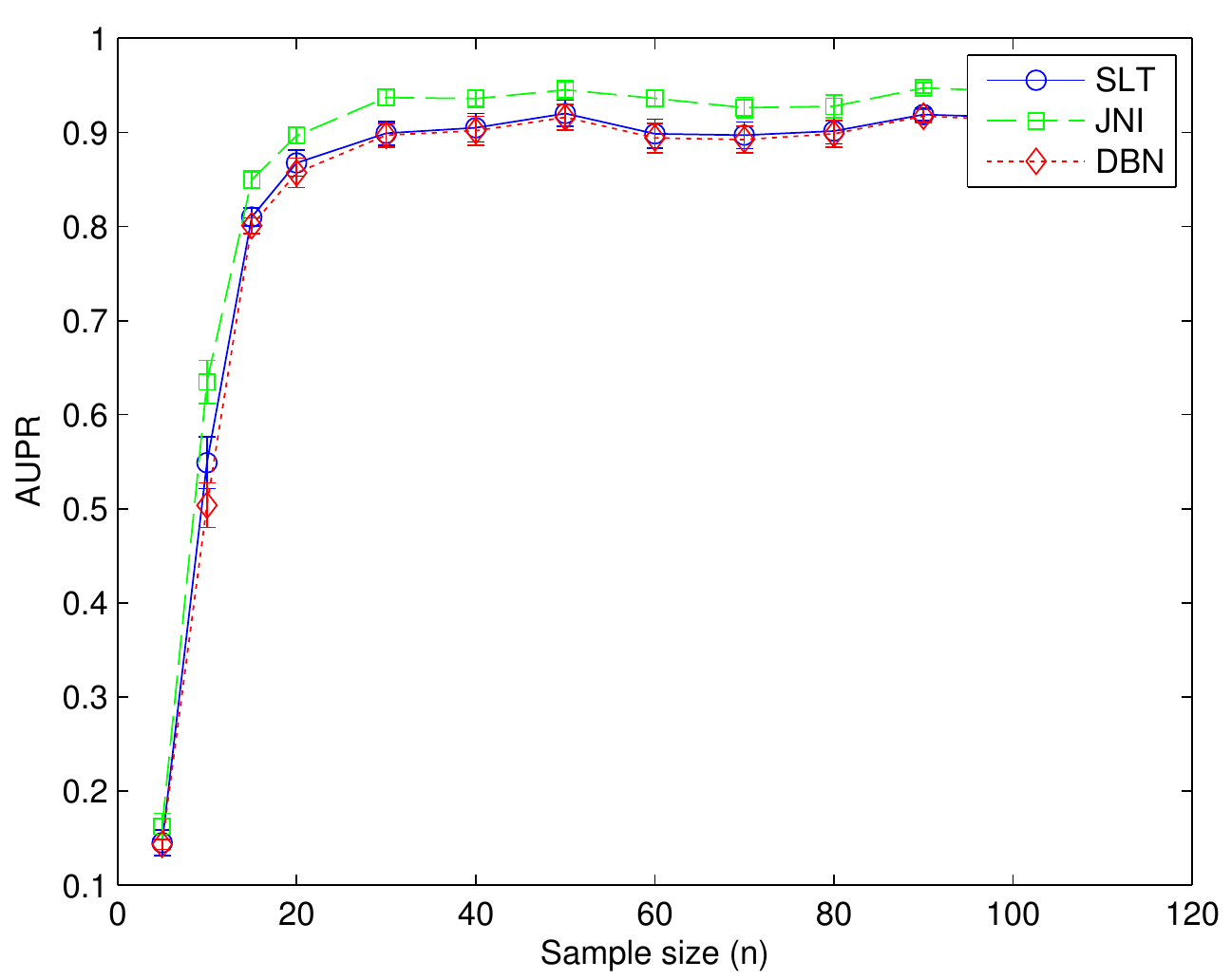}
\includegraphics[width = 0.24\textwidth]{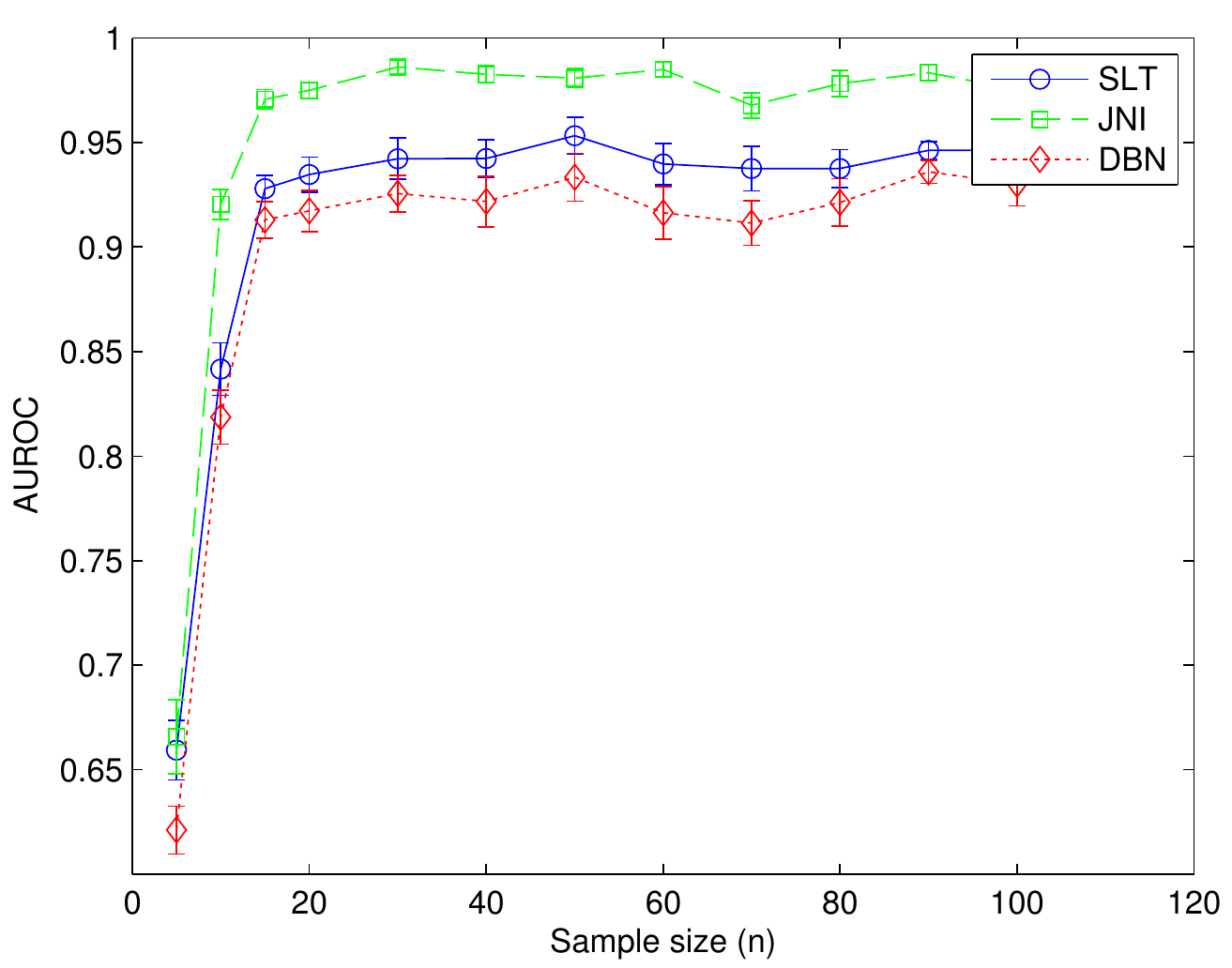}
\includegraphics[width = 0.24\textwidth]{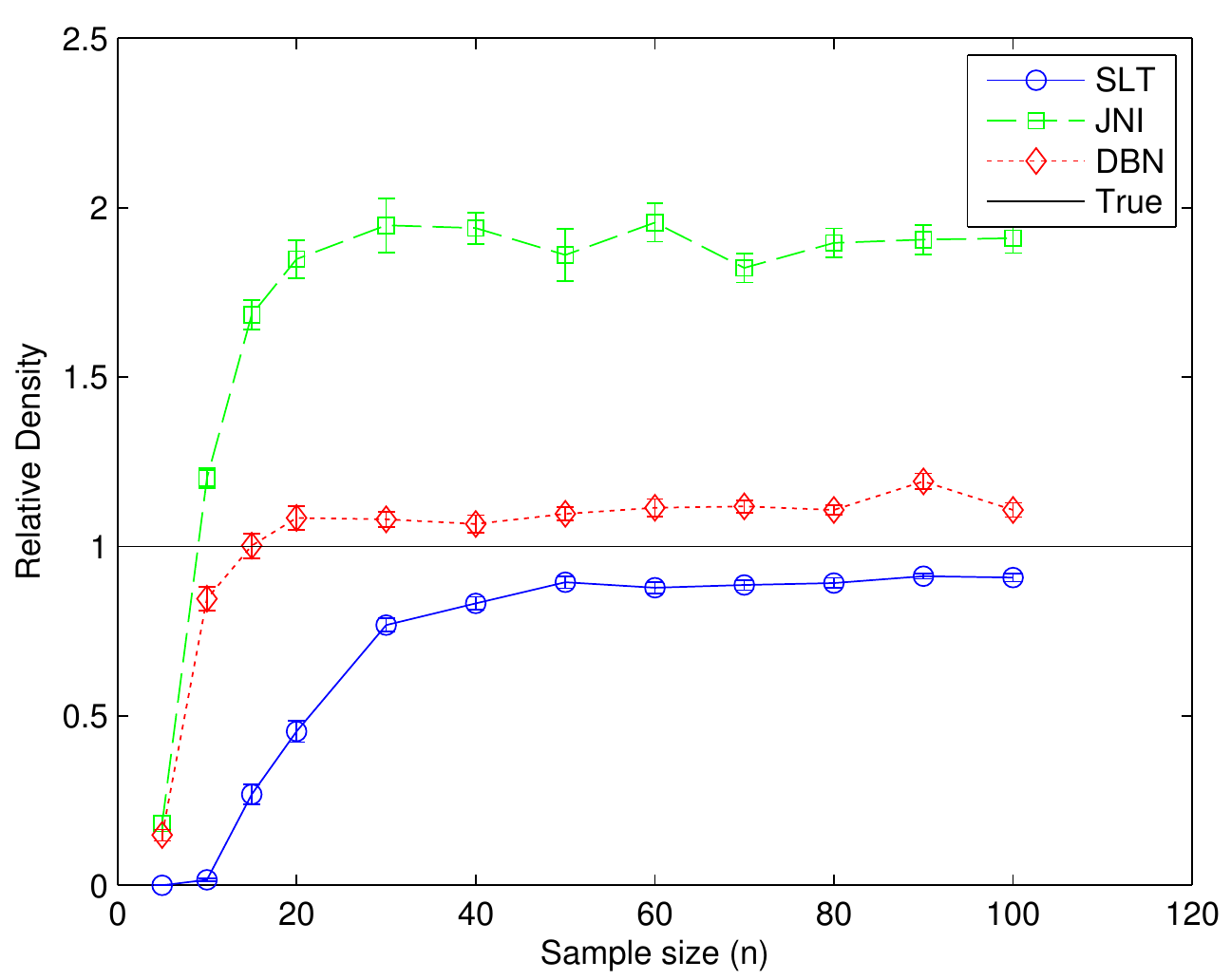}
\includegraphics[width = 0.24\textwidth]{Figures/vary_n_2_L1_Error.pdf}

\includegraphics[width = 0.24\textwidth]{Figures/vary_n_2_Matthews_Correlation_Coefficient.pdf}
\includegraphics[width = 0.24\textwidth]{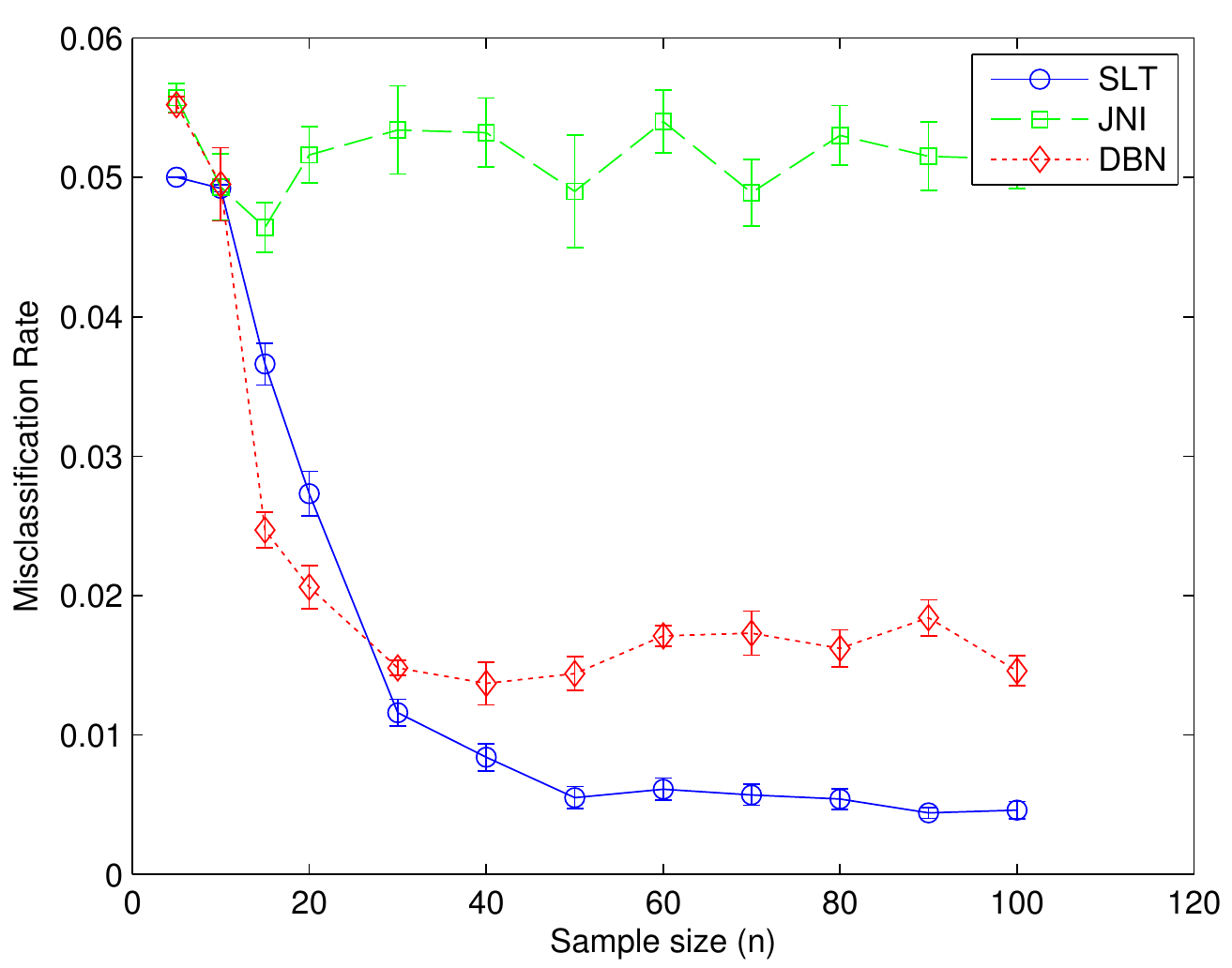}
\includegraphics[width = 0.24\textwidth]{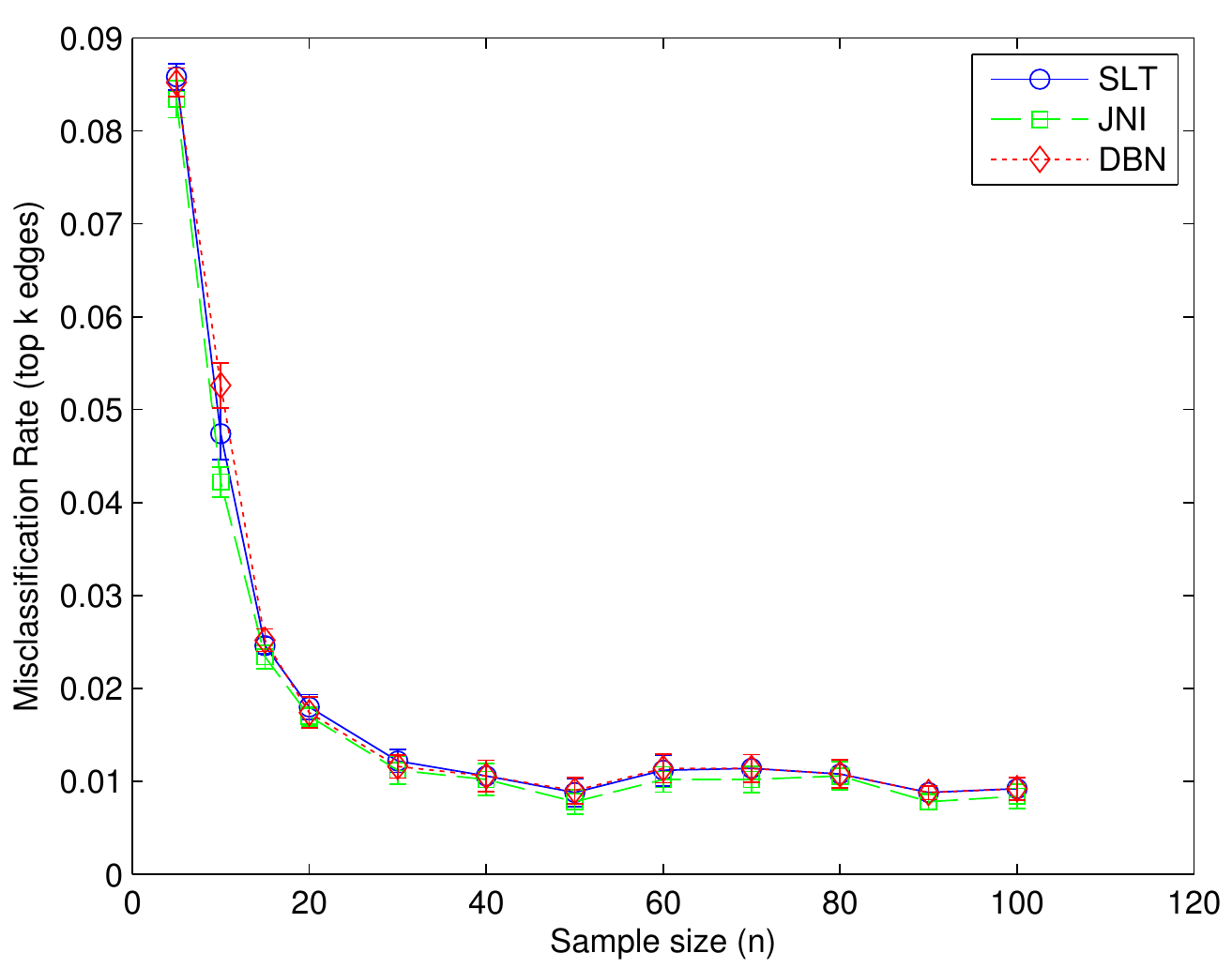}
\includegraphics[width = 0.24\textwidth]{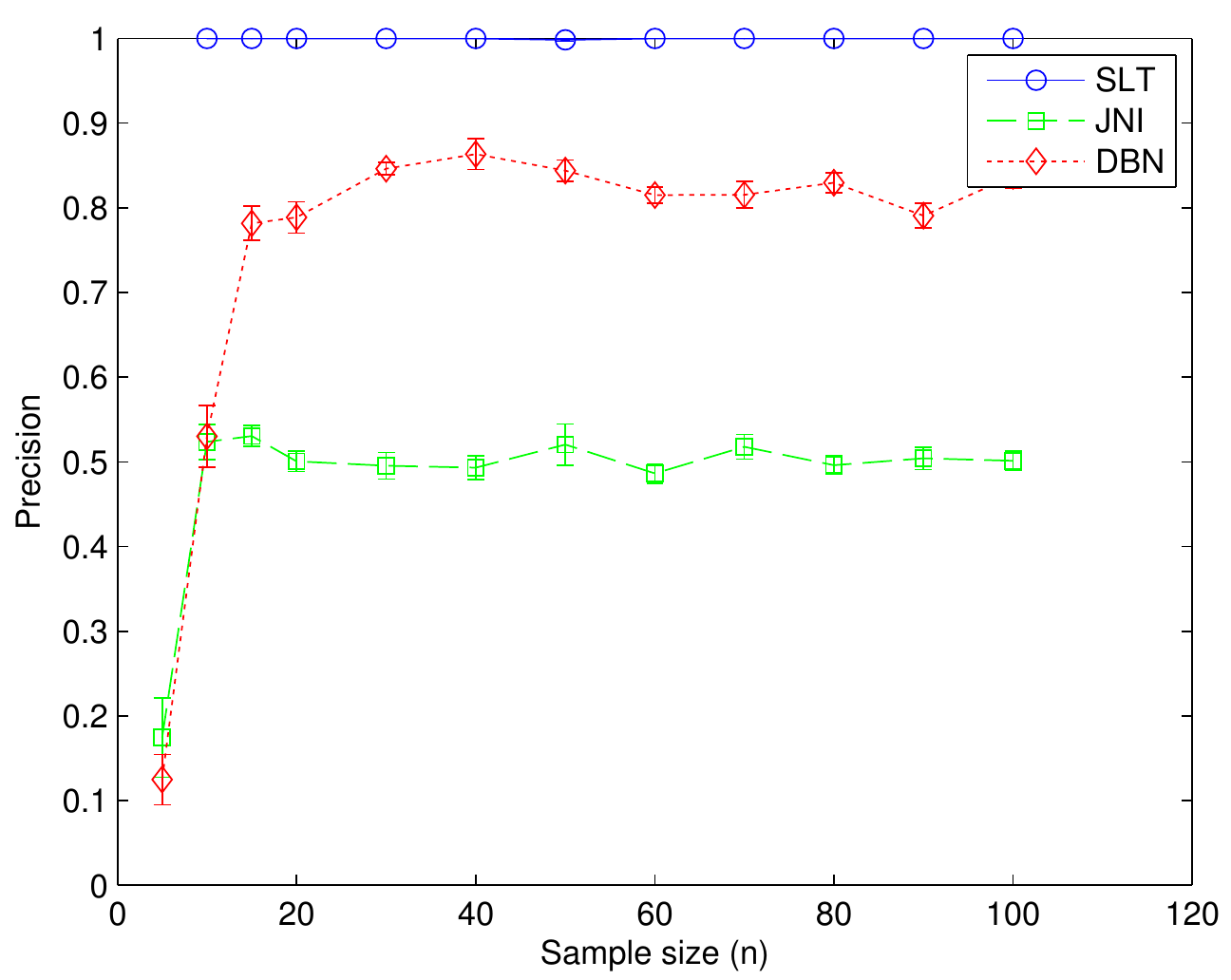}

\includegraphics[width = 0.24\textwidth]{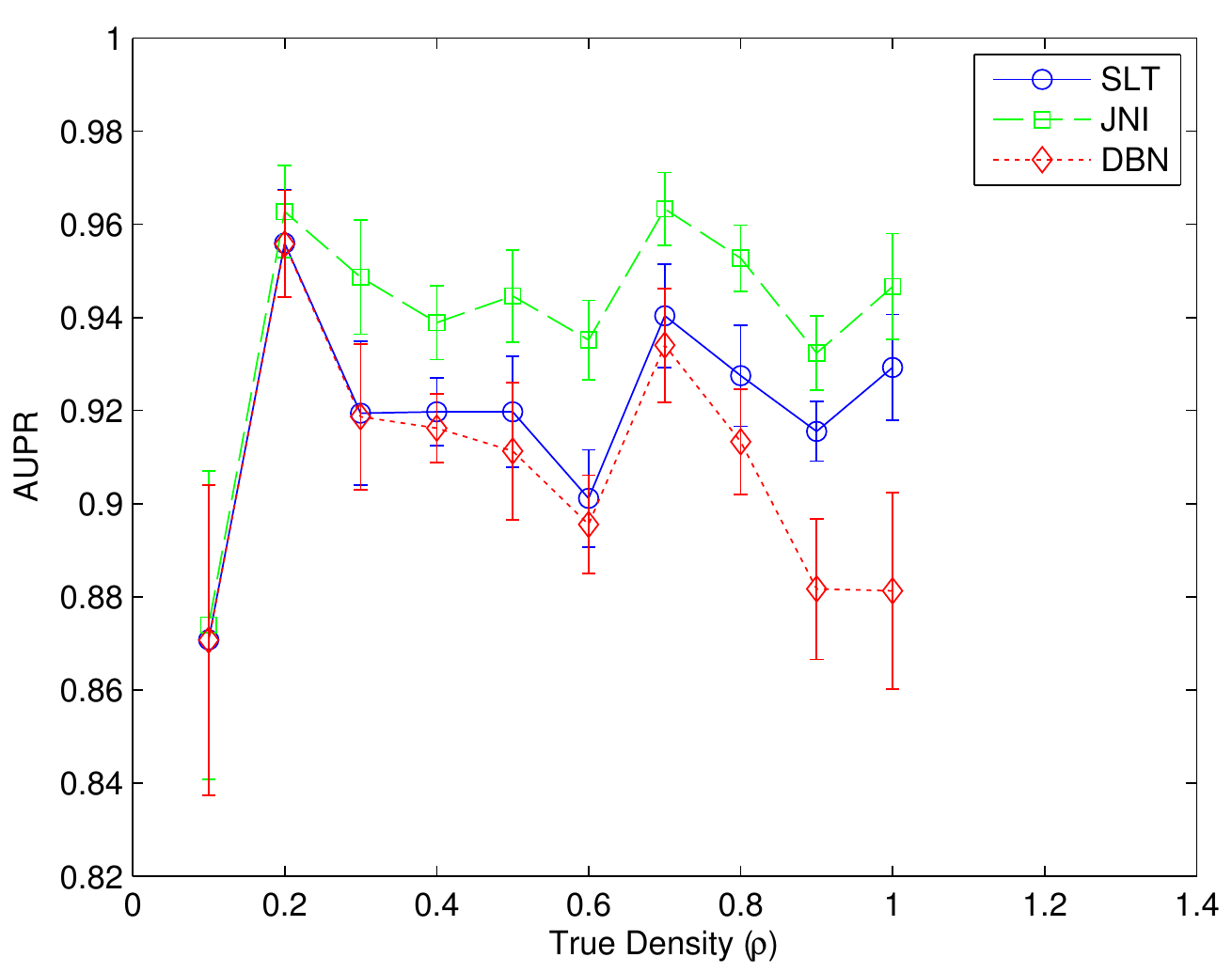}
\includegraphics[width = 0.24\textwidth]{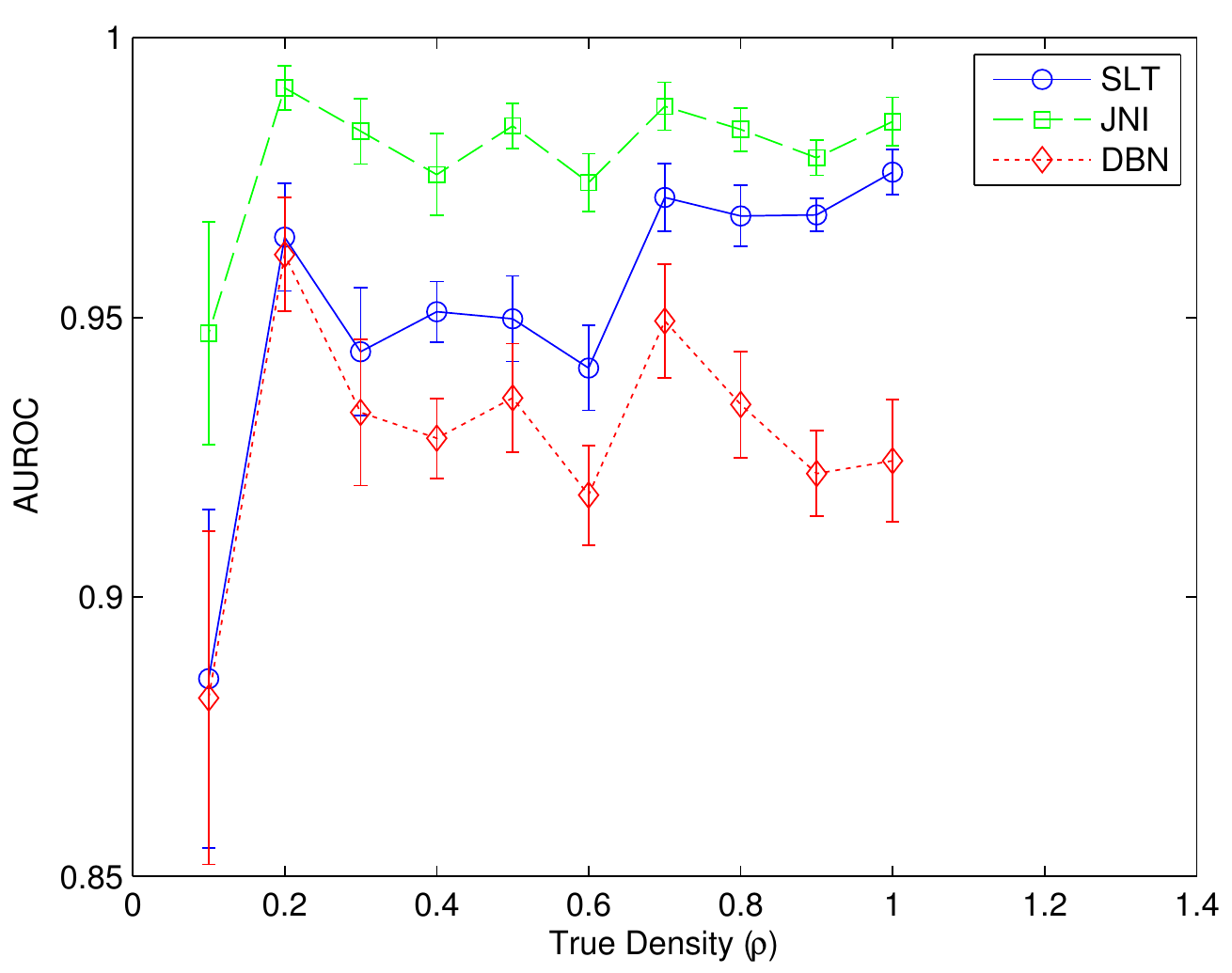}
\includegraphics[width = 0.24\textwidth]{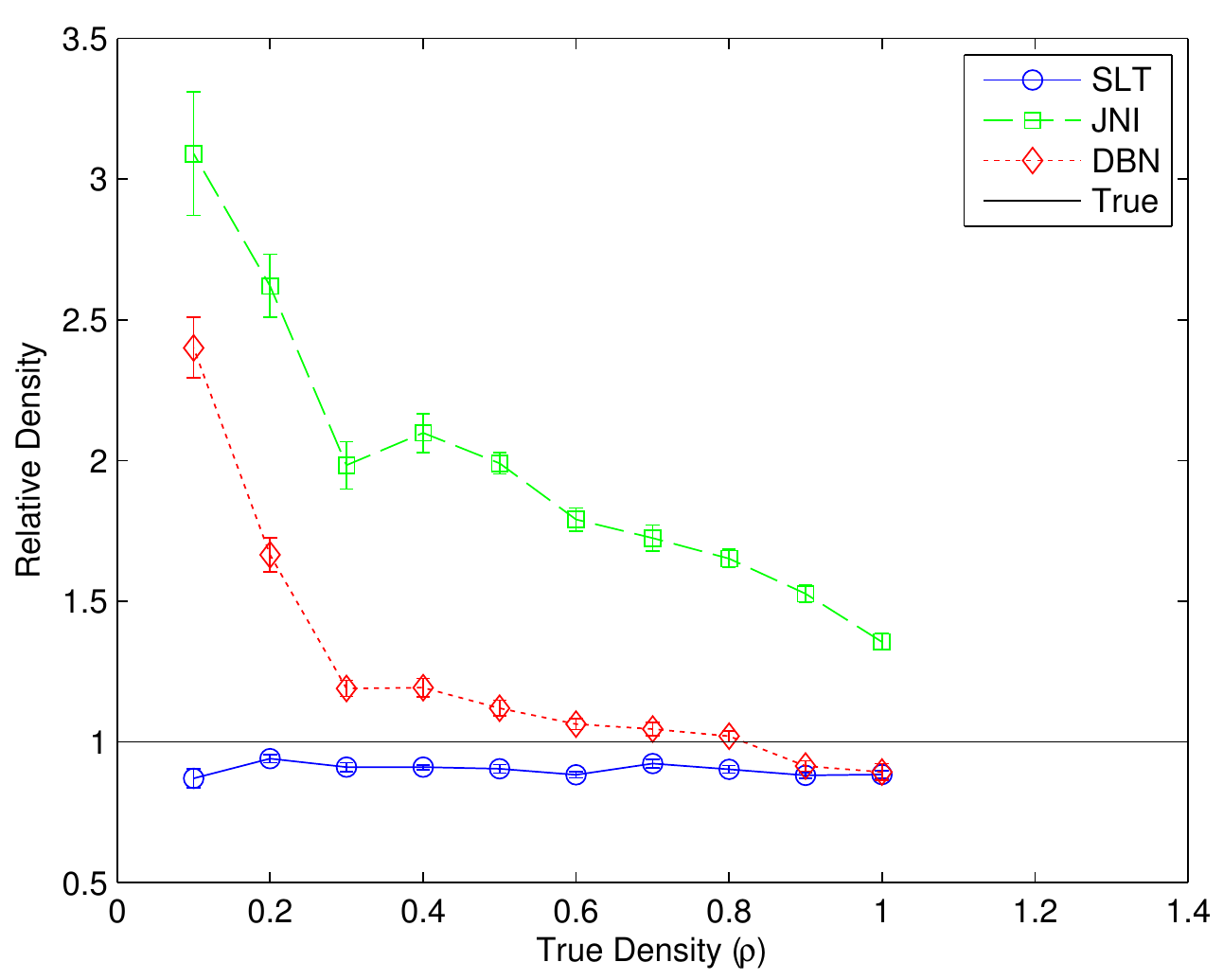}
\includegraphics[width = 0.24\textwidth]{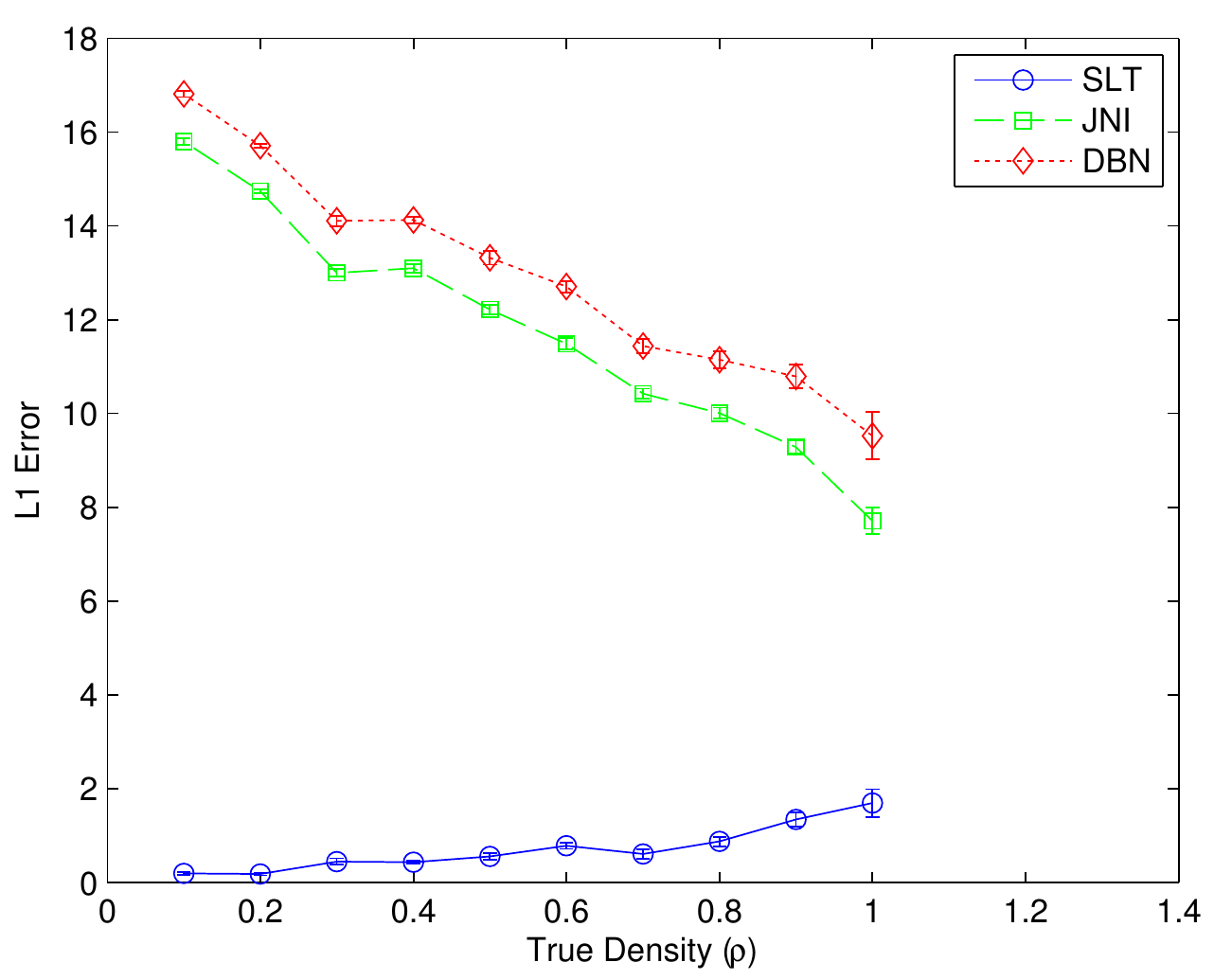}

\includegraphics[width = 0.24\textwidth]{Figures/vary_sparsity_2_Matthews_Correlation_Coefficient.pdf}
\includegraphics[width = 0.24\textwidth]{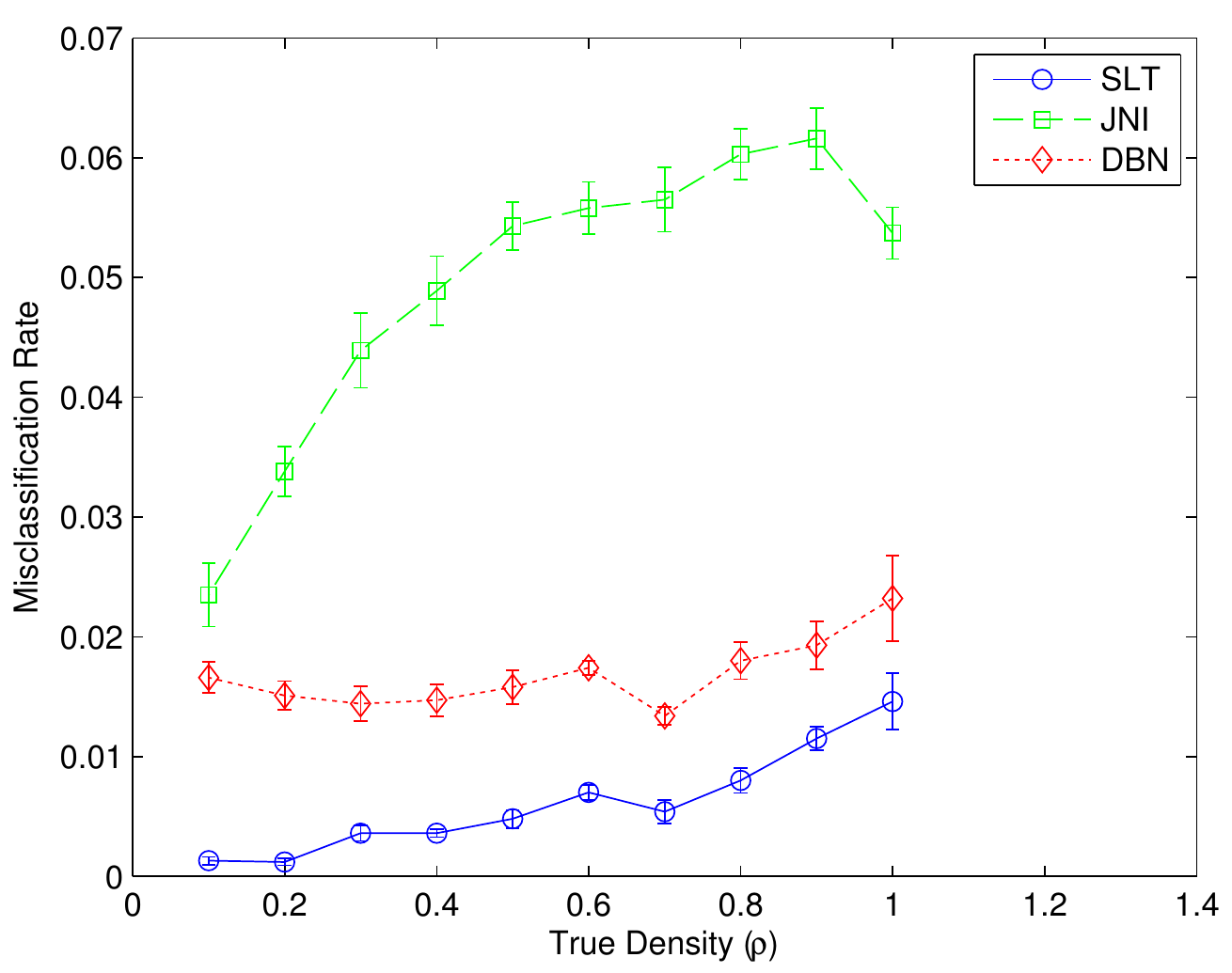}
\includegraphics[width = 0.24\textwidth]{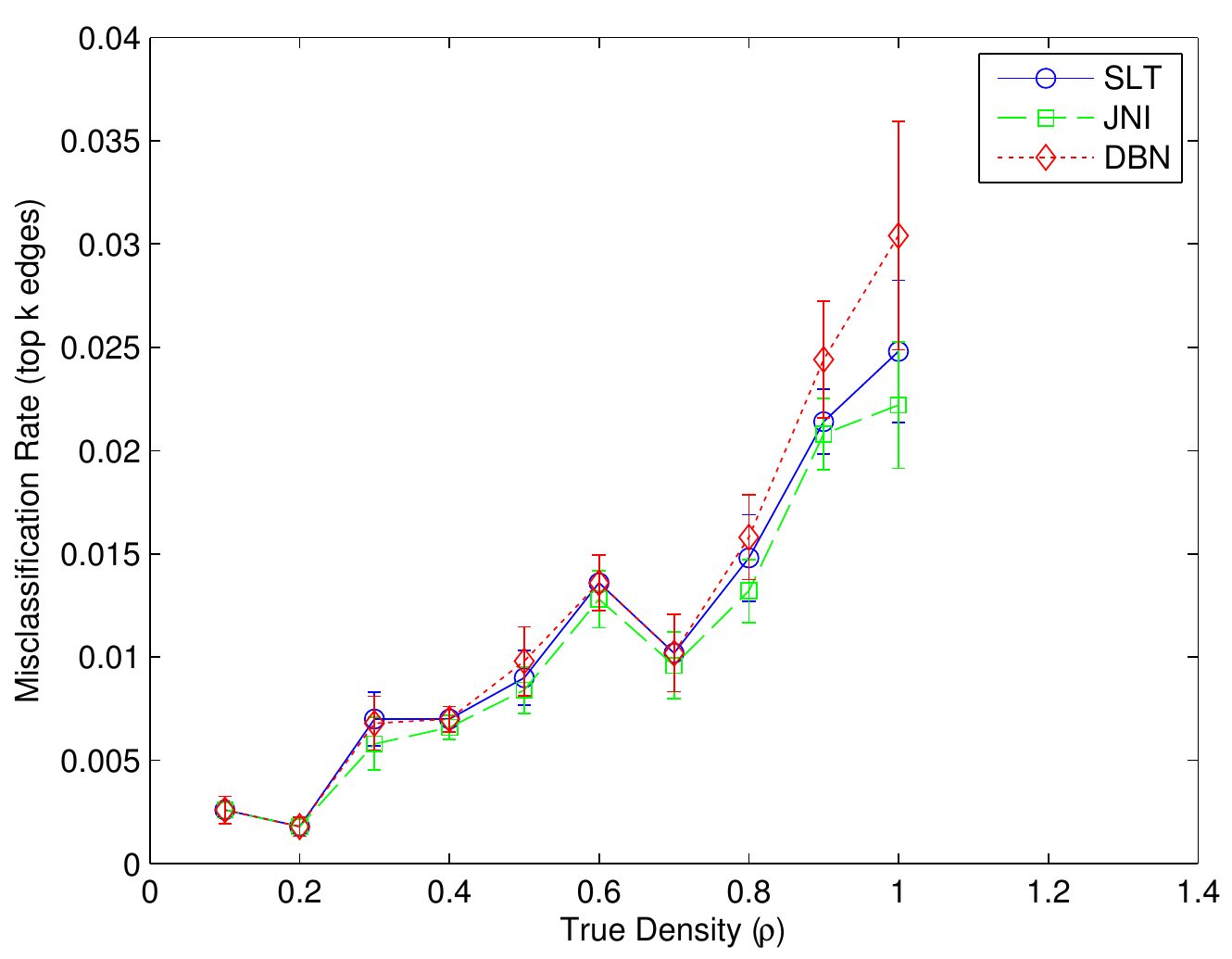}
\includegraphics[width = 0.24\textwidth]{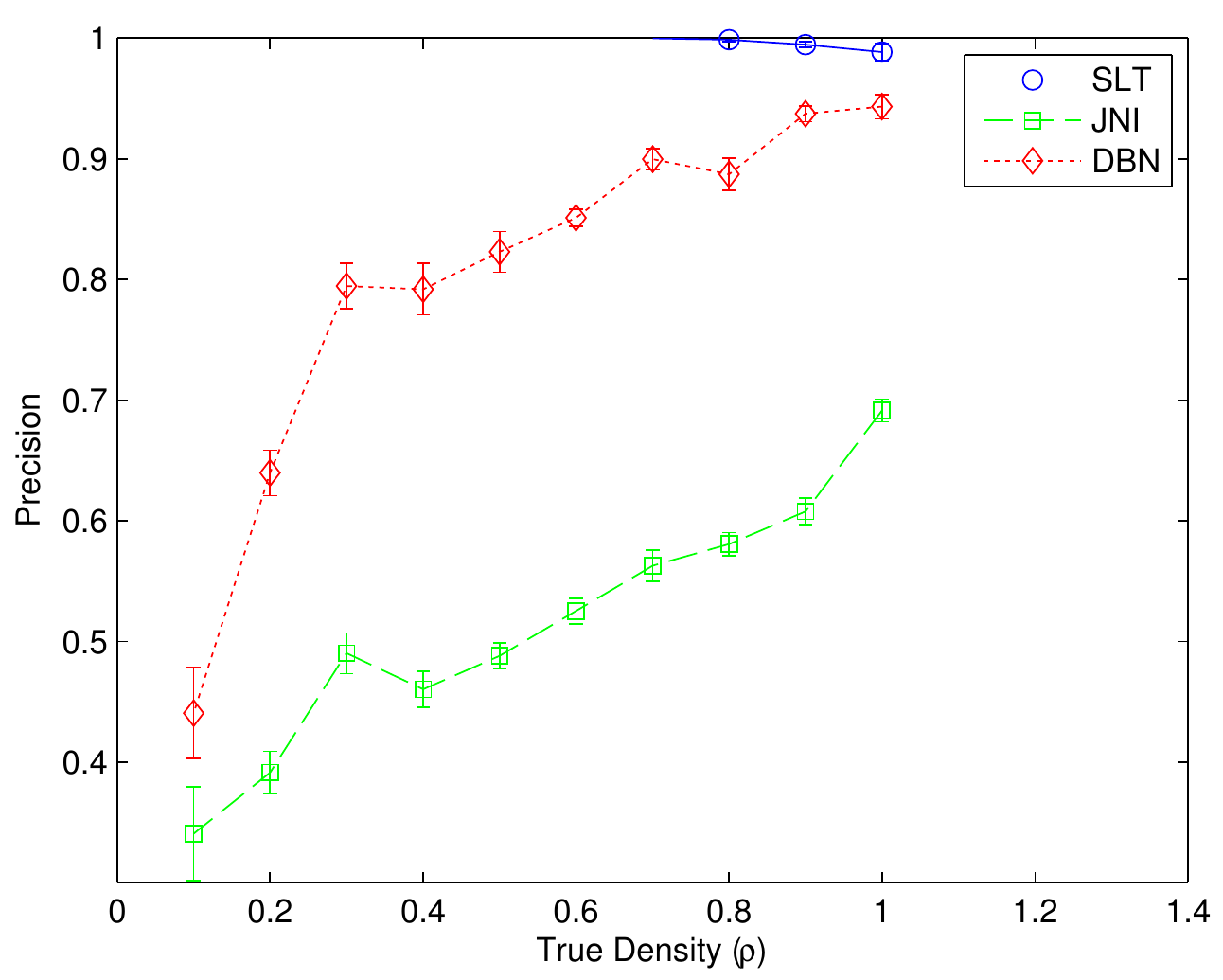}

\caption{Results on simulated data generated from a 2-tier SLT with weakly exchangeable structure [Network estimators: ``SLT'' = structure learning trees; ``JNI'' = joint network inference; ``DBN'' = inference for each network independently. For each estimator we considered both thresholded and un-thresholded adjacency matrices. Performance scores: ``AUROC'' = area under the receiver operating characteristic curve; ``AUPR'' = area under the precision-recall curve; ``L1 Error'' = $\ell_1$ distance from the true adjacency matrices to the inferred weighted adjacency matrices; ``top k edges'' = the $\rho P$ most probable edges. Performance scores were averaged over all 10 data-generating networks and all 10 datasets; error bars denote standard errors of mean performance over datasets. We considered both varying $n$ for fixed $\rho = 0.5$ and varying $\rho$ for fixed $n=60$.]}
\label{sim res2}
\end{figure*}

\begin{figure*}[t!]
\centering
\includegraphics[width = 0.24\textwidth]{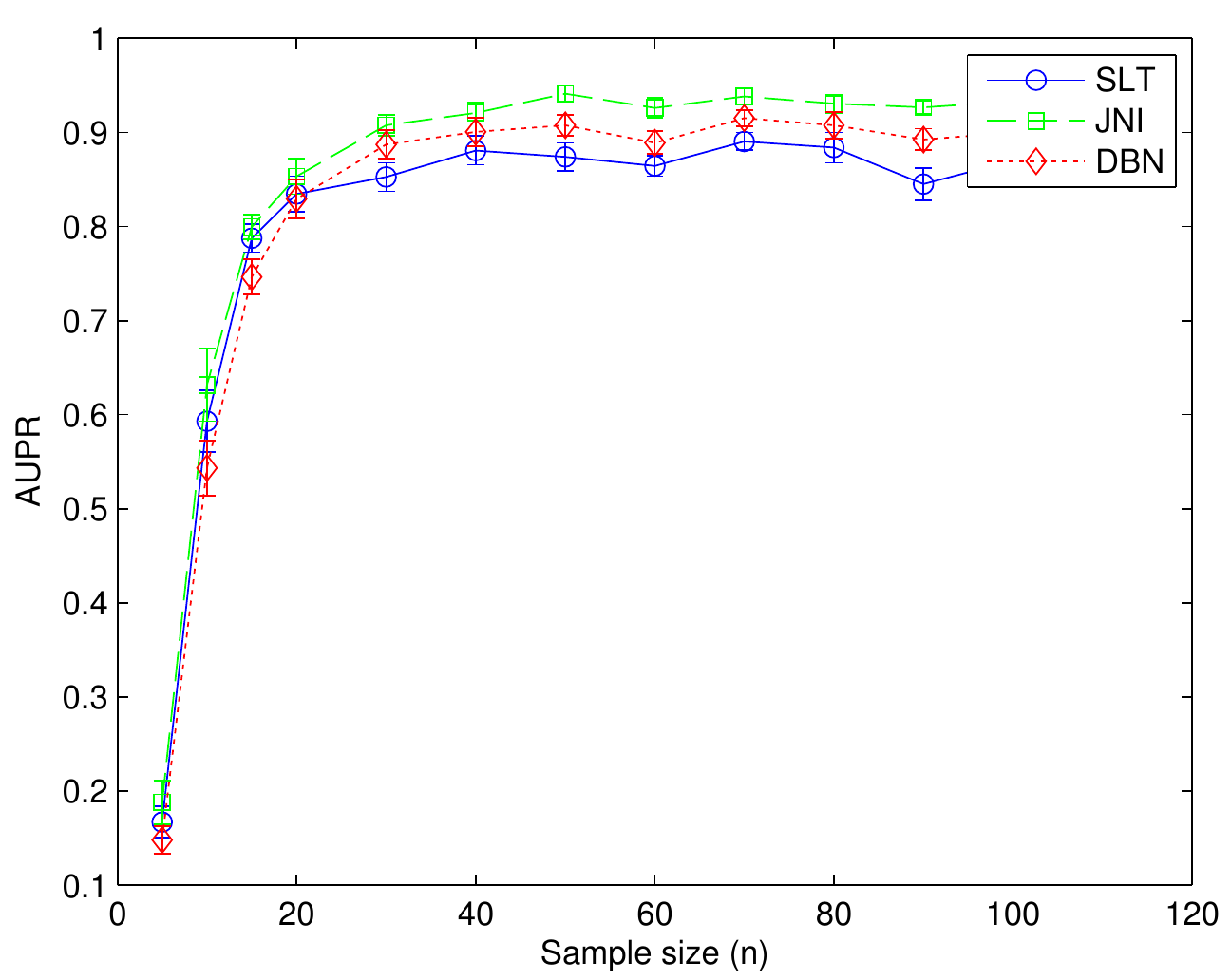}
\includegraphics[width = 0.24\textwidth]{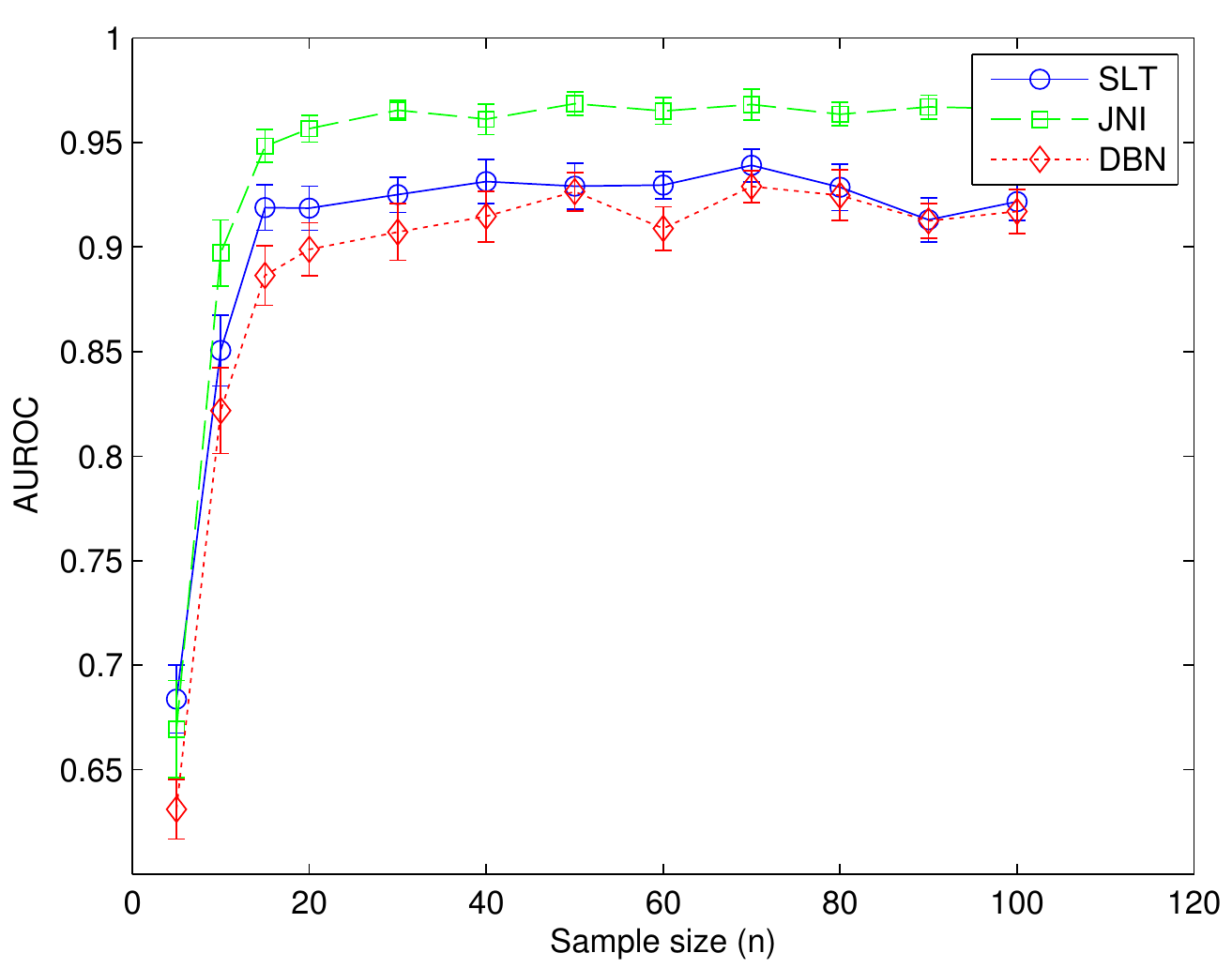}
\includegraphics[width = 0.24\textwidth]{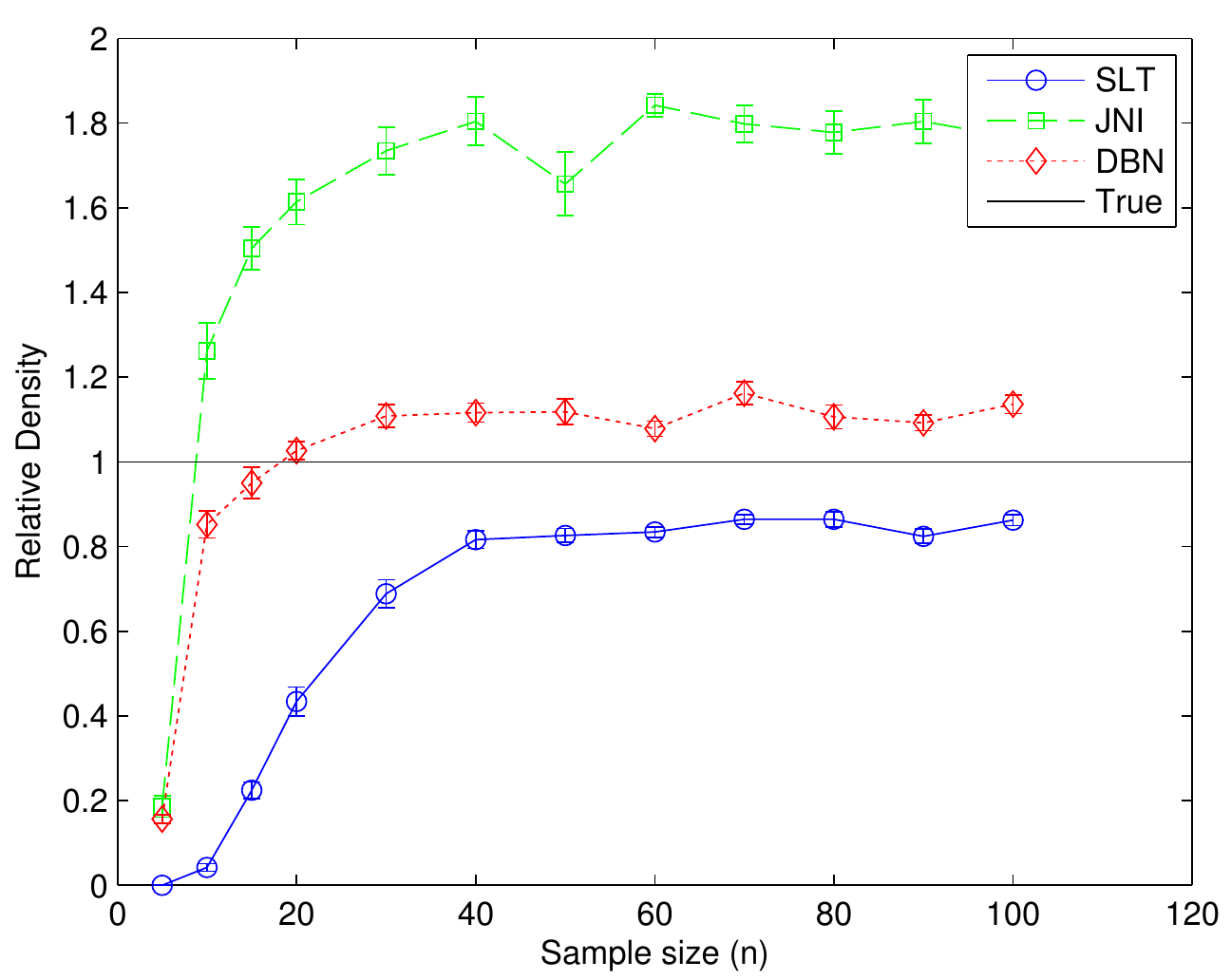}
\includegraphics[width = 0.24\textwidth]{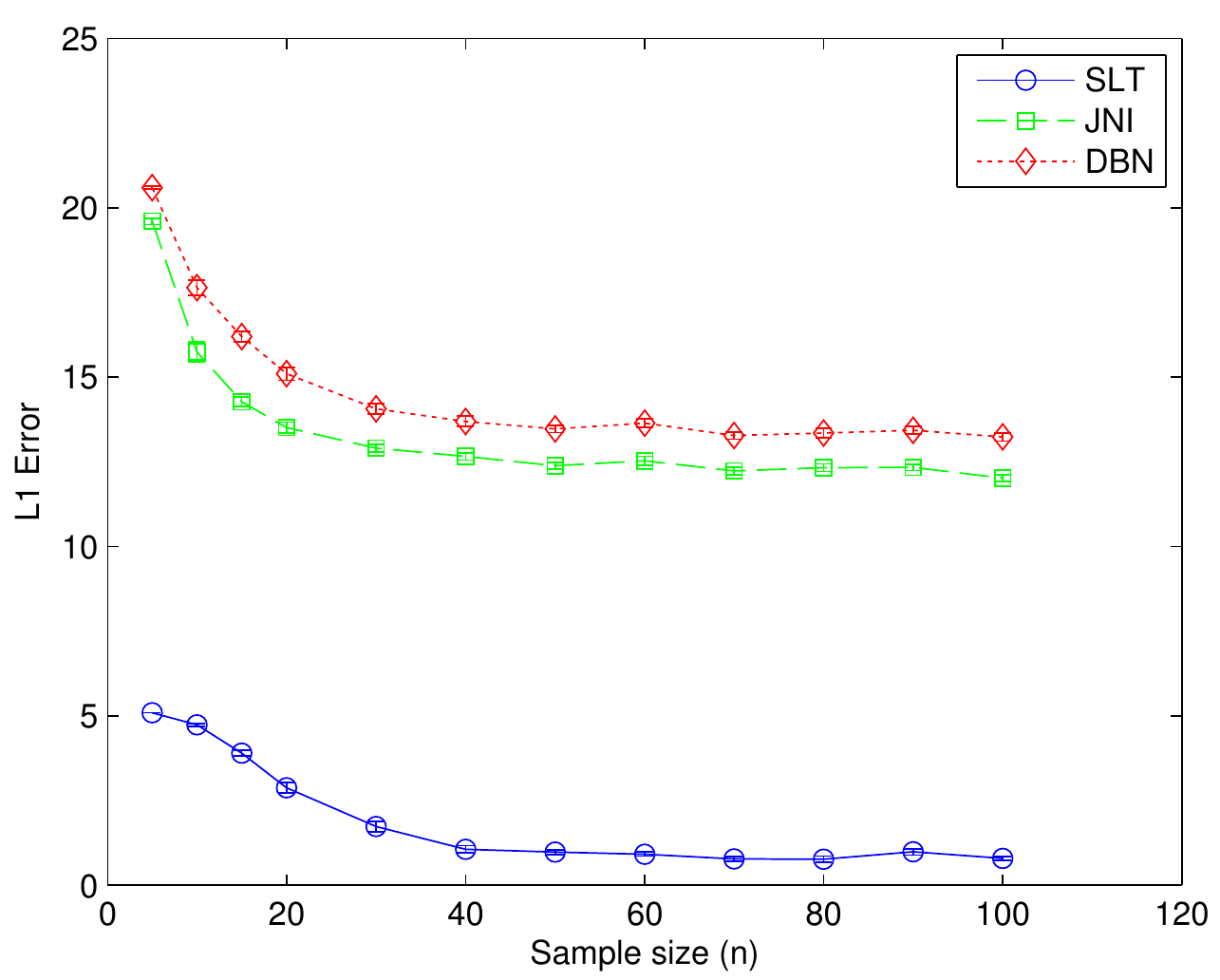}

\includegraphics[width = 0.24\textwidth]{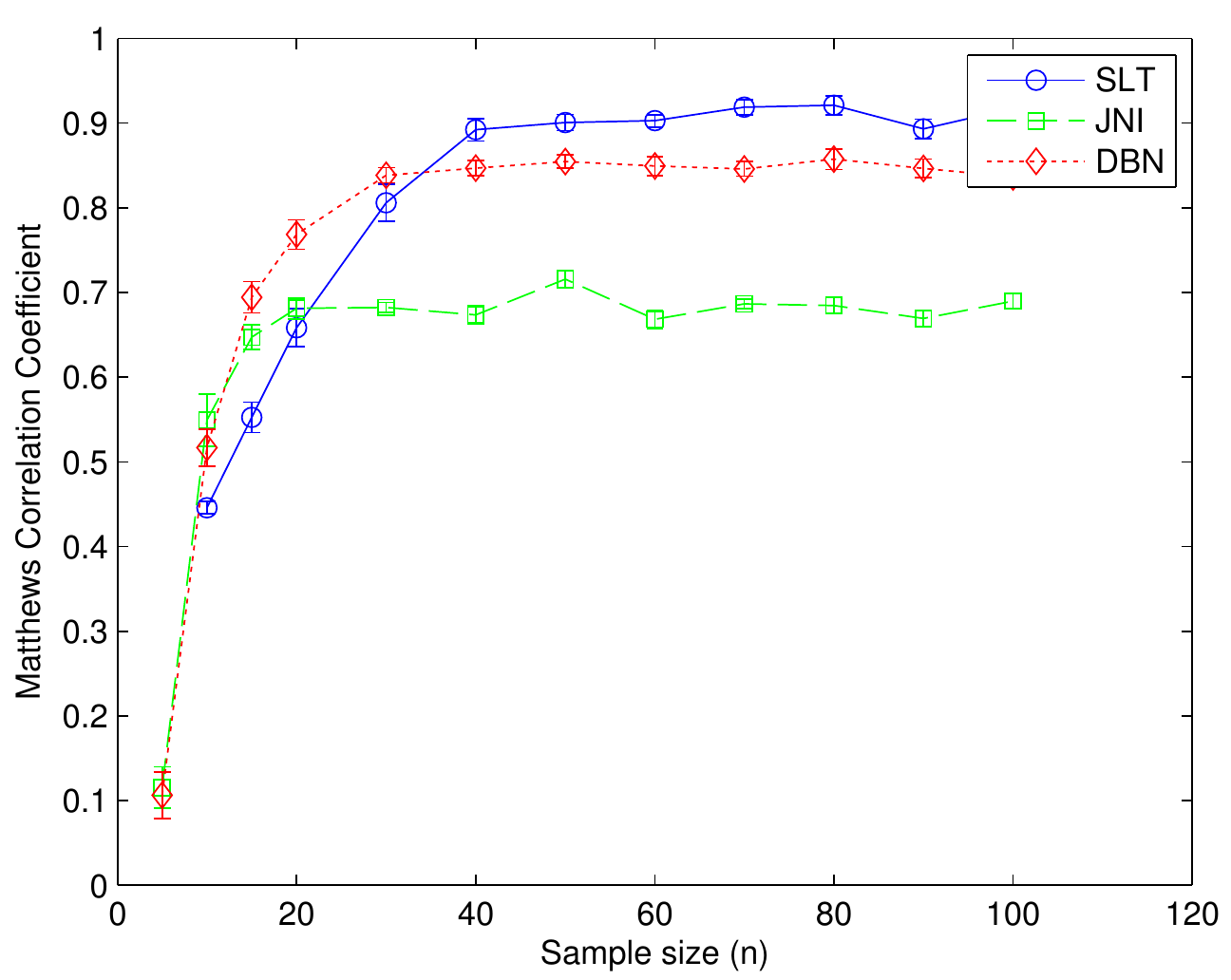}
\includegraphics[width = 0.24\textwidth]{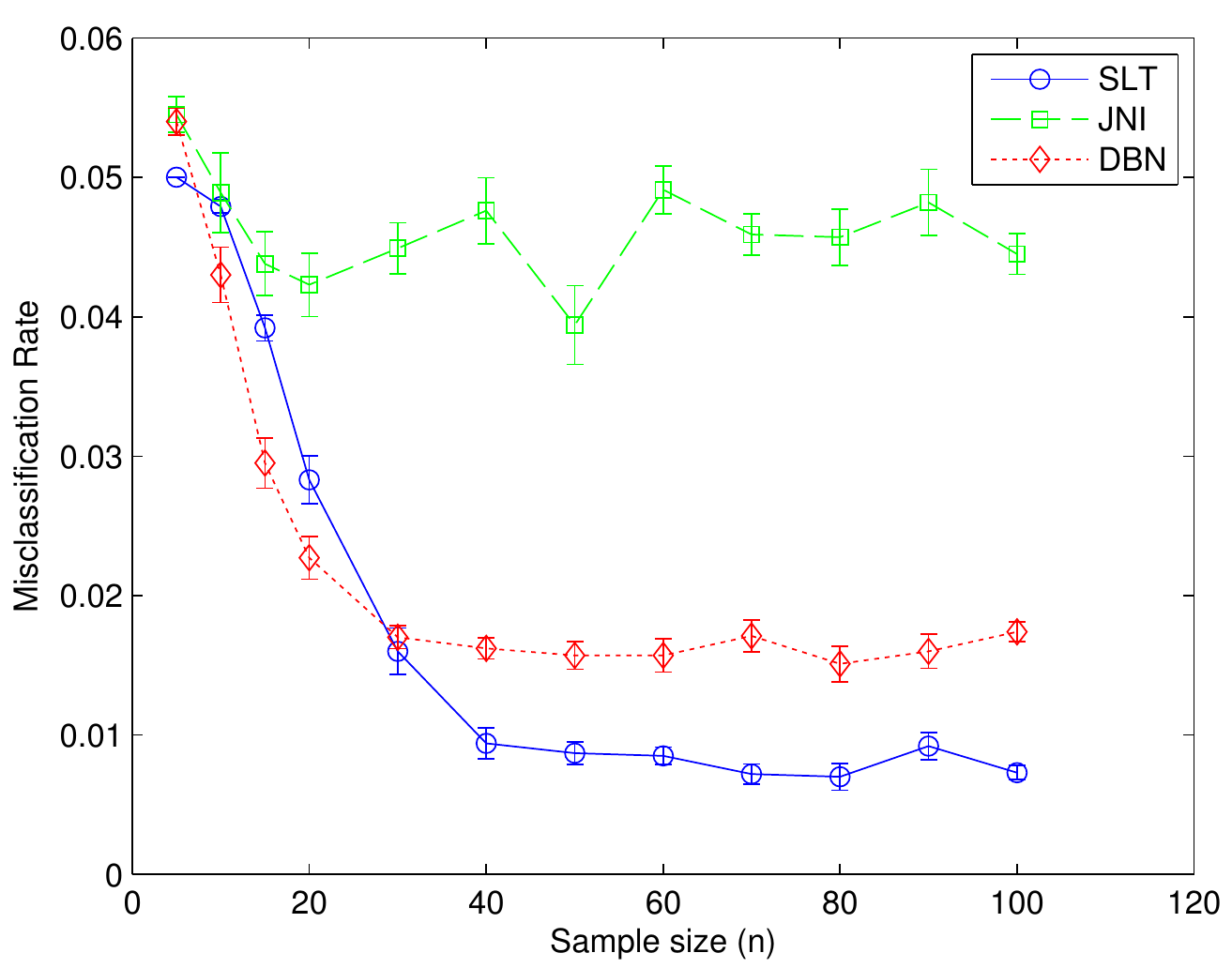}
\includegraphics[width = 0.24\textwidth]{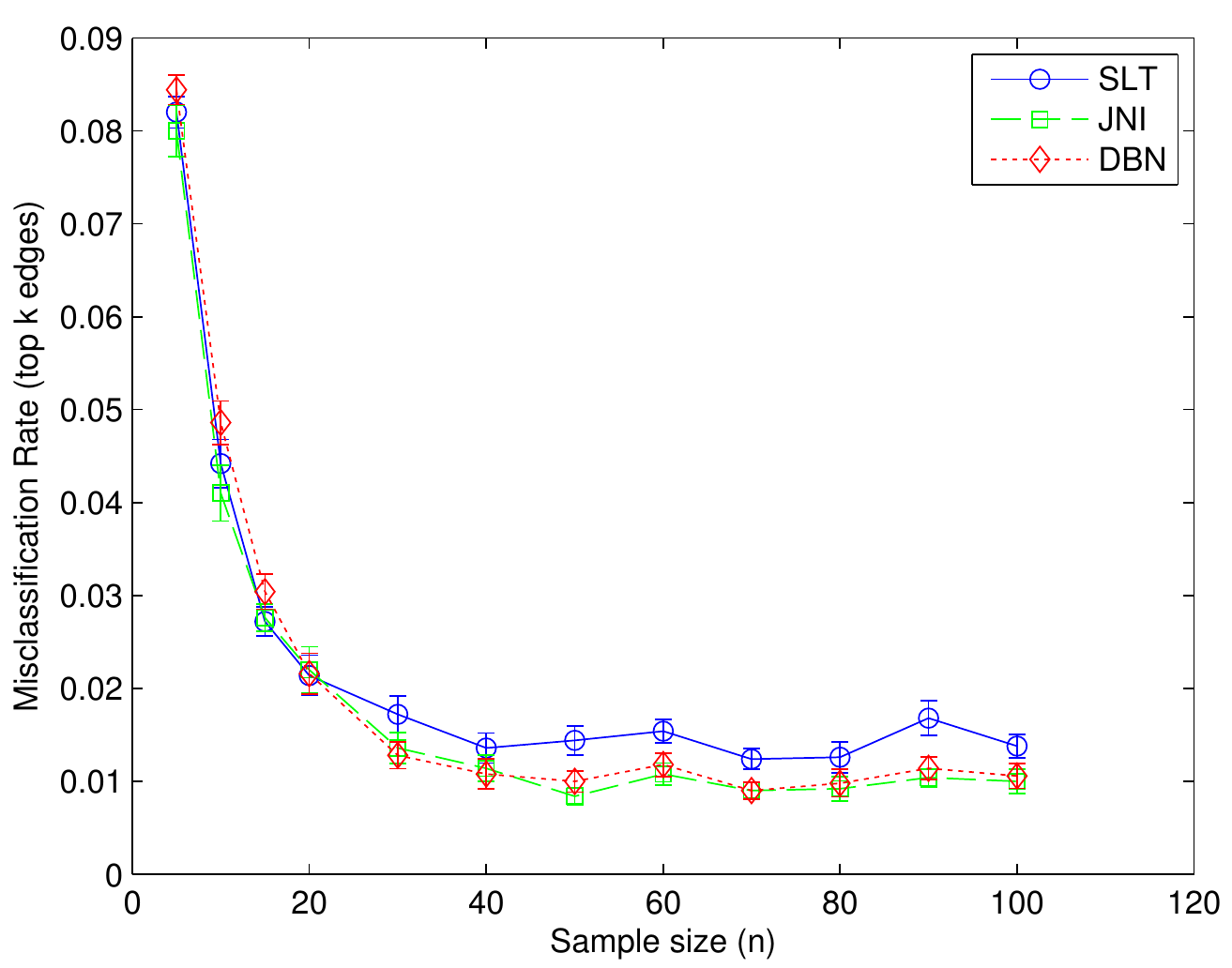}
\includegraphics[width = 0.24\textwidth]{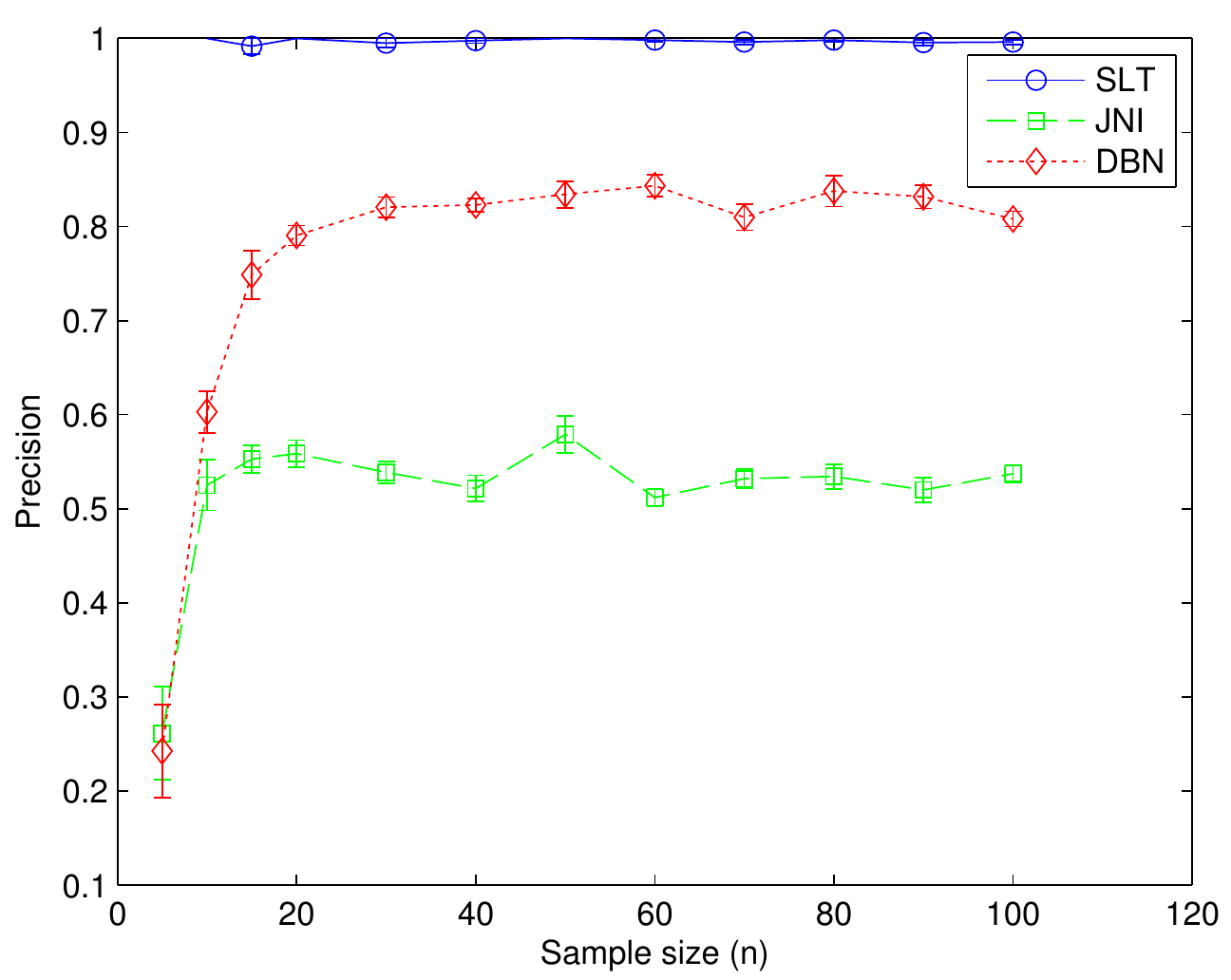}

\includegraphics[width = 0.24\textwidth]{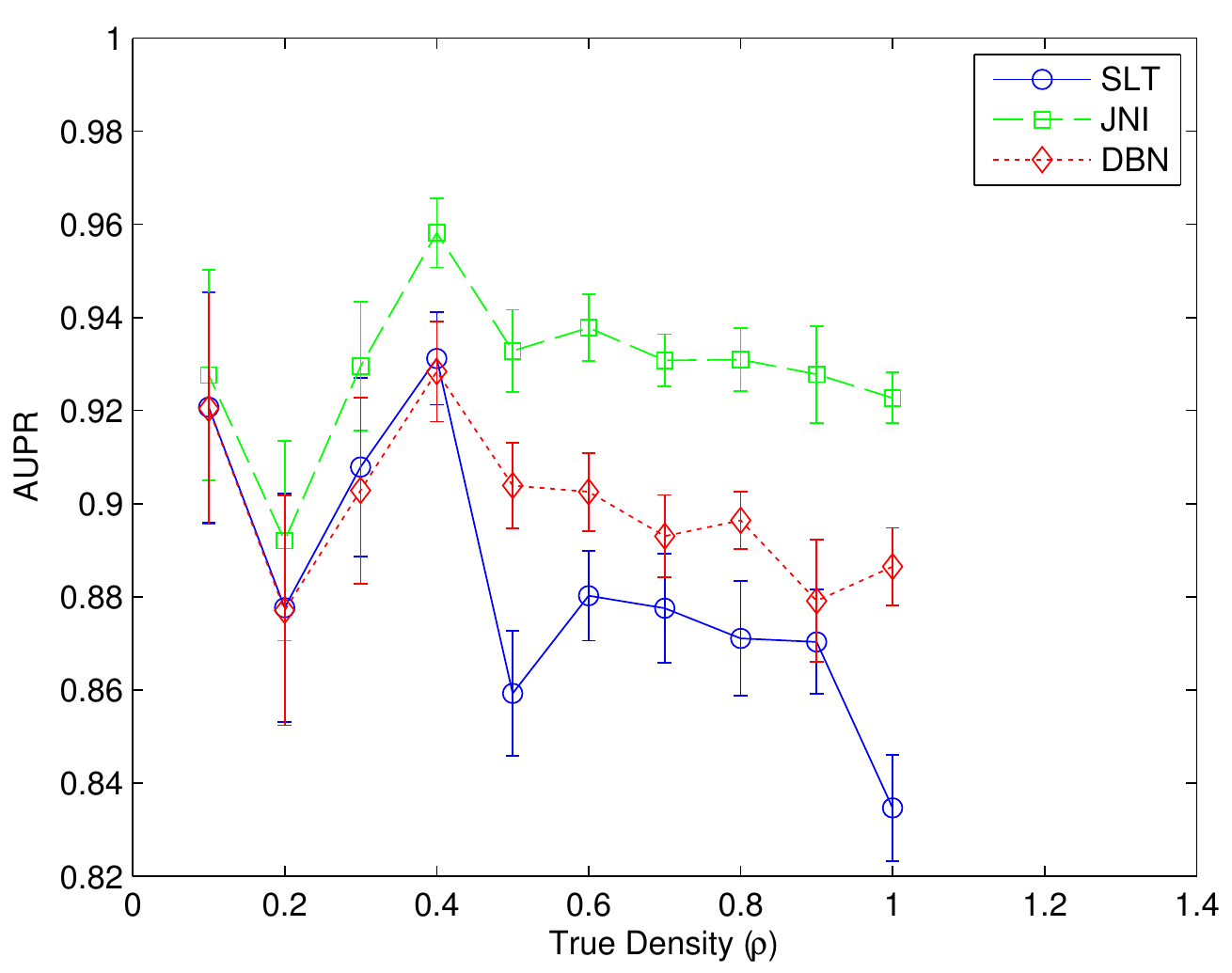}
\includegraphics[width = 0.24\textwidth]{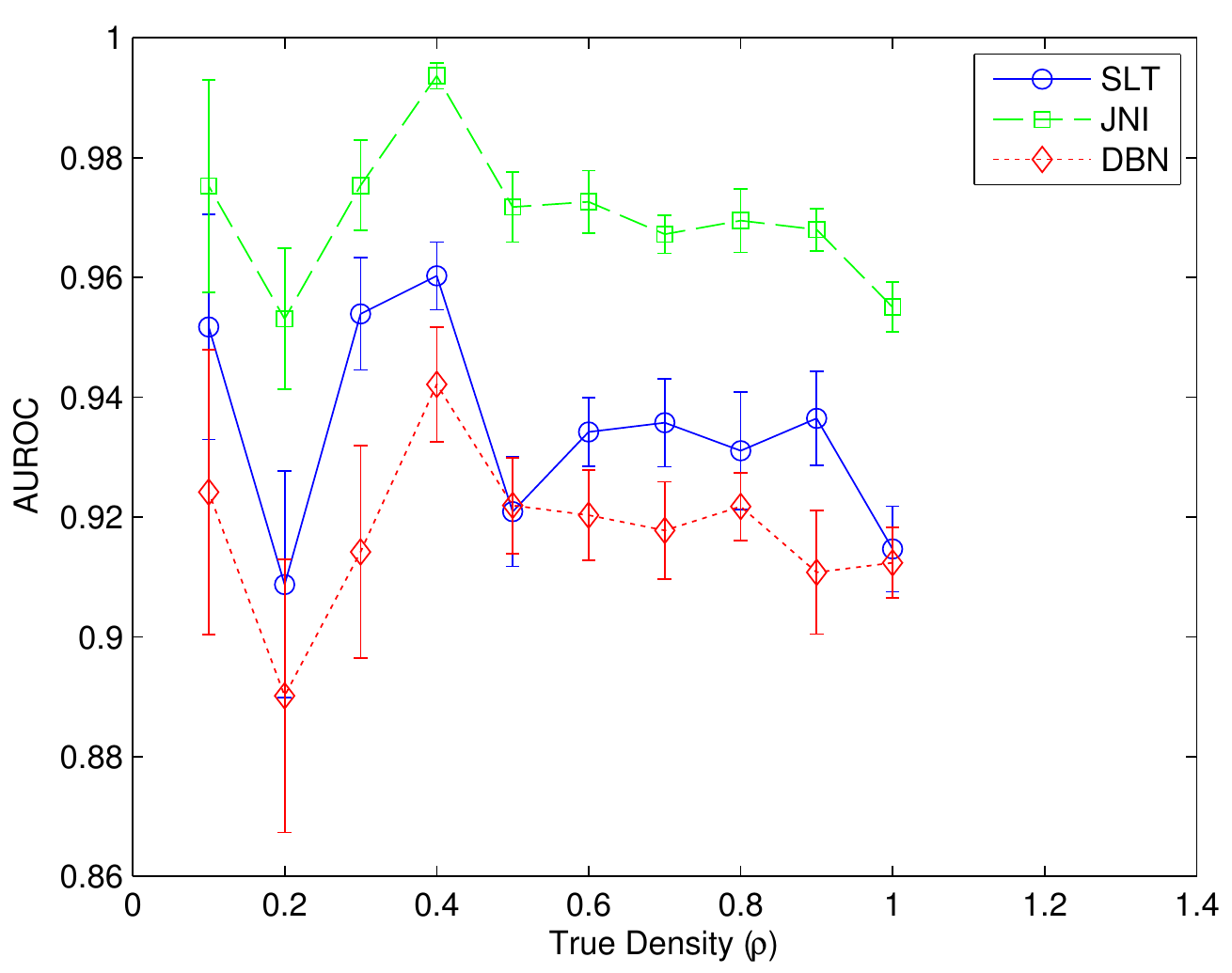}
\includegraphics[width = 0.24\textwidth]{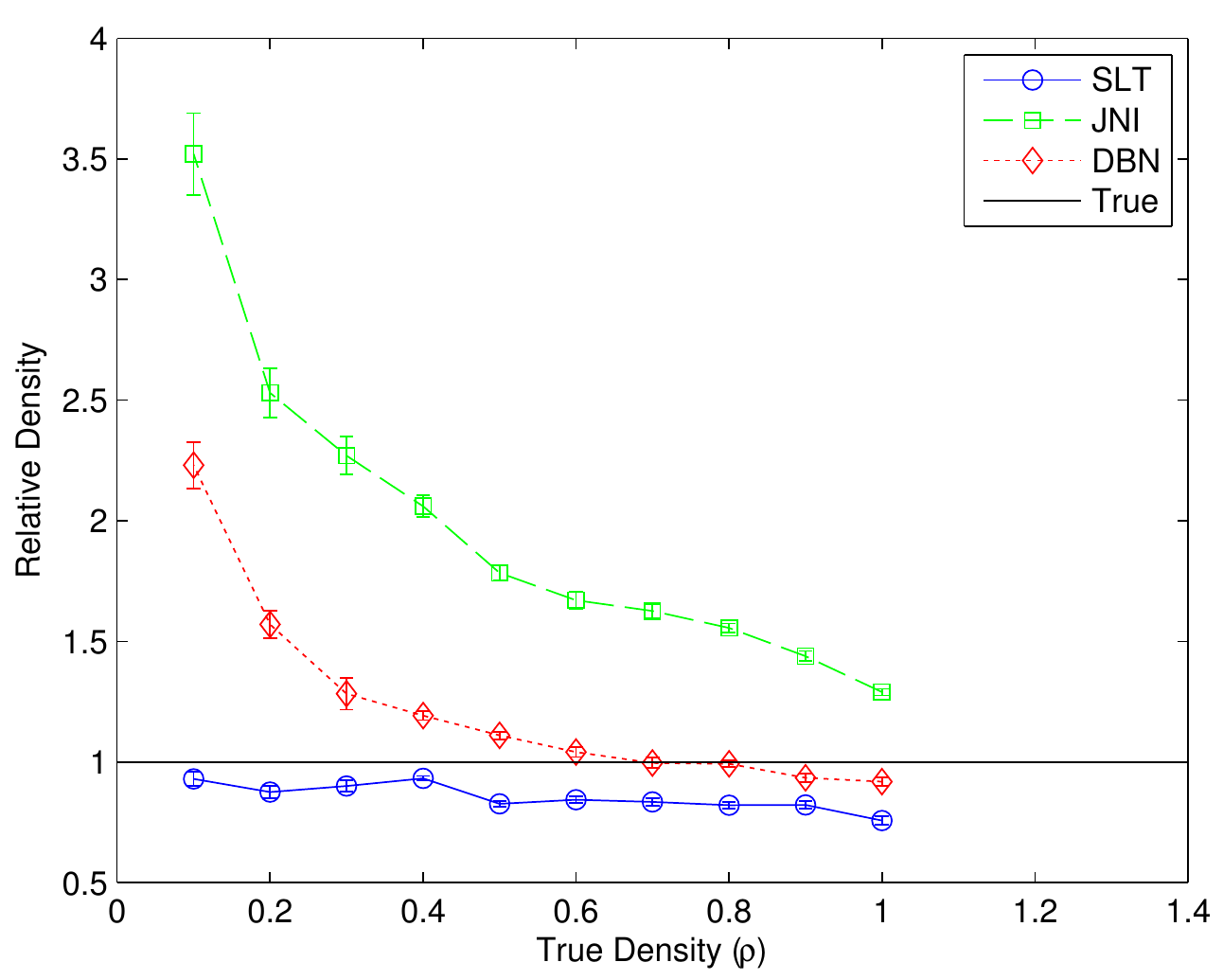}
\includegraphics[width = 0.24\textwidth]{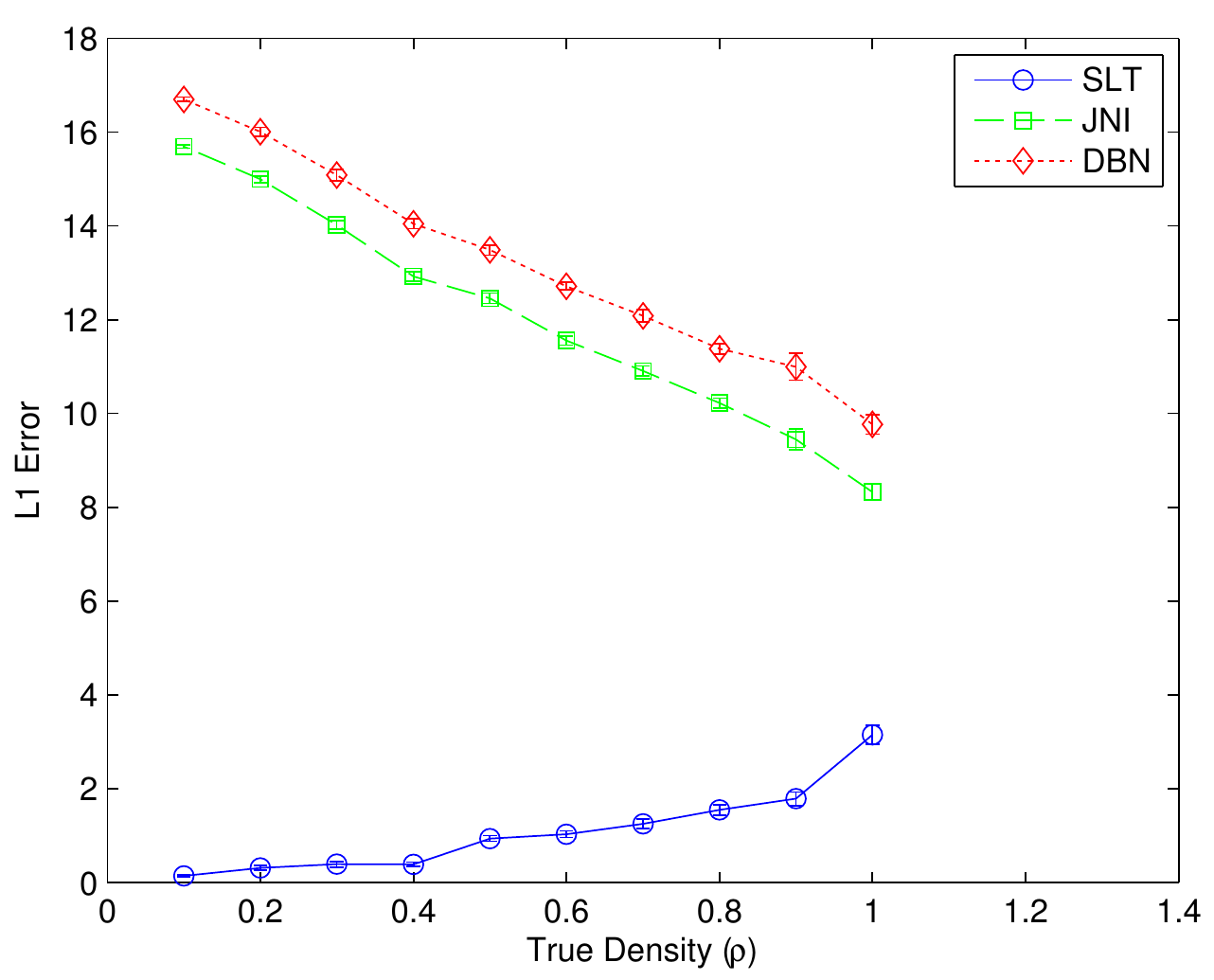}

\includegraphics[width = 0.24\textwidth]{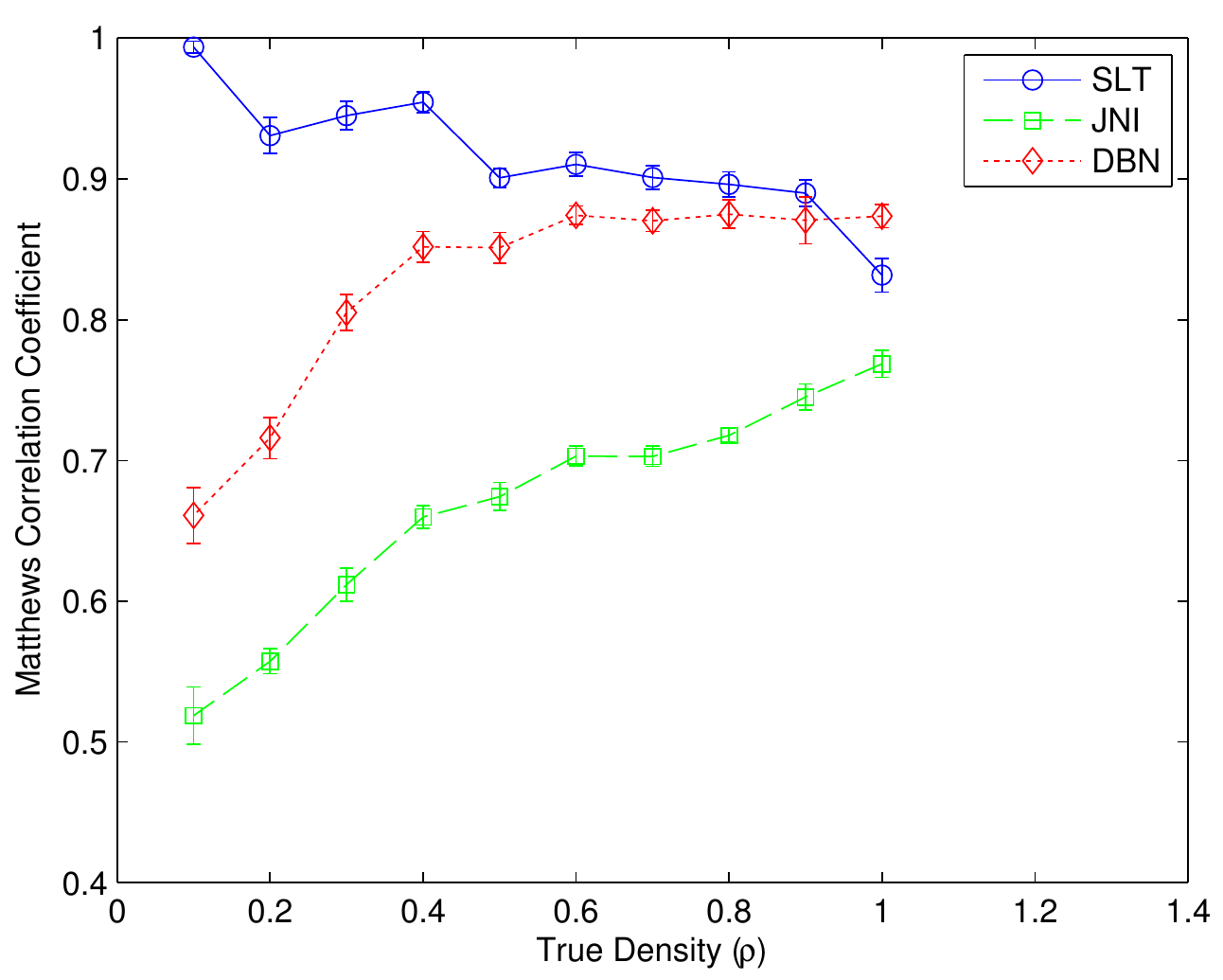}
\includegraphics[width = 0.24\textwidth]{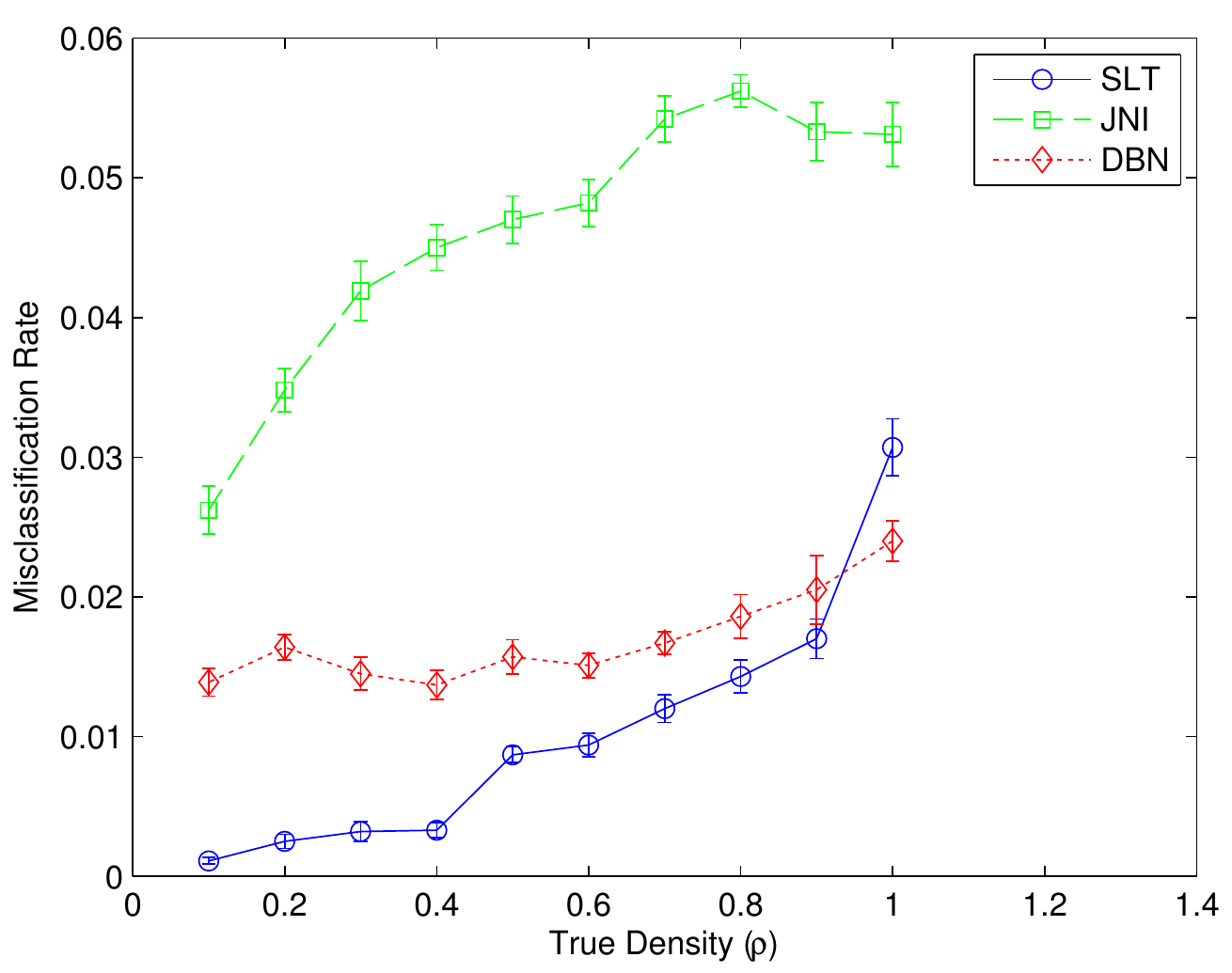}
\includegraphics[width = 0.24\textwidth]{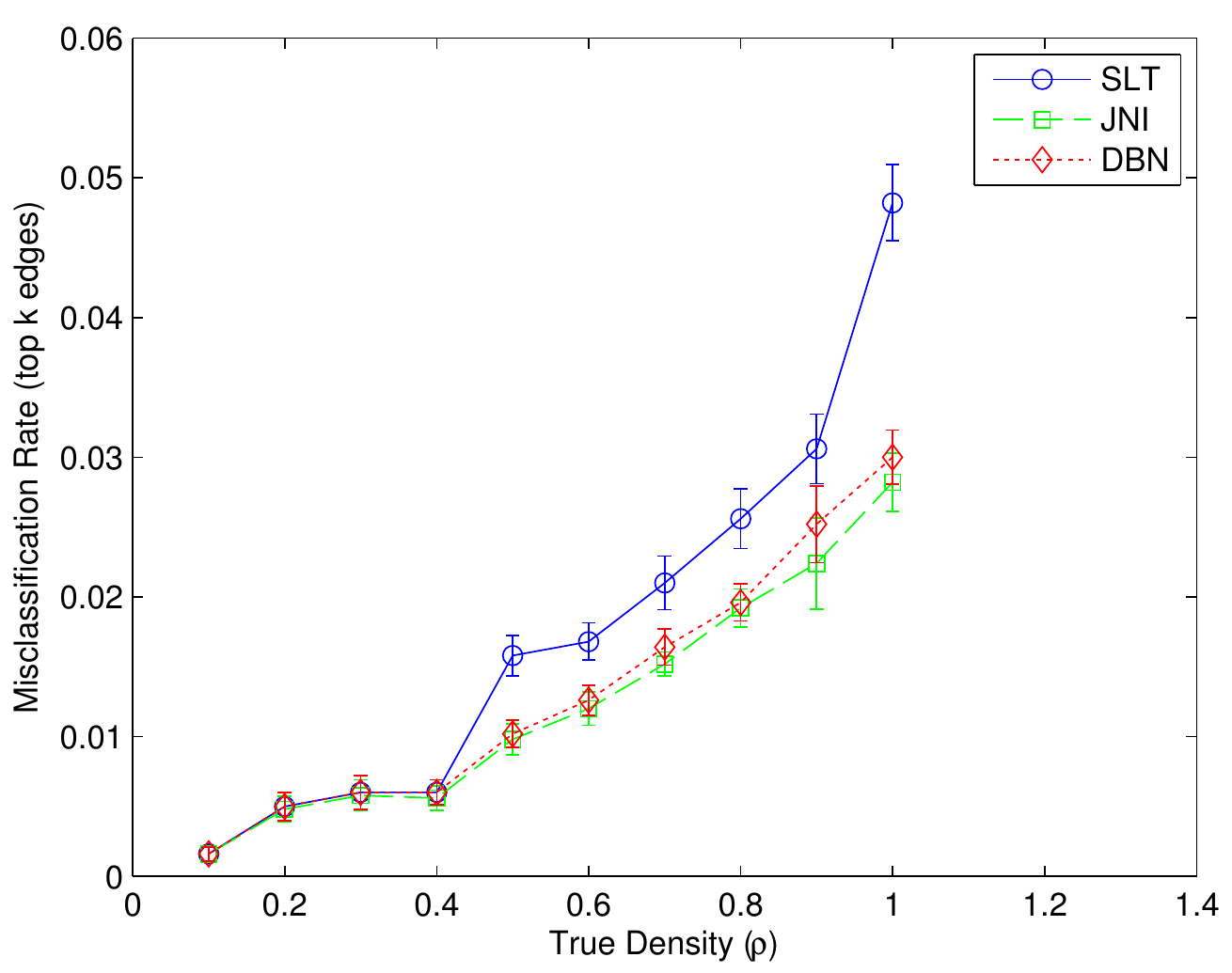}
\includegraphics[width = 0.24\textwidth]{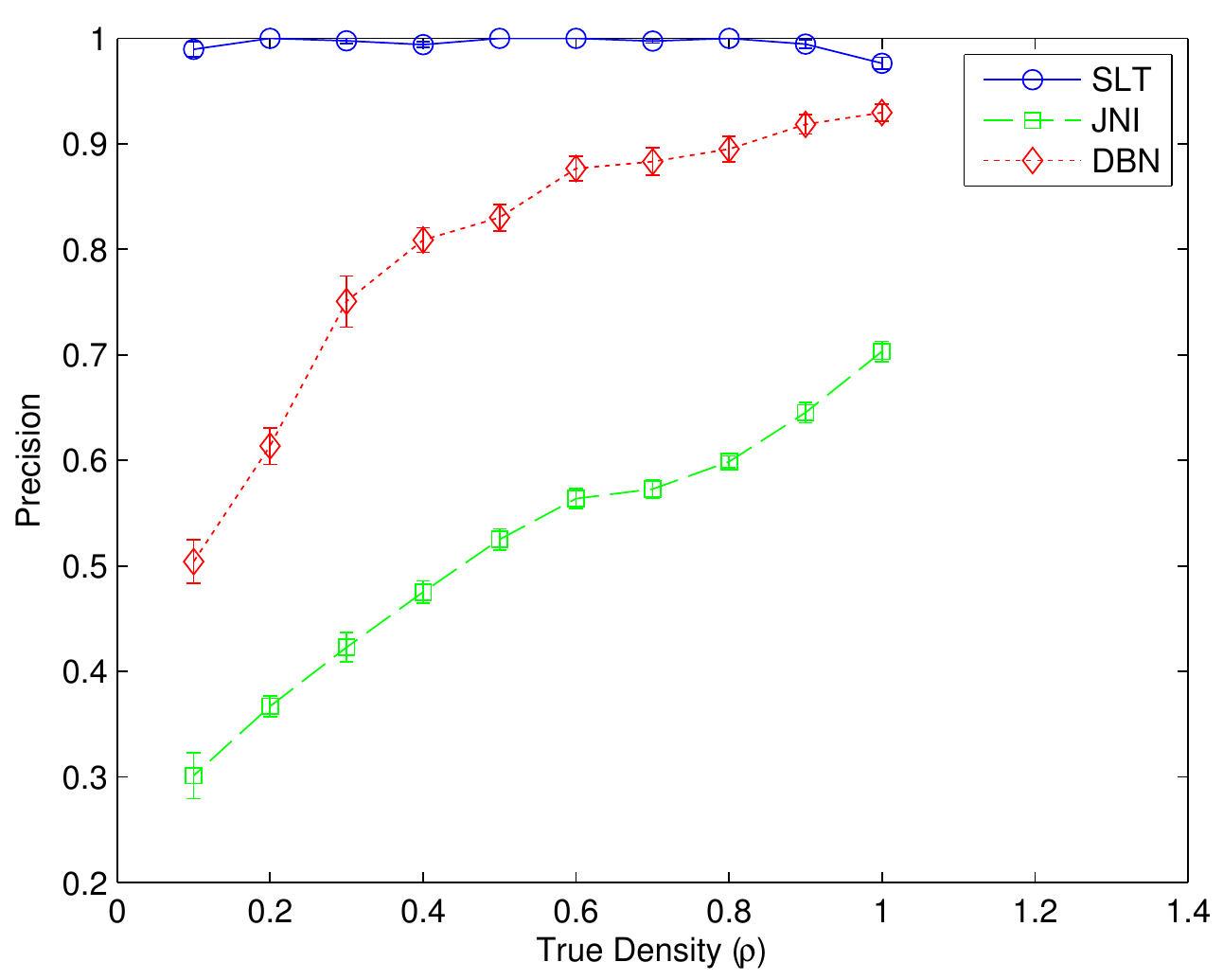}

\caption{Results on simulated data generated from a fully exchangeable SLT. [Network estimators: ``SLT'' = structure learning trees (based on 2 tiers); ``JNI'' = joint network inference; ``DBN'' = inference for each network independently. For each estimator we considered both thresholded and un-thresholded adjacency matrices. Performance scores: ``AUROC'' = area under the receiver operating characteristic curve; ``AUPR'' = area under the precision-recall curve; ``L1 Error'' = $\ell_1$ distance from the true adjacency matrices to the inferred weighted adjacency matrices; ``top k edges'' = the $\rho P$ most probable edges. Performance scores were averaged over all 10 data-generating networks and all 10 datasets; error bars denote standard errors of mean performance over datasets. We considered both varying $n$ for fixed $\rho = 0.5$ and varying $\rho$ for fixed $n=60$.]}
\label{sim res3}
\end{figure*}

Next, we  relaxed the distinct sub-group architecture that likely favours SLT by allowing $\overline{G_{11}}$, $\overline{G_{12}}$ to share edges (``weakly exchangeable'' regime (2); SFig. \ref{ex2}).
Results (Fig. \ref{reg2} and SFig. \ref{sim res2}) in this regime closely mirrored those of the disjoint sub-group regime, suggesting that SLTs offer improved estimation in more realistic weakly exchangeable settings.
However in the fully exchangeable regime (3) (SFig. \ref{ex3}) there was a decrease in performance of SLT with respect to JNI as quantified by AUPR, AUROC and the misclassification rate among top ranked edges (SFig. \ref{sim res3}).

\begin{figure*}[t!]
\centering
\includegraphics[width = 0.24\textwidth]{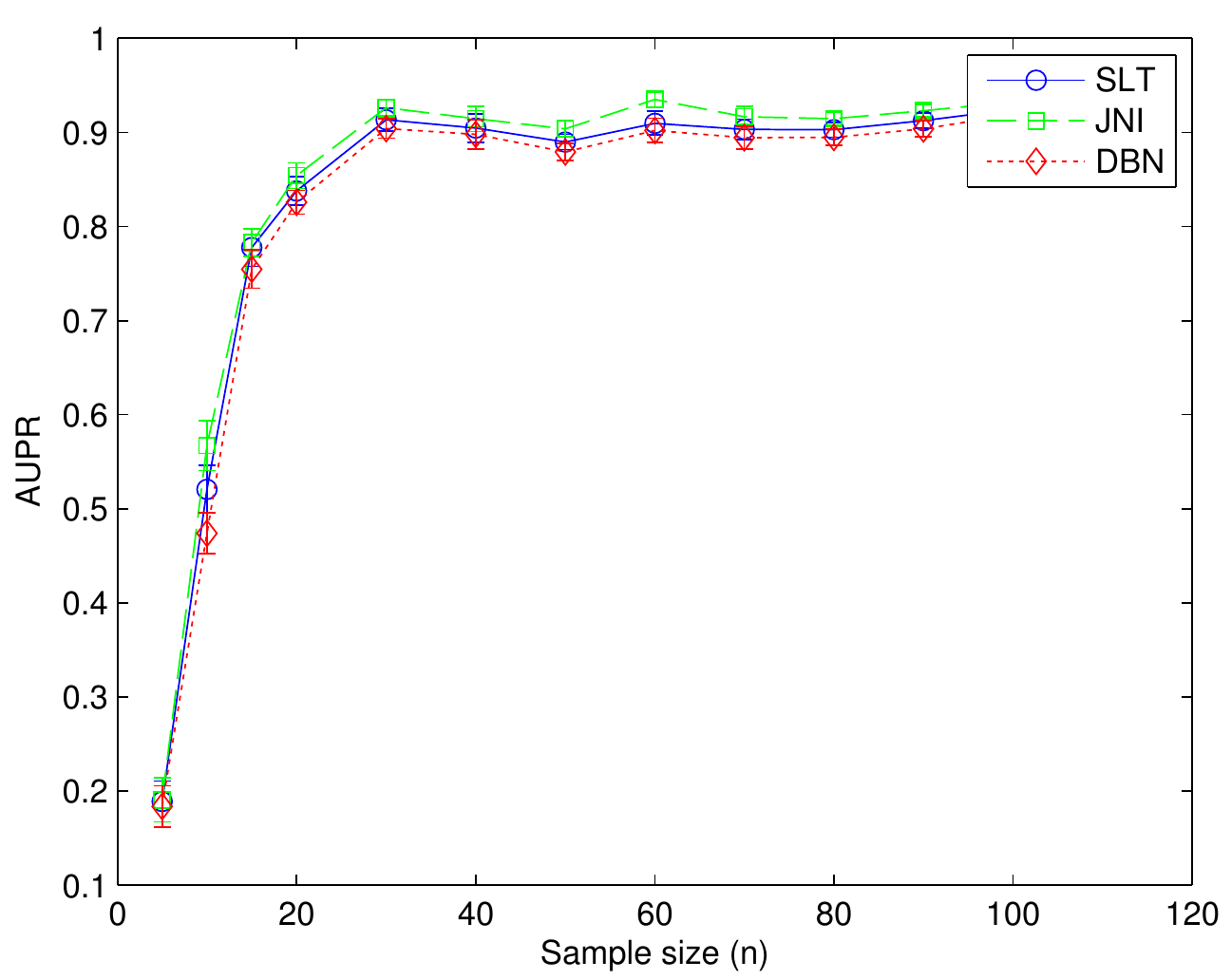}
\includegraphics[width = 0.24\textwidth]{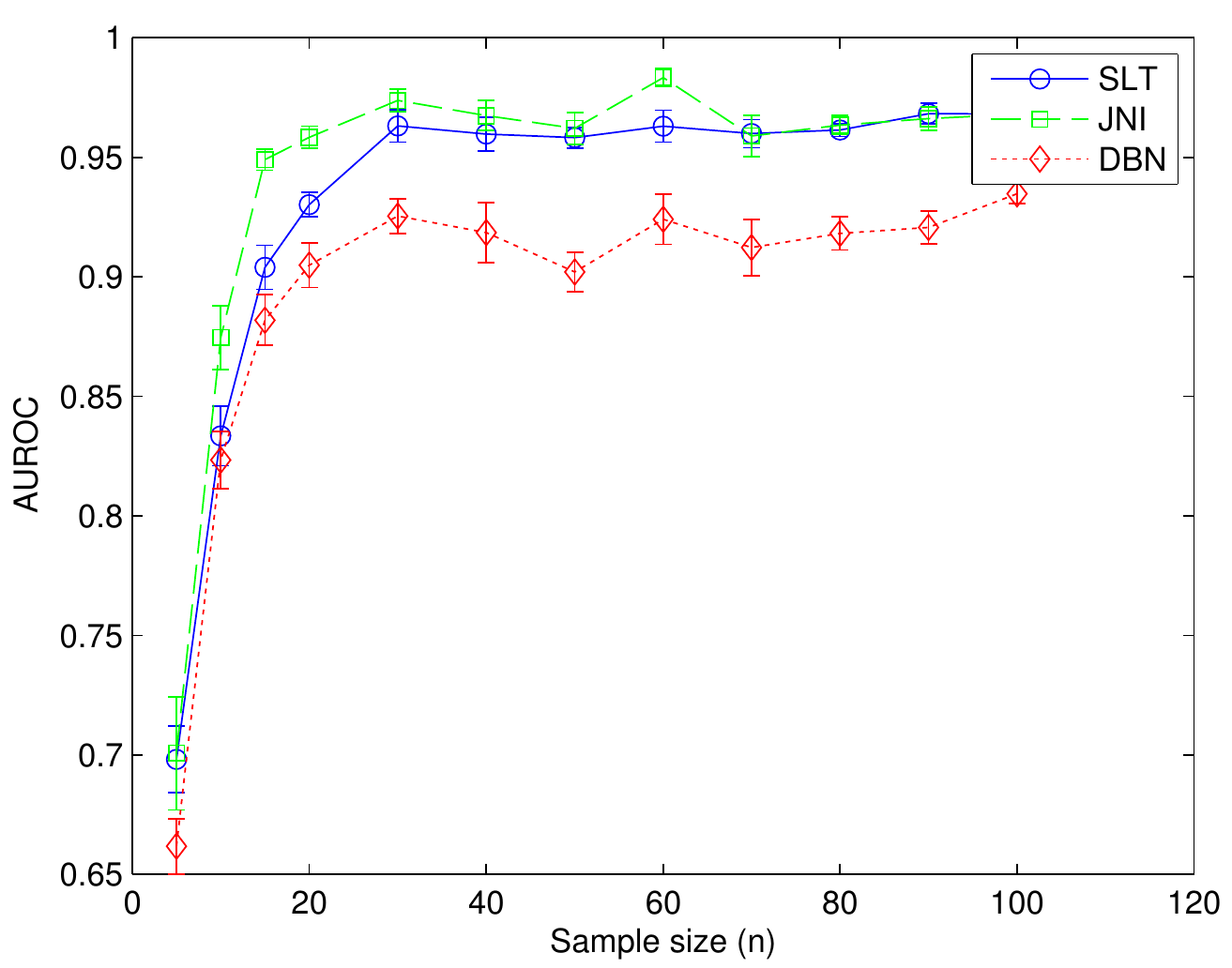}
\includegraphics[width = 0.24\textwidth]{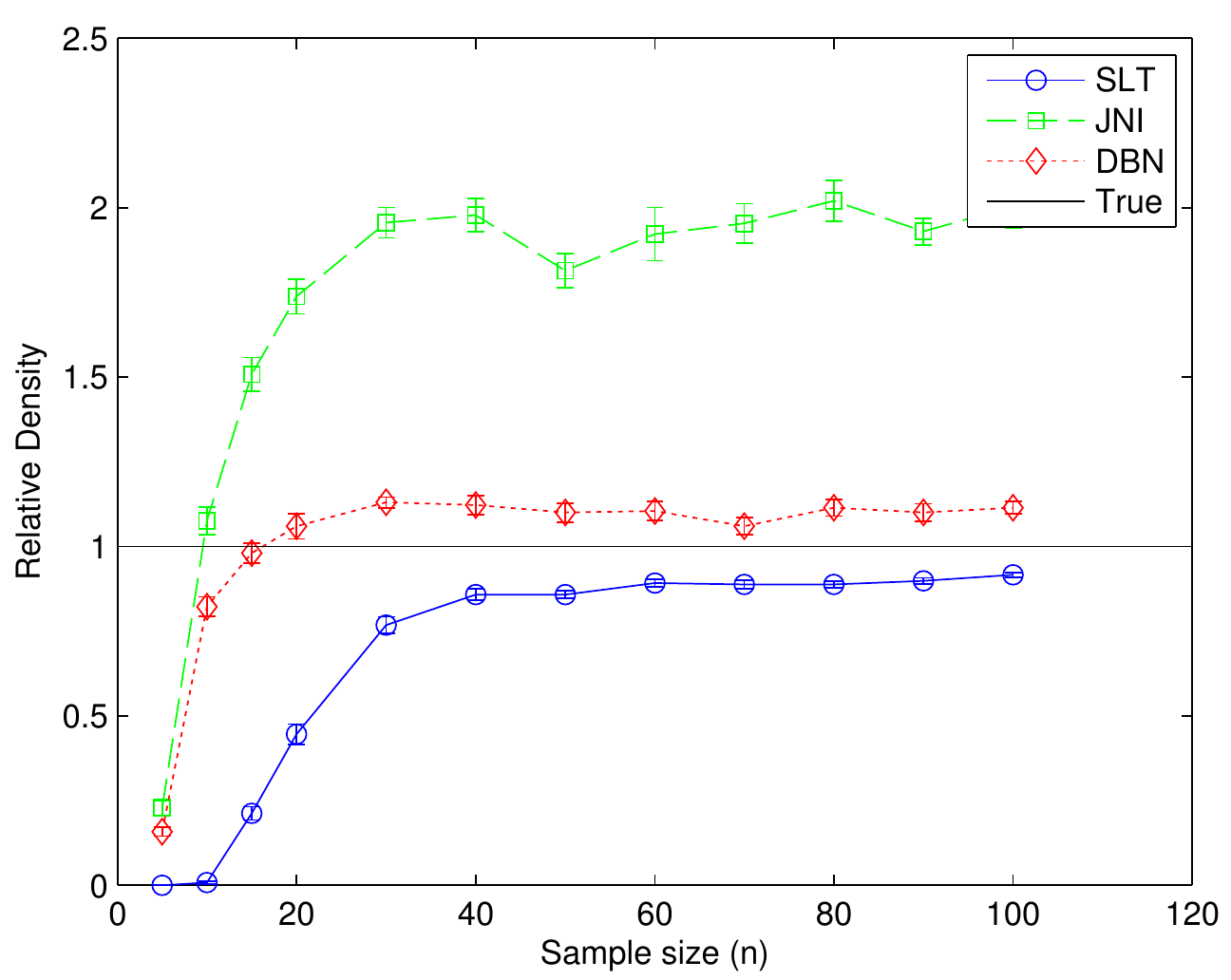}
\includegraphics[width = 0.24\textwidth]{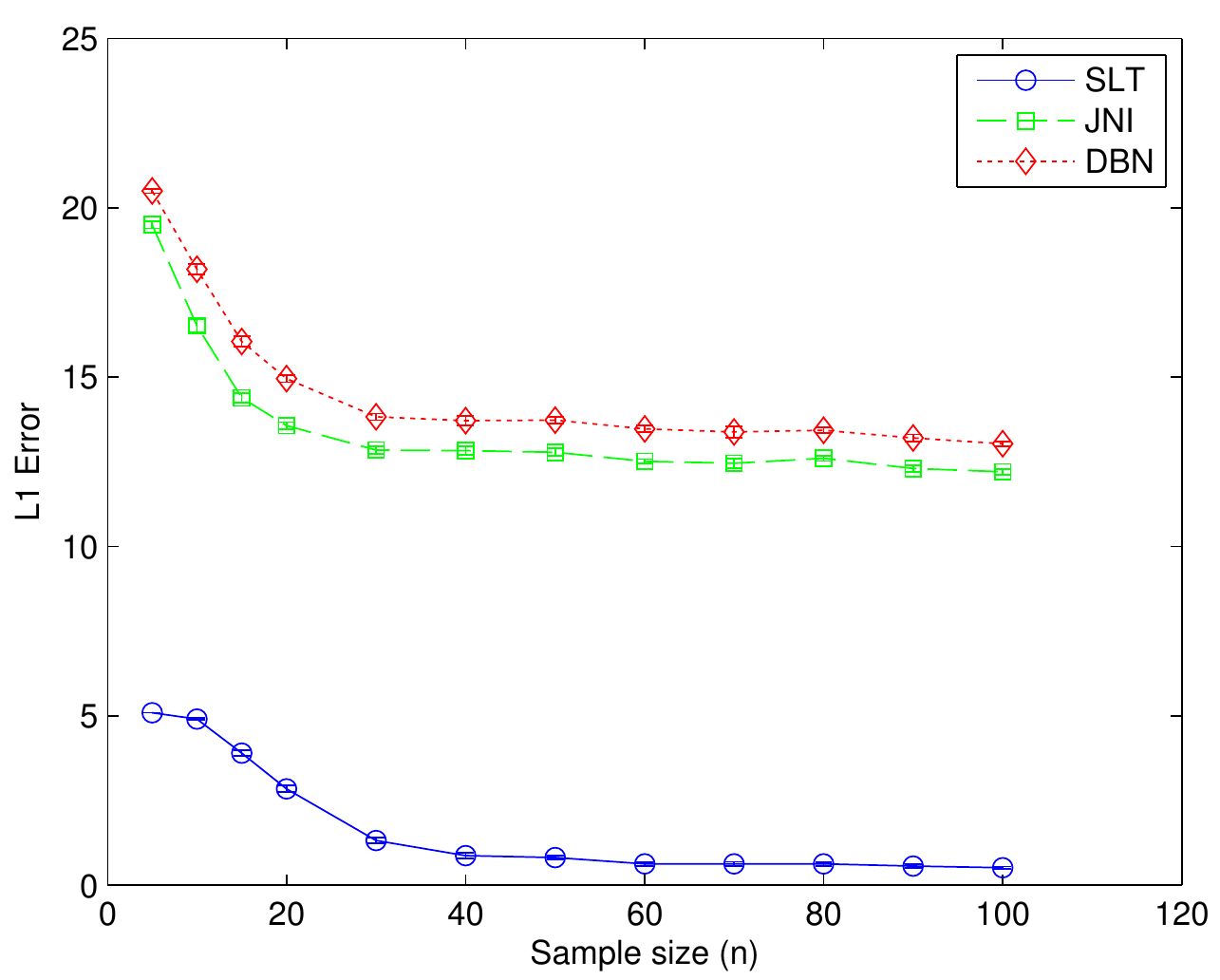}

\includegraphics[width = 0.24\textwidth]{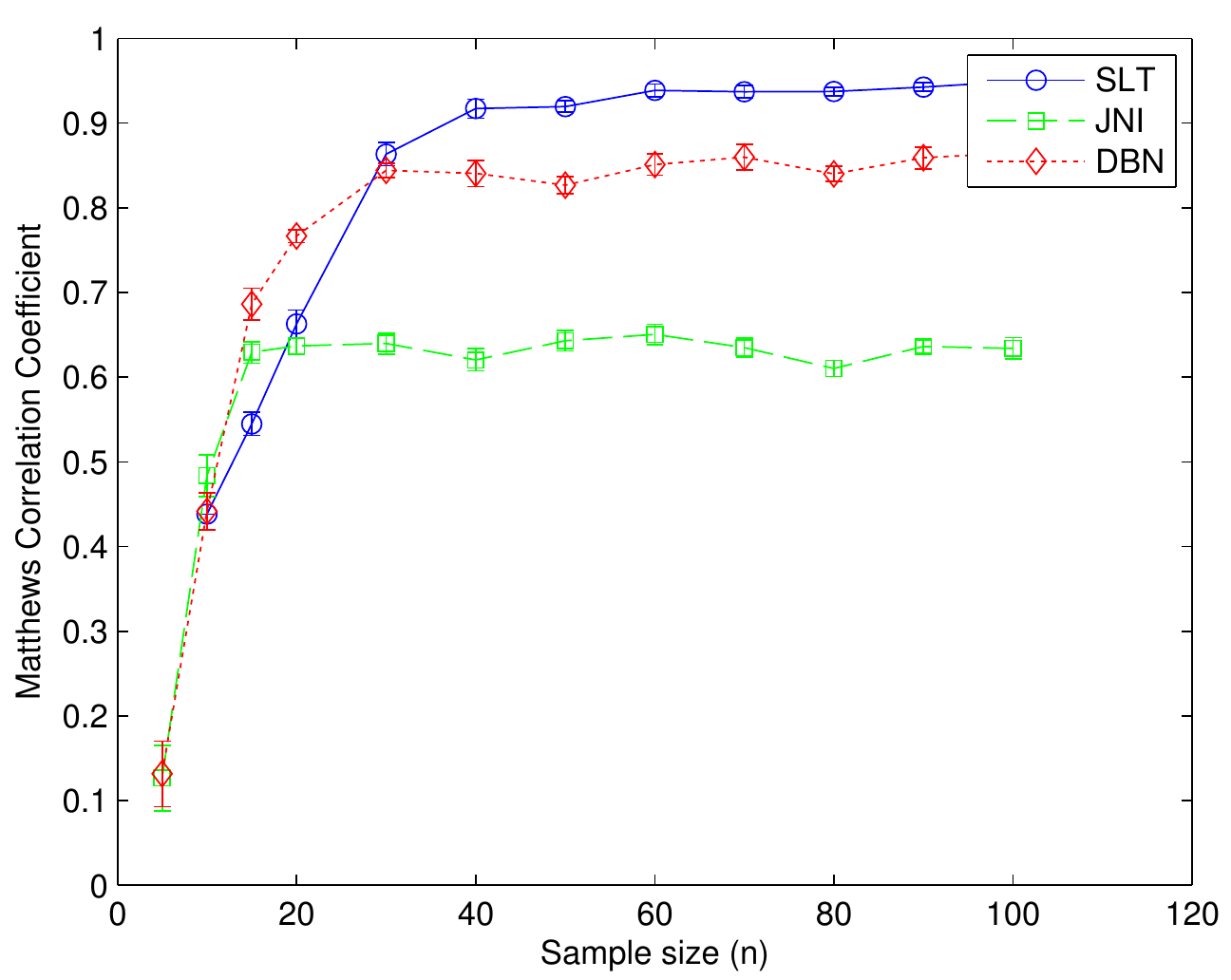}
\includegraphics[width = 0.24\textwidth]{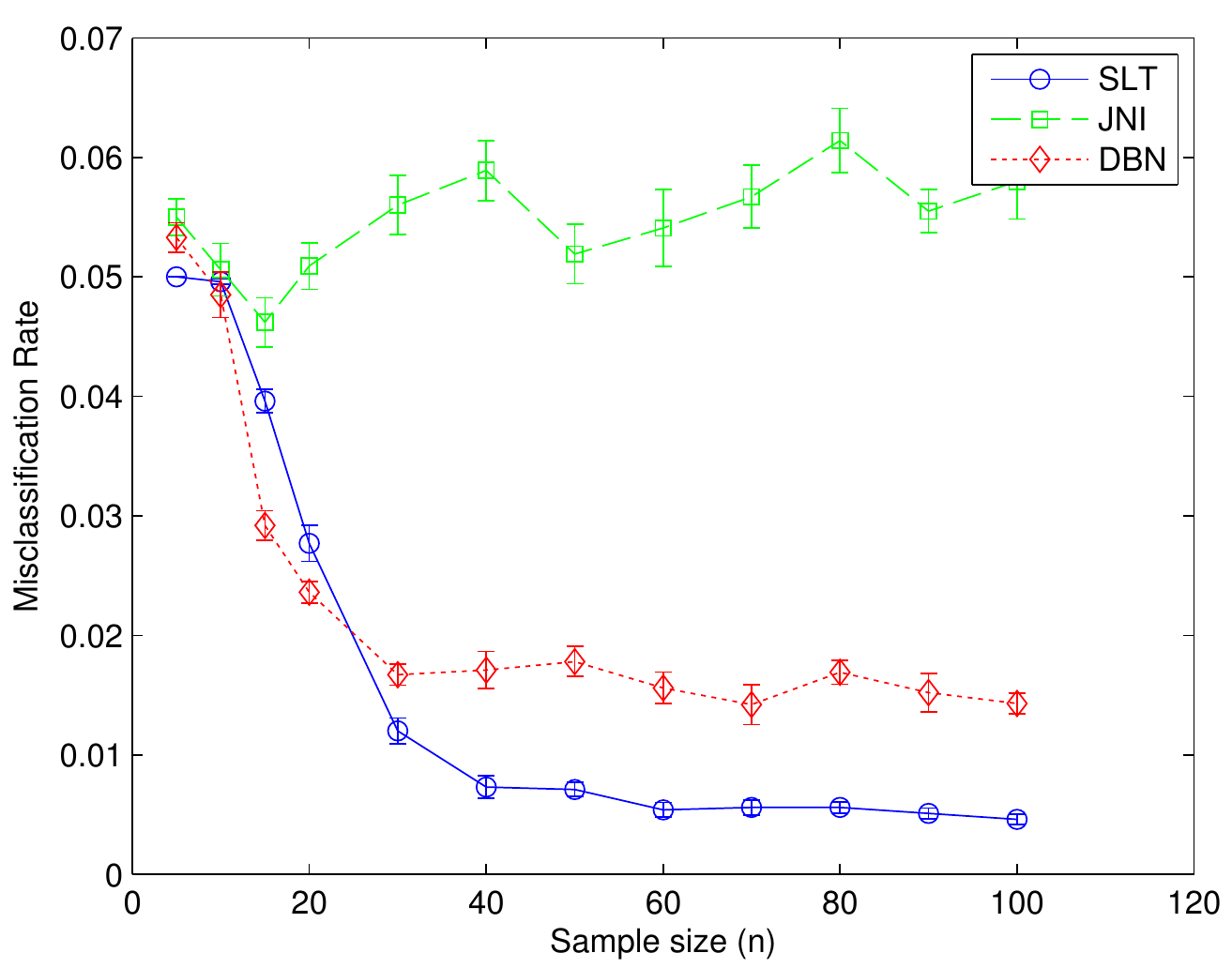}
\includegraphics[width = 0.24\textwidth]{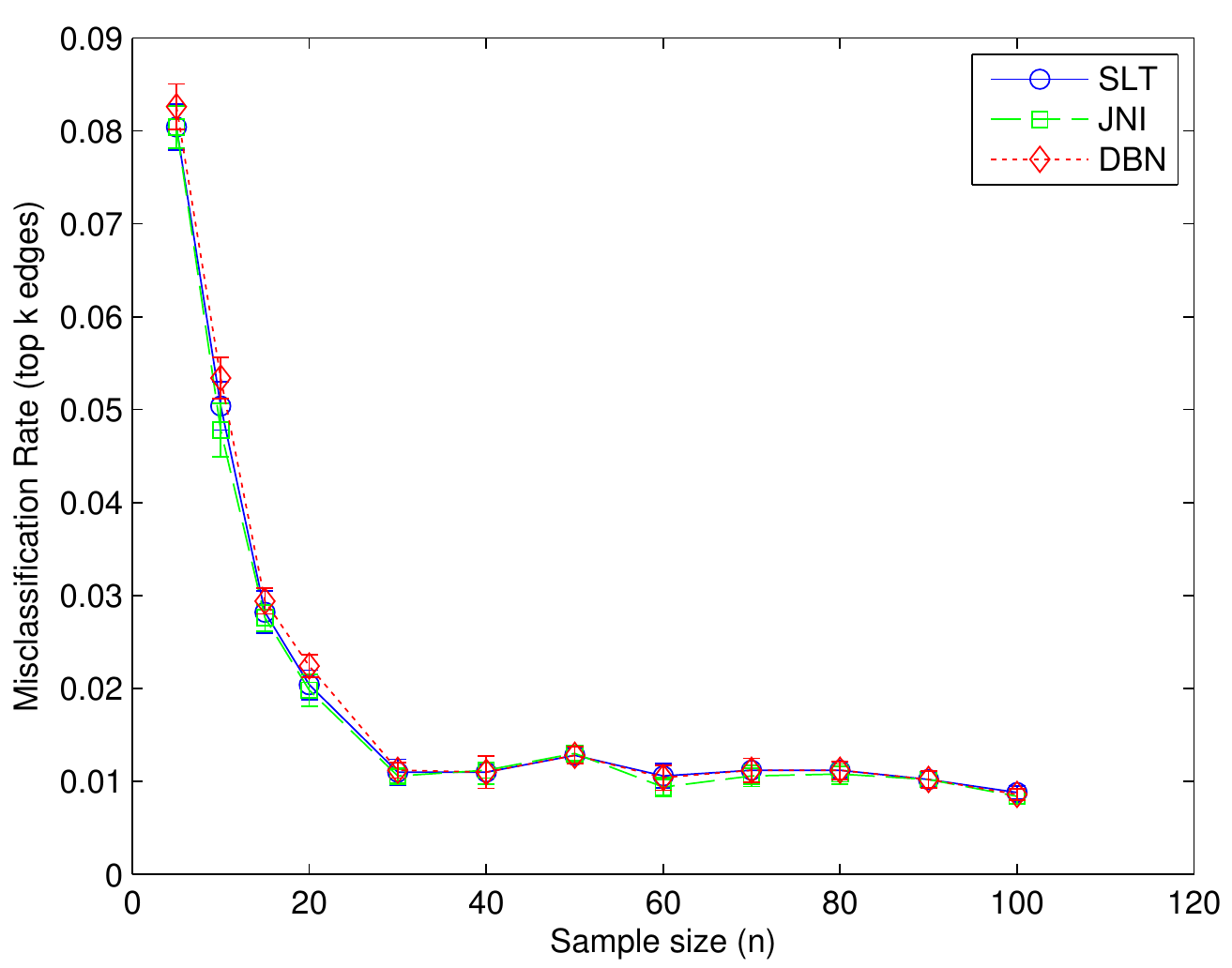}
\includegraphics[width = 0.24\textwidth]{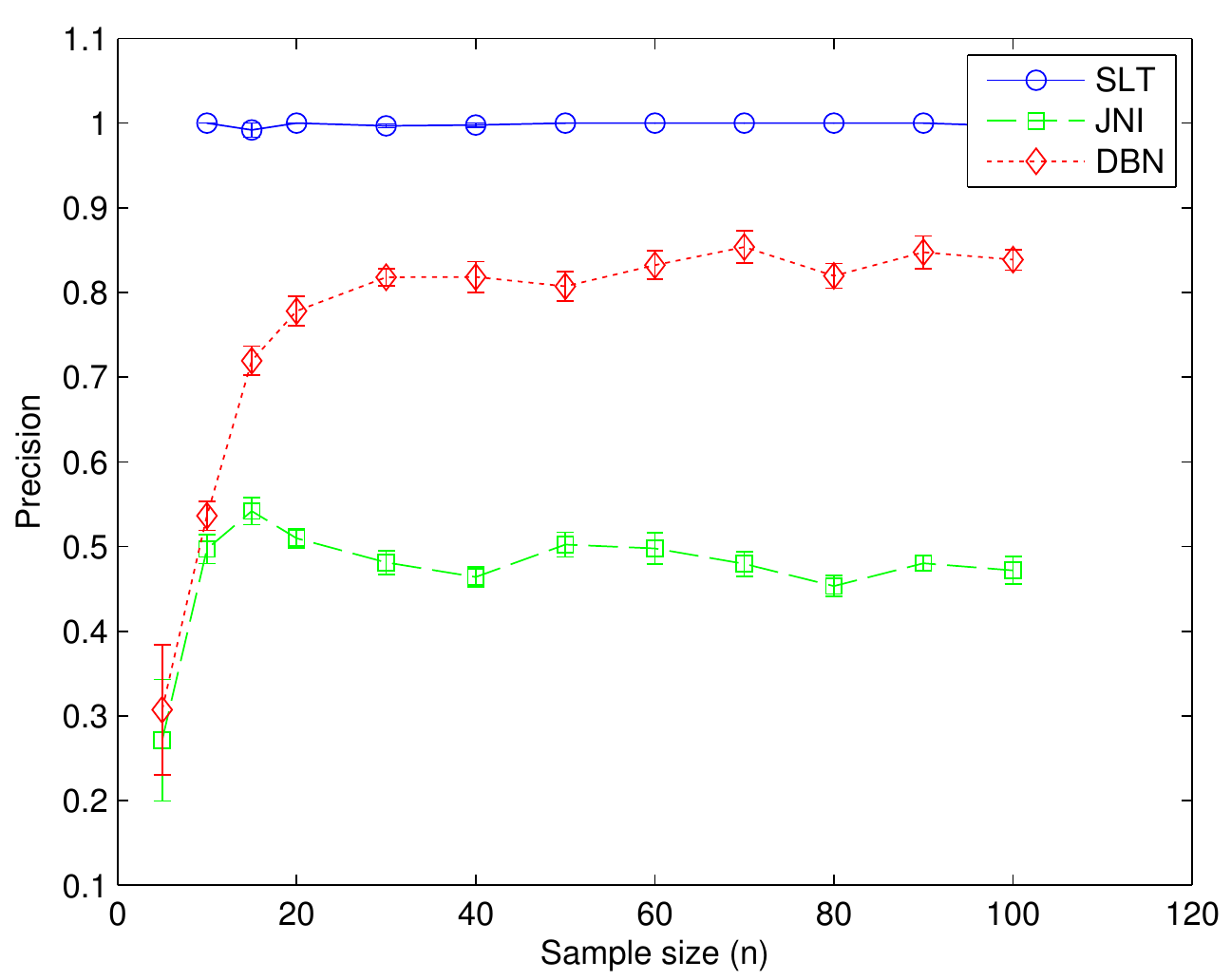}

\includegraphics[width = 0.24\textwidth]{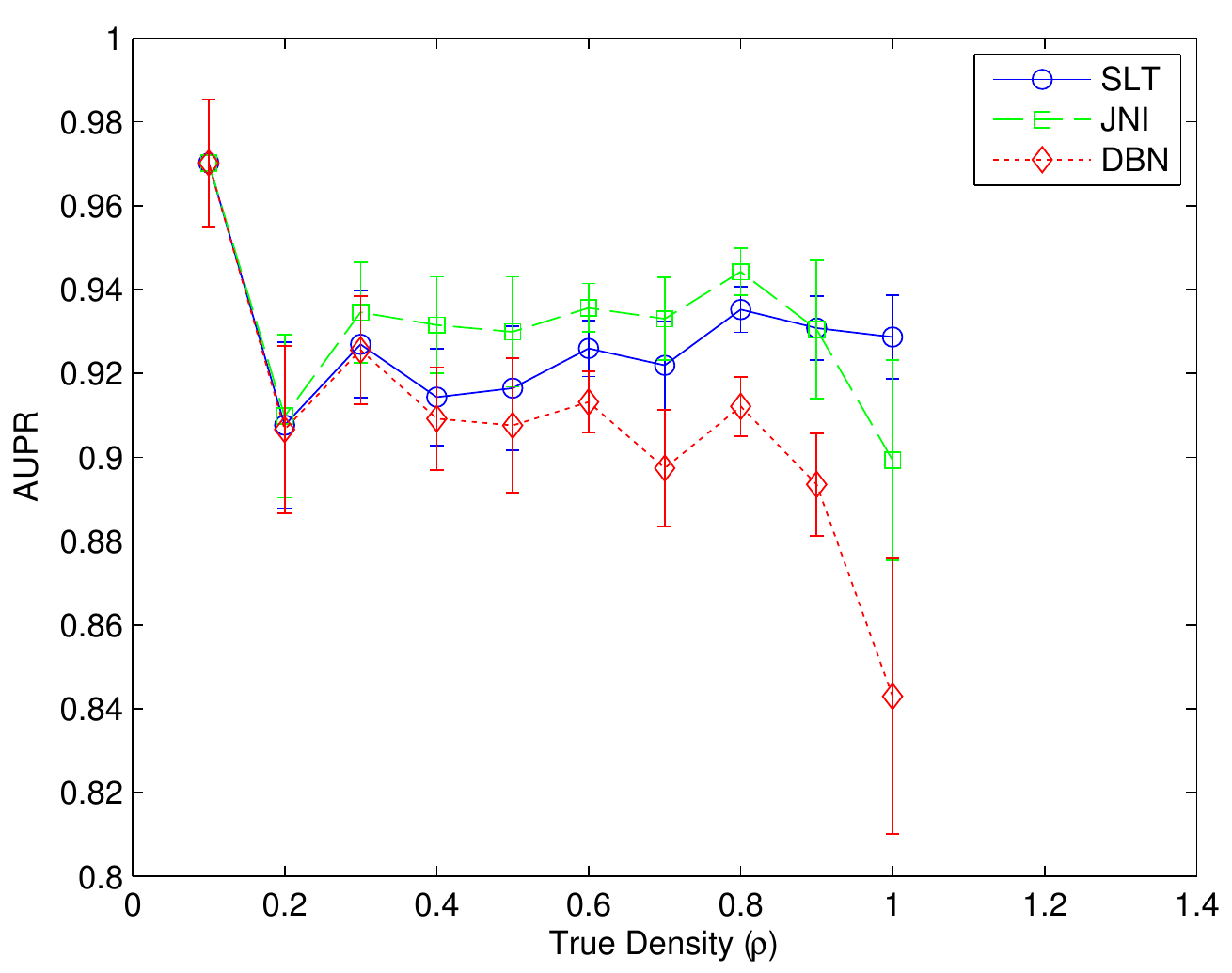}
\includegraphics[width = 0.24\textwidth]{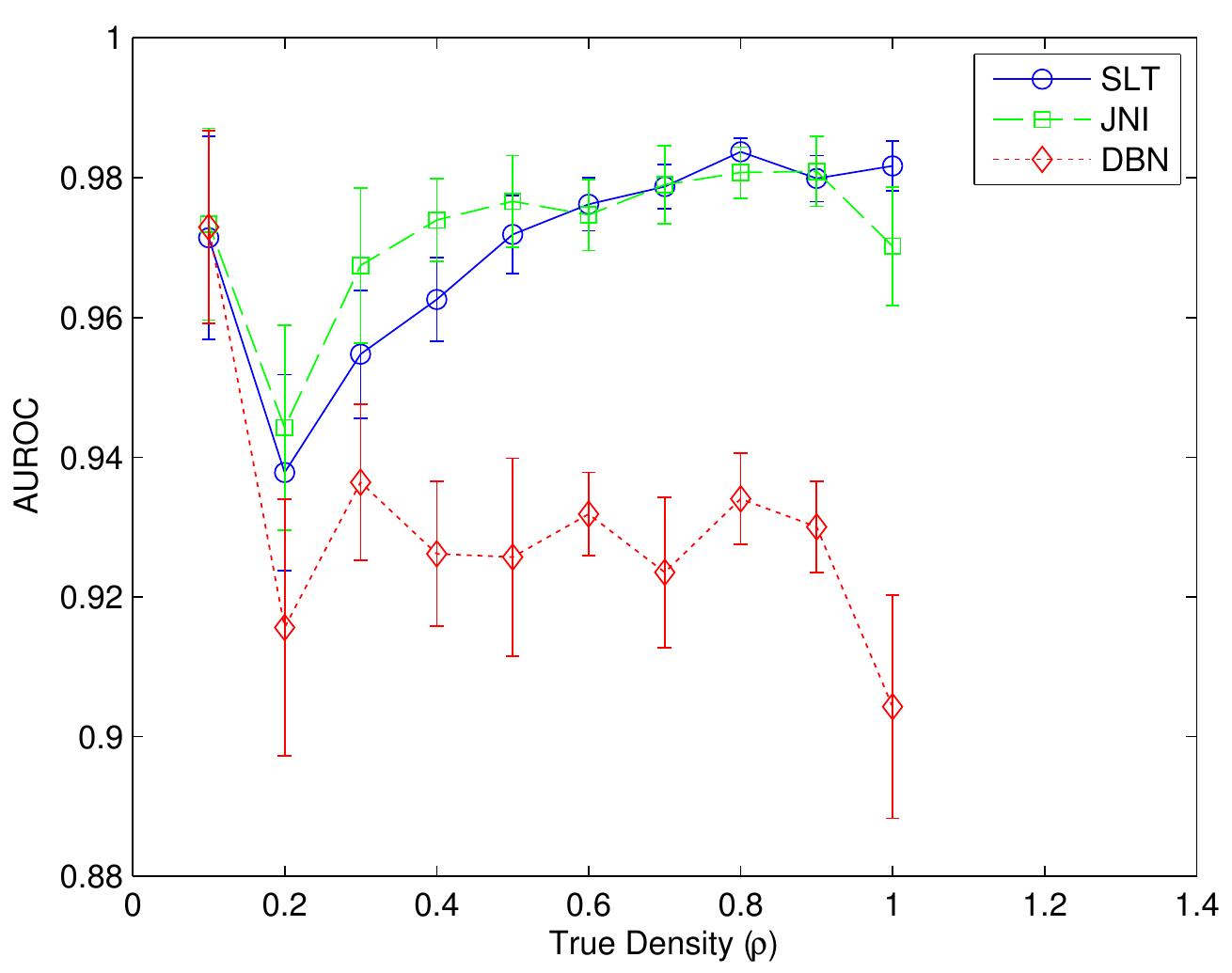}
\includegraphics[width = 0.24\textwidth]{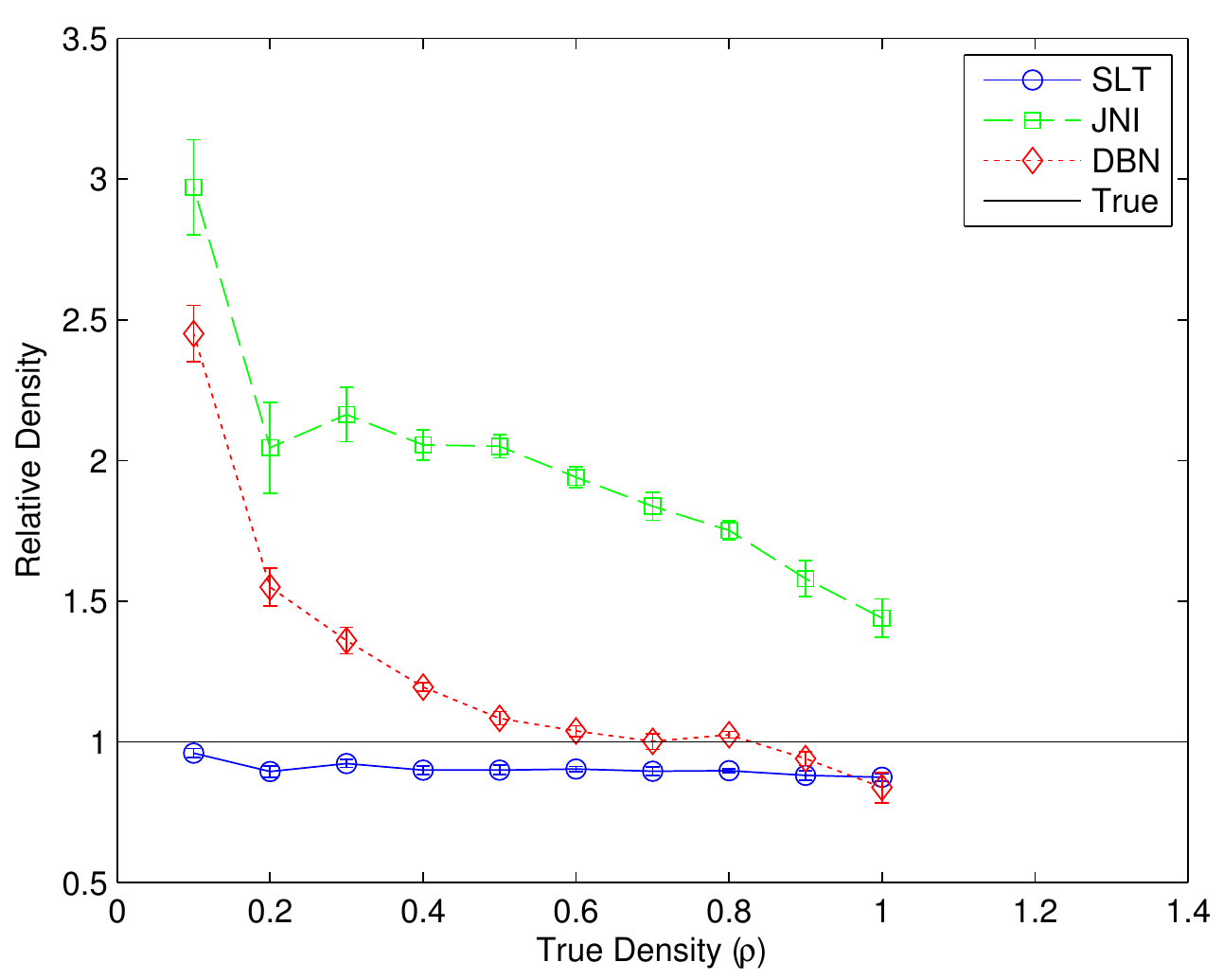}
\includegraphics[width = 0.24\textwidth]{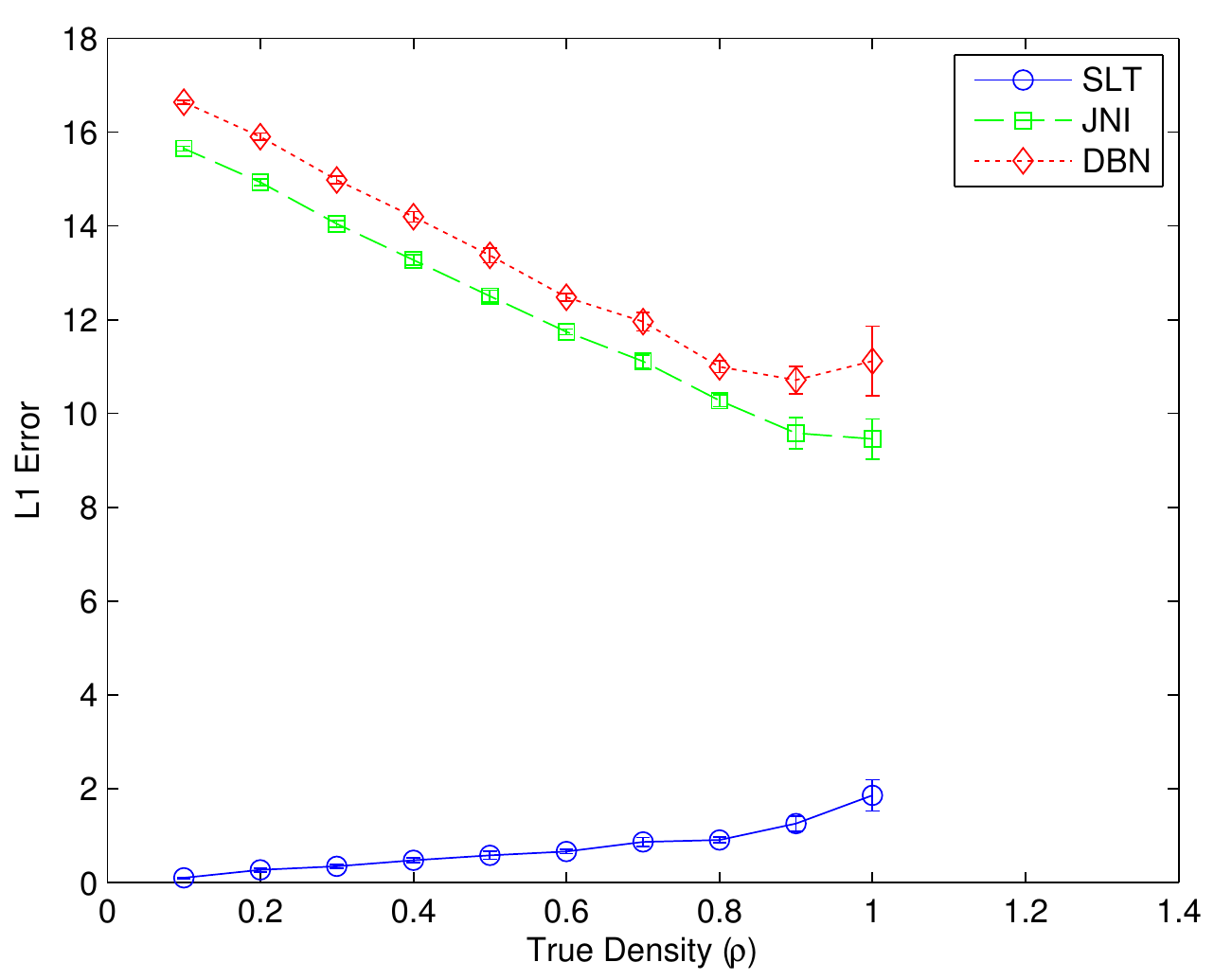}

\includegraphics[width = 0.24\textwidth]{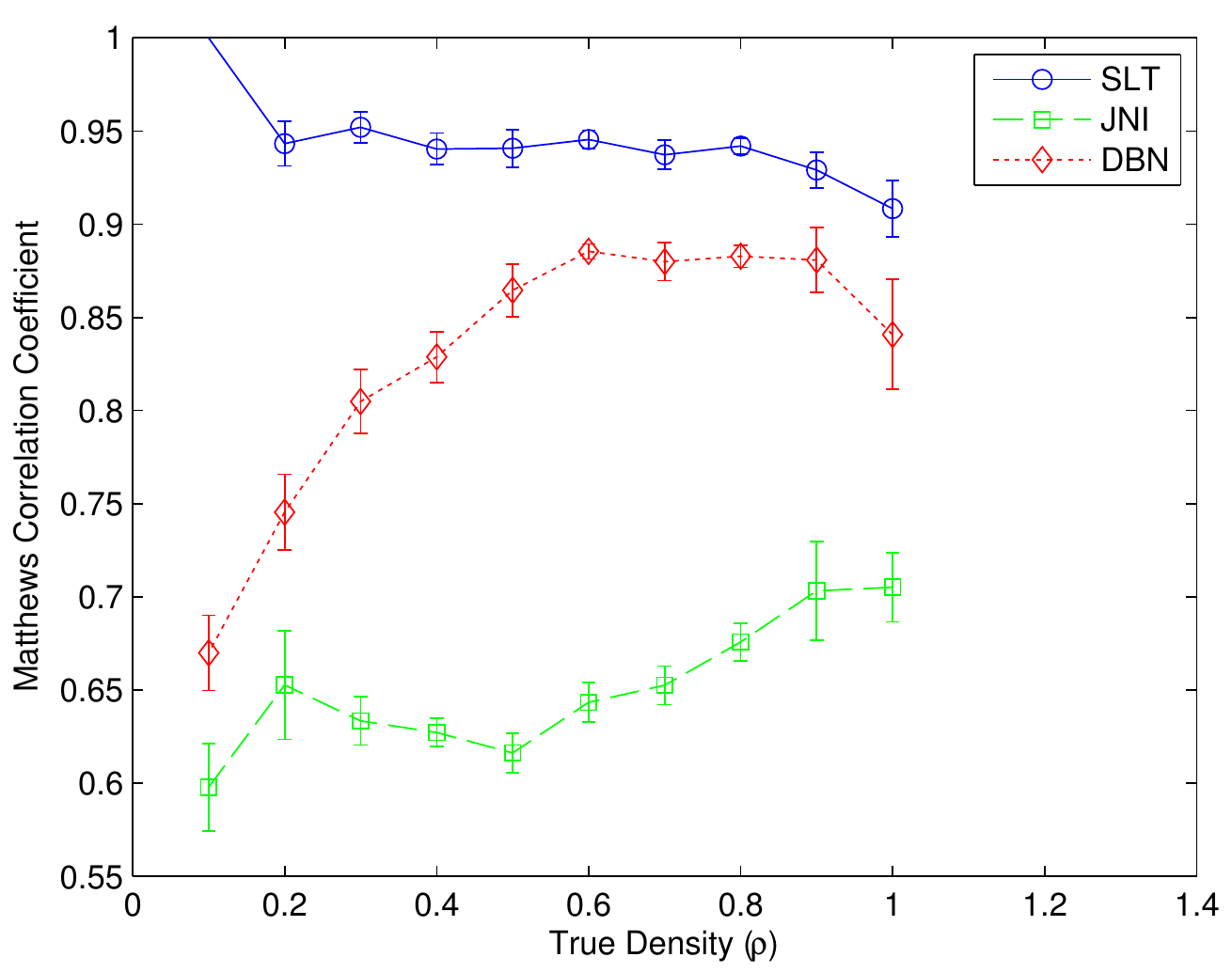}
\includegraphics[width = 0.24\textwidth]{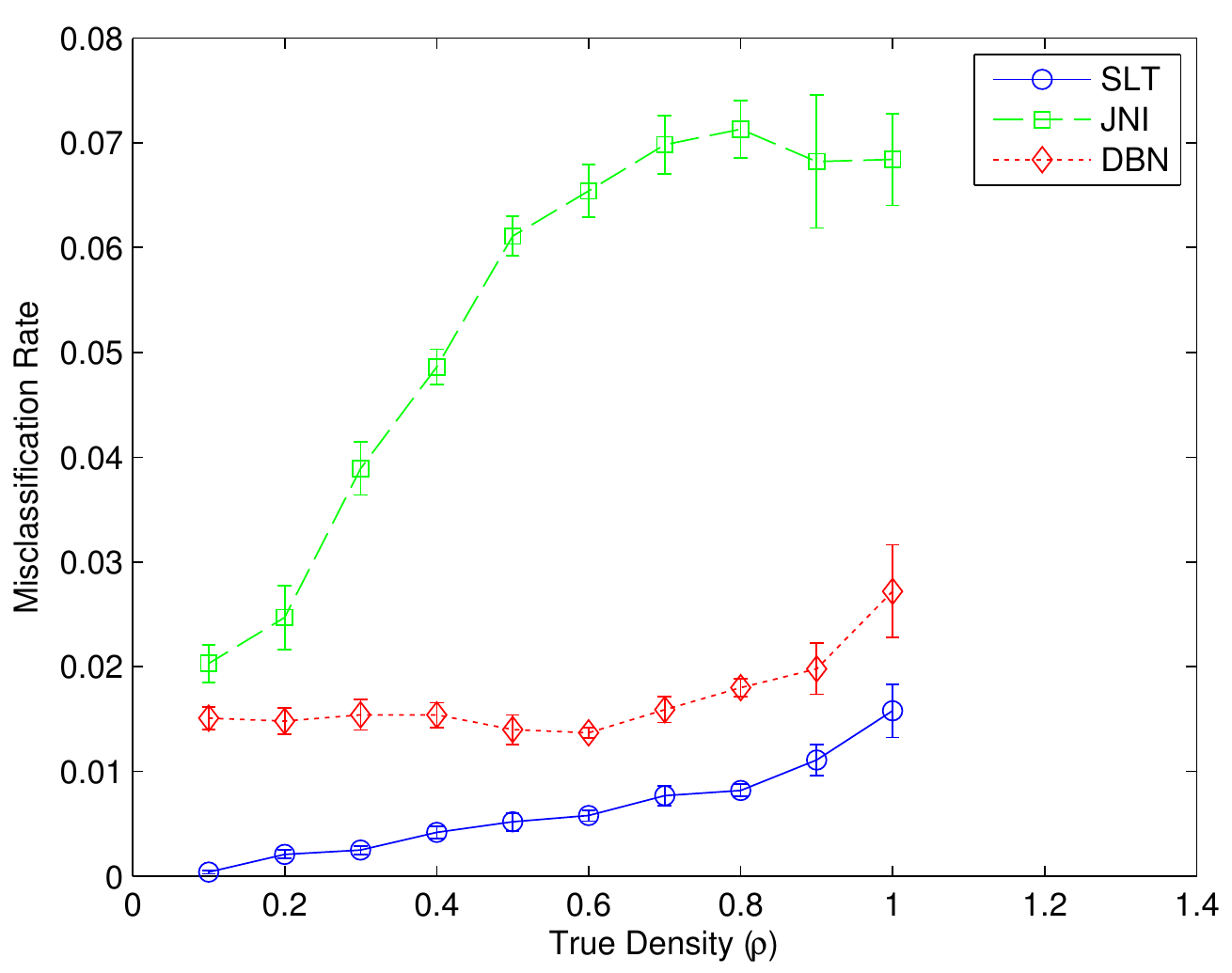}
\includegraphics[width = 0.24\textwidth]{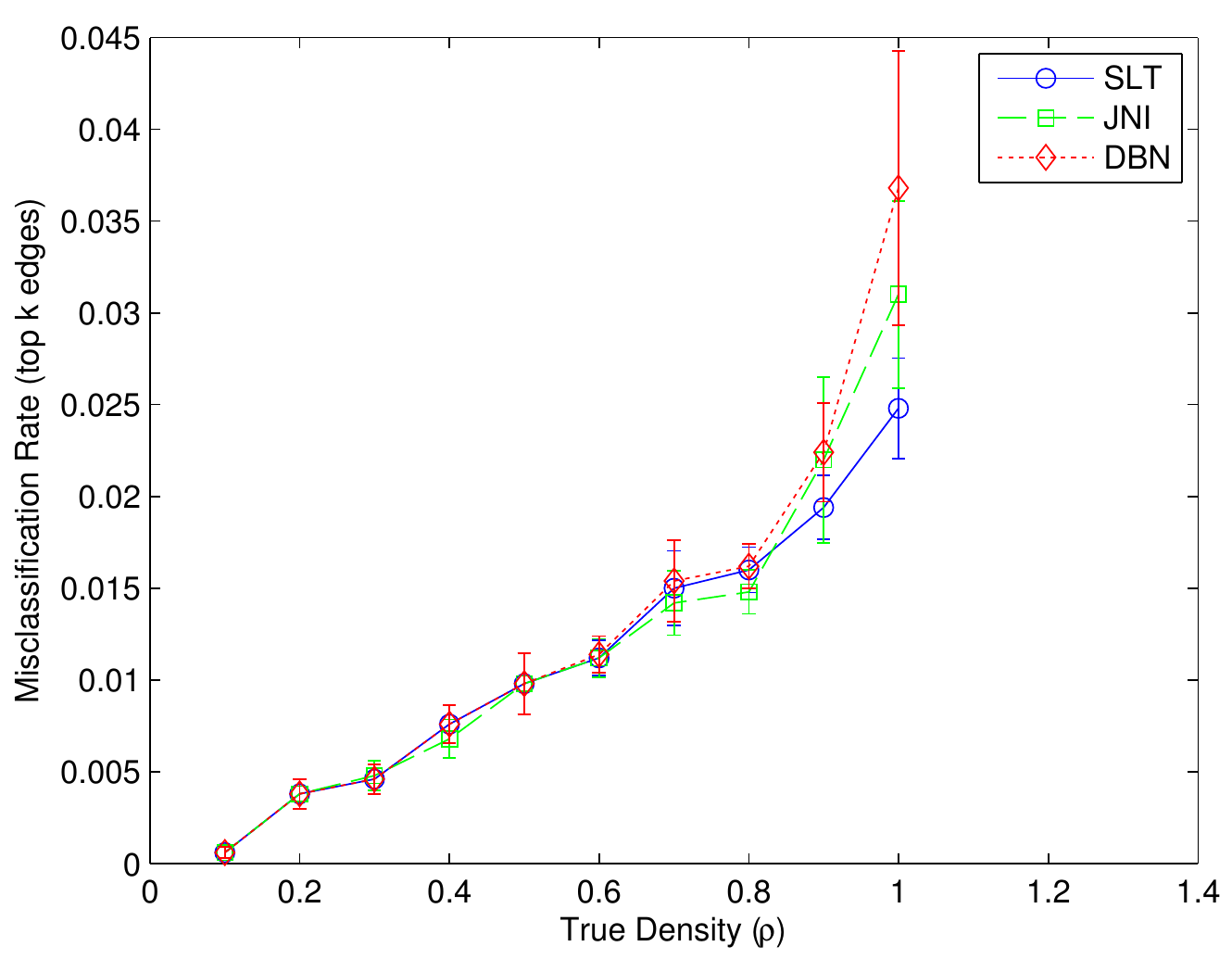}
\includegraphics[width = 0.24\textwidth]{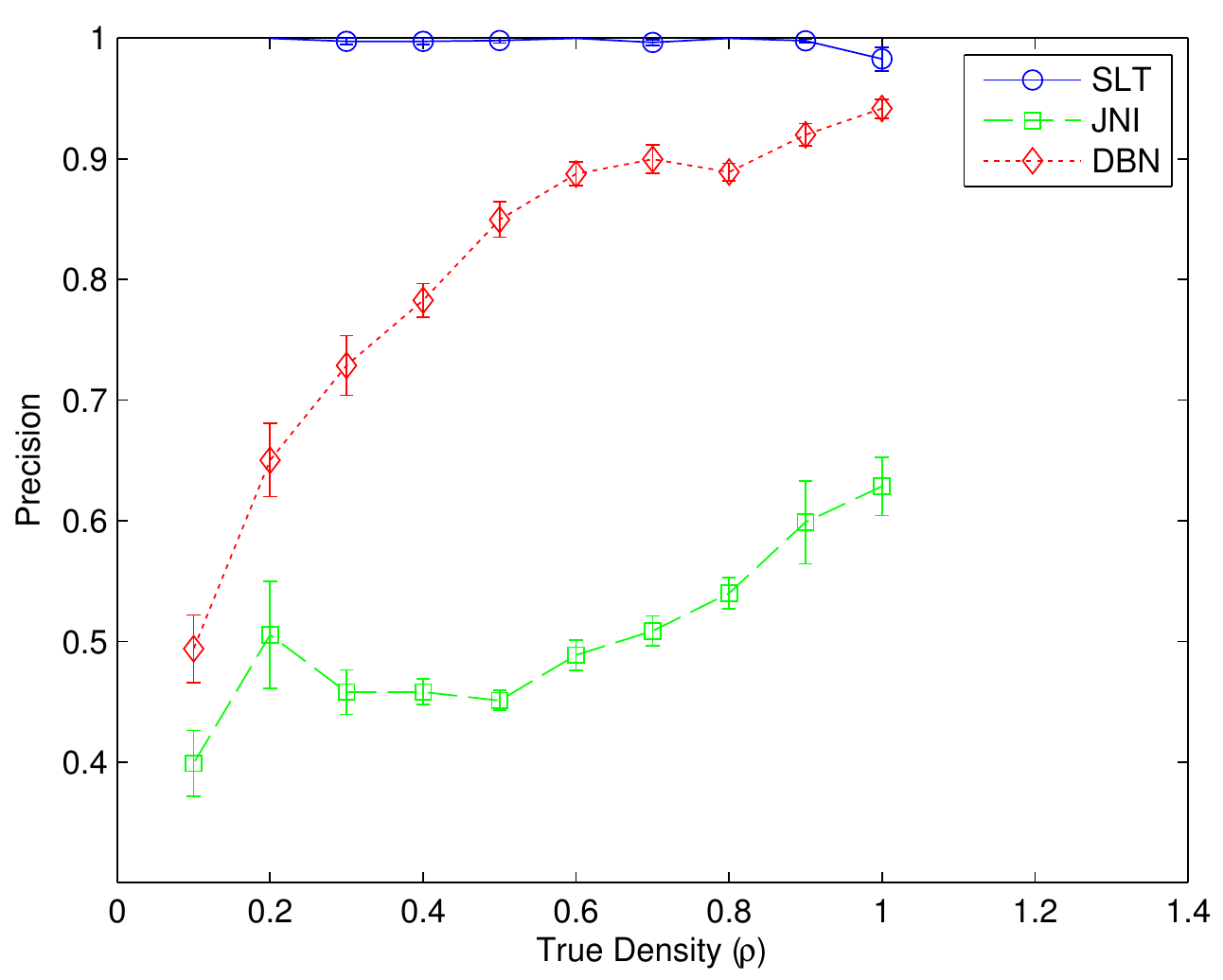}

\caption{Results on simulated data generated from a 2-tier SLT with misspecified tree structure. [Network estimators: ``SLT'' = structure learning trees; ``JNI'' = joint network inference; ``DBN'' = inference for each network independently. For each estimator we considered both thresholded and un-thresholded adjacency matrices. Performance scores: ``AUROC'' = area under the receiver operating characteristic curve; ``AUPR'' = area under the precision-recall curve; ``L1 Error'' = $\ell_1$ distance from the true adjacency matrices to the inferred weighted adjacency matrices; ``top k edges'' = the $\rho P$ most probable edges. Performance scores were averaged over all 10 data-generating networks and all 10 datasets; error bars denote standard errors of mean performance over datasets. We considered both varying $n$ for fixed $\rho = 0.5$ and varying $\rho$ for fixed $n=60$.]}
\label{sim res4}
\end{figure*}

\begin{figure*}[t!]
\centering
\includegraphics[width = 0.24\textwidth]{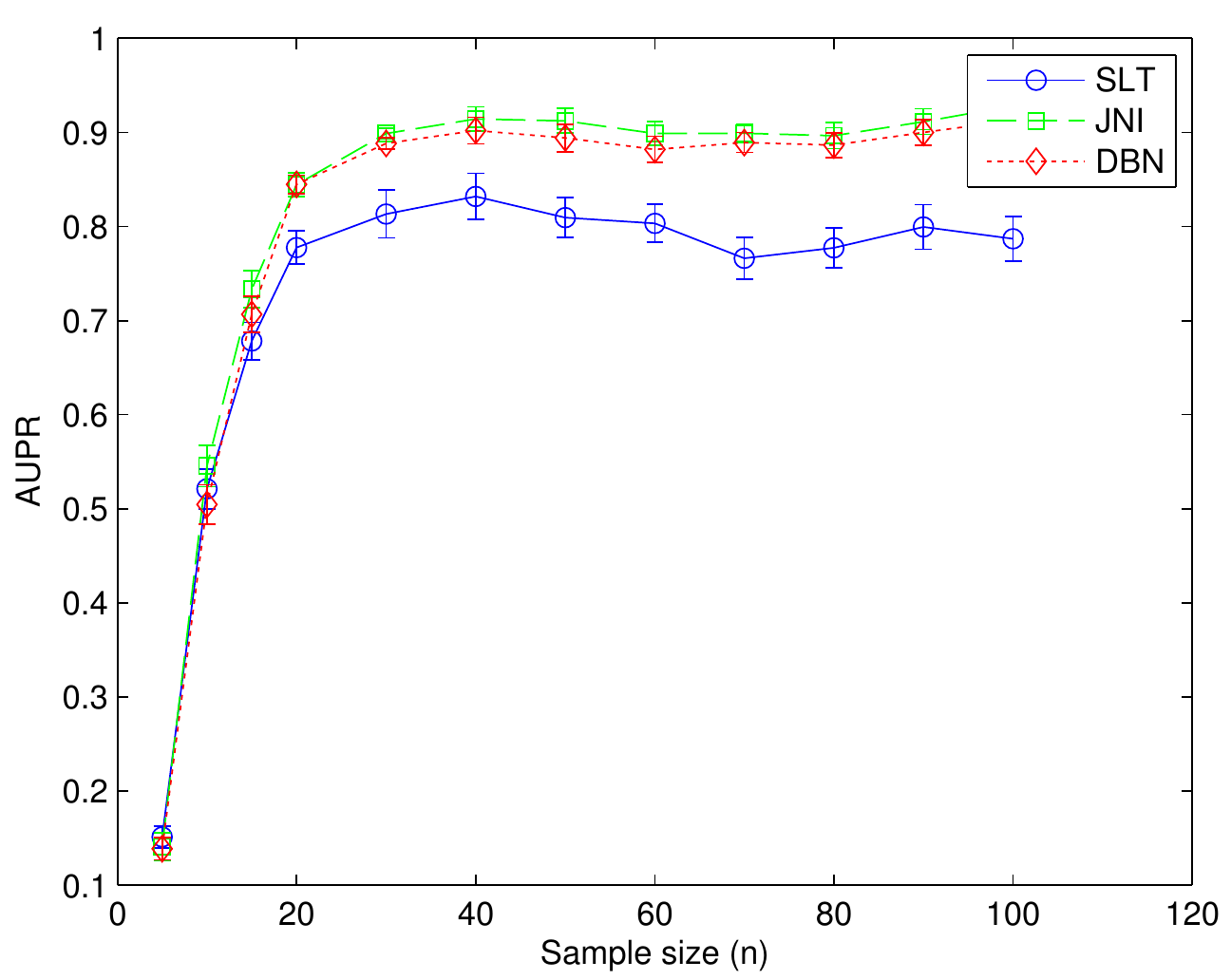}
\includegraphics[width = 0.24\textwidth]{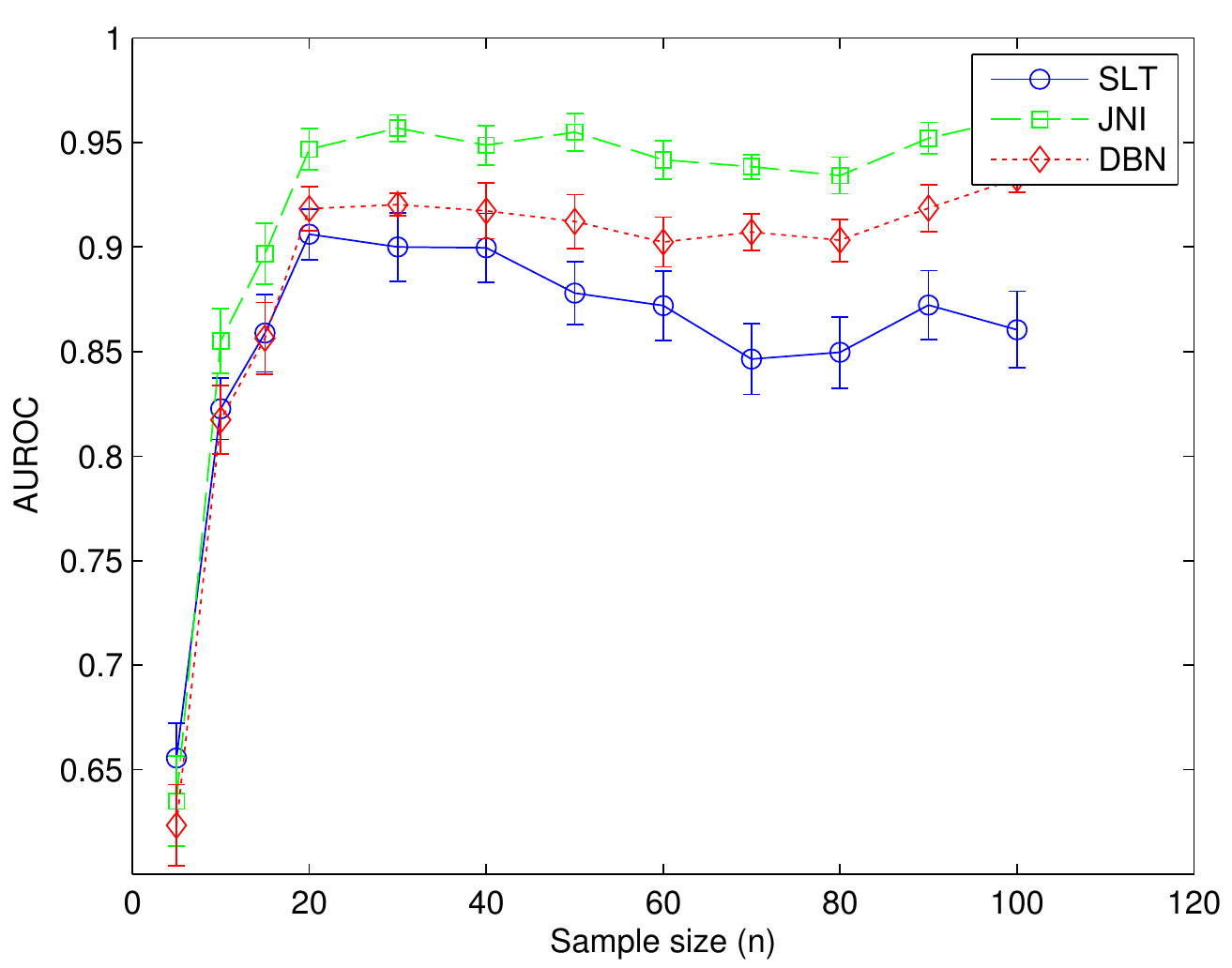}
\includegraphics[width = 0.24\textwidth]{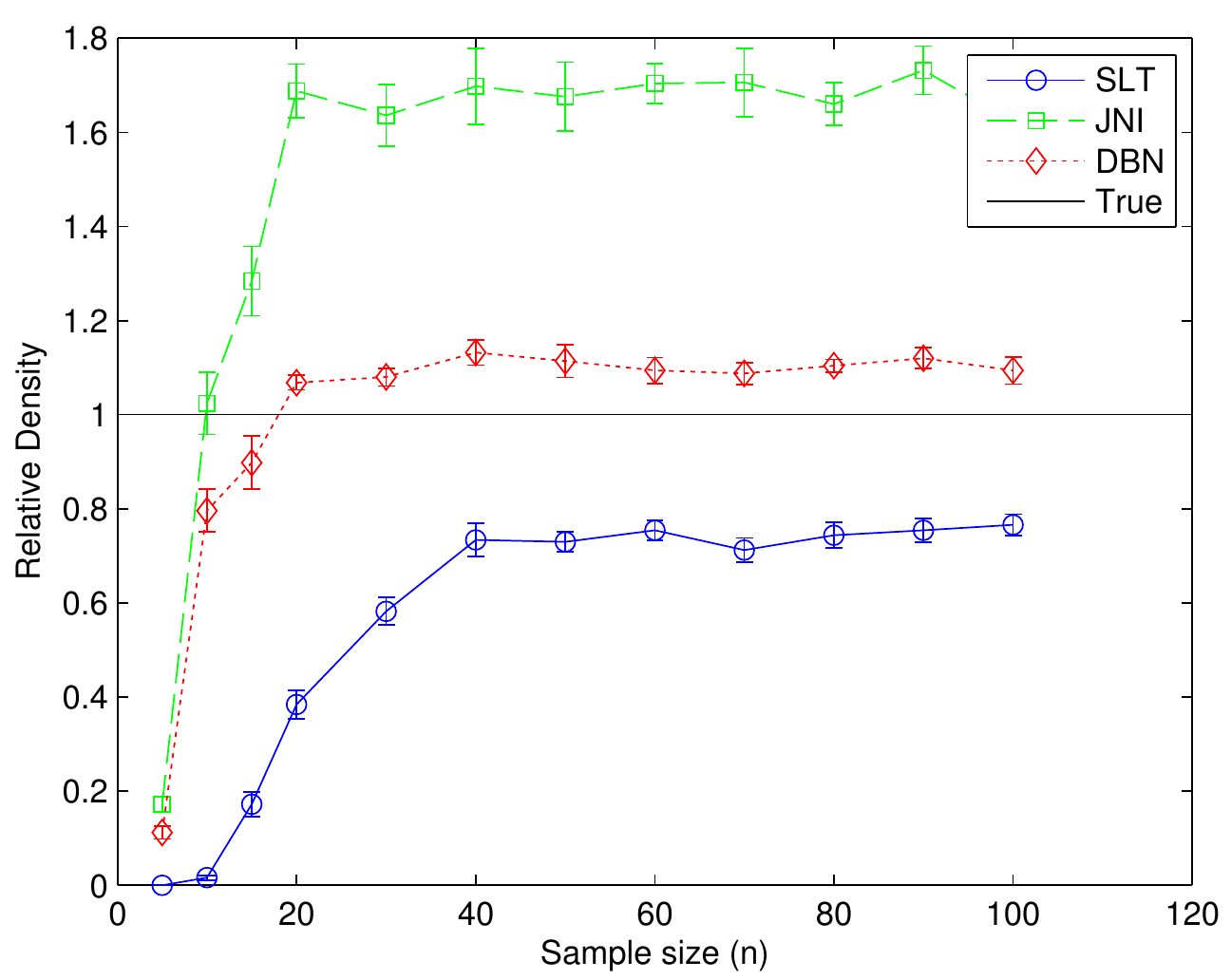}
\includegraphics[width = 0.24\textwidth]{Figures/vary_n_5_L1_Error.pdf}

\includegraphics[width = 0.24\textwidth]{Figures/vary_n_5_Matthews_Correlation_Coefficient.pdf}
\includegraphics[width = 0.24\textwidth]{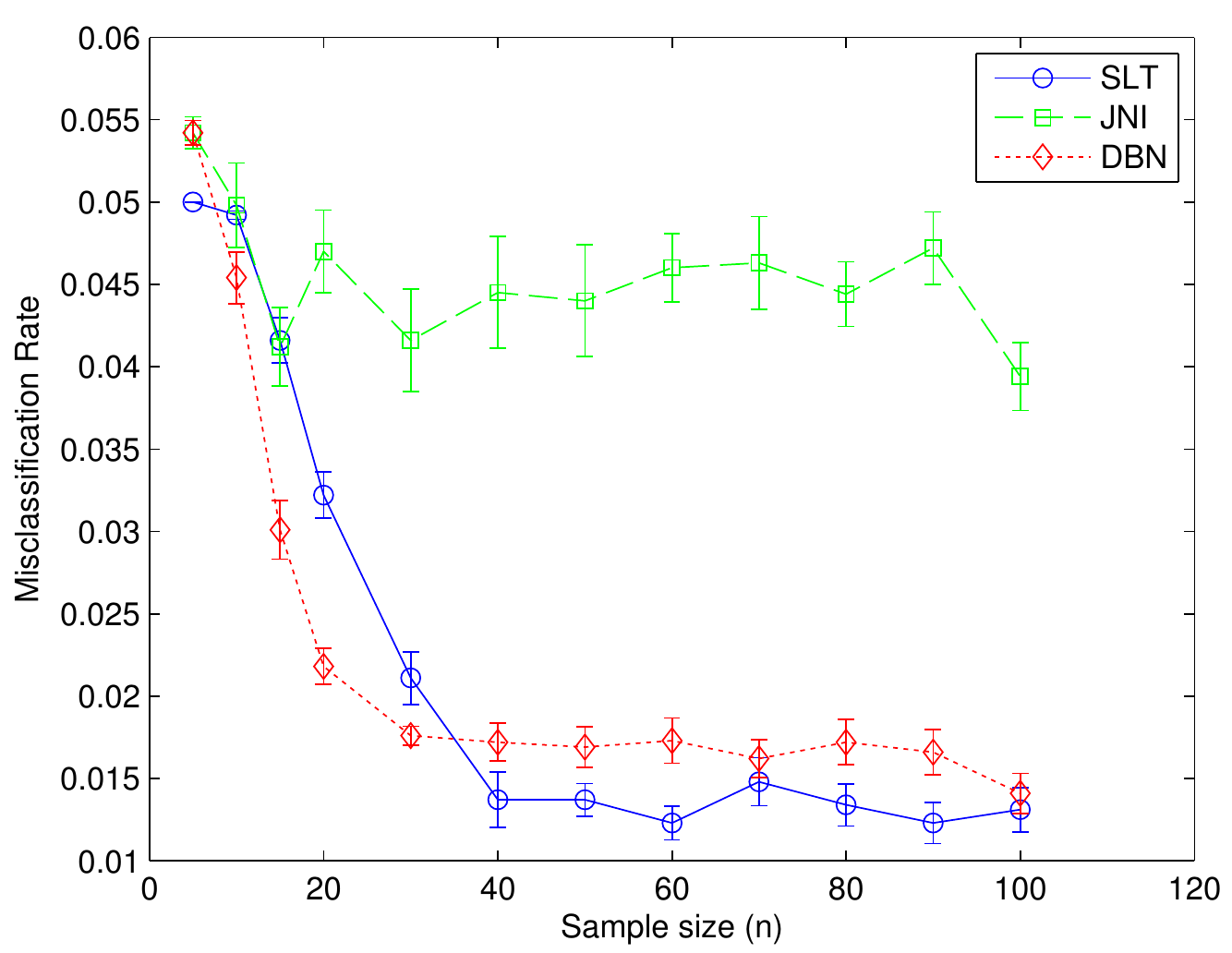}
\includegraphics[width = 0.24\textwidth]{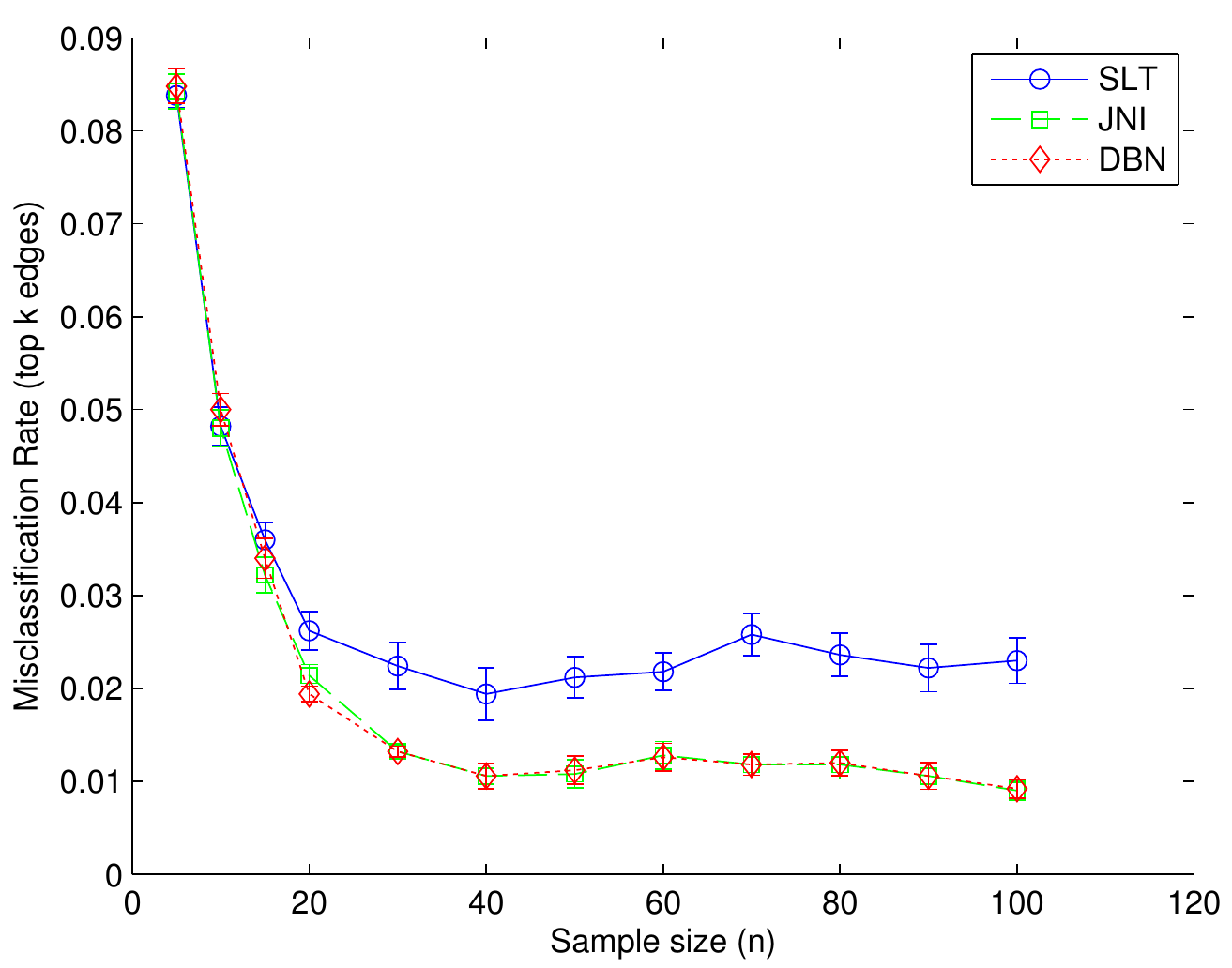}
\includegraphics[width = 0.24\textwidth]{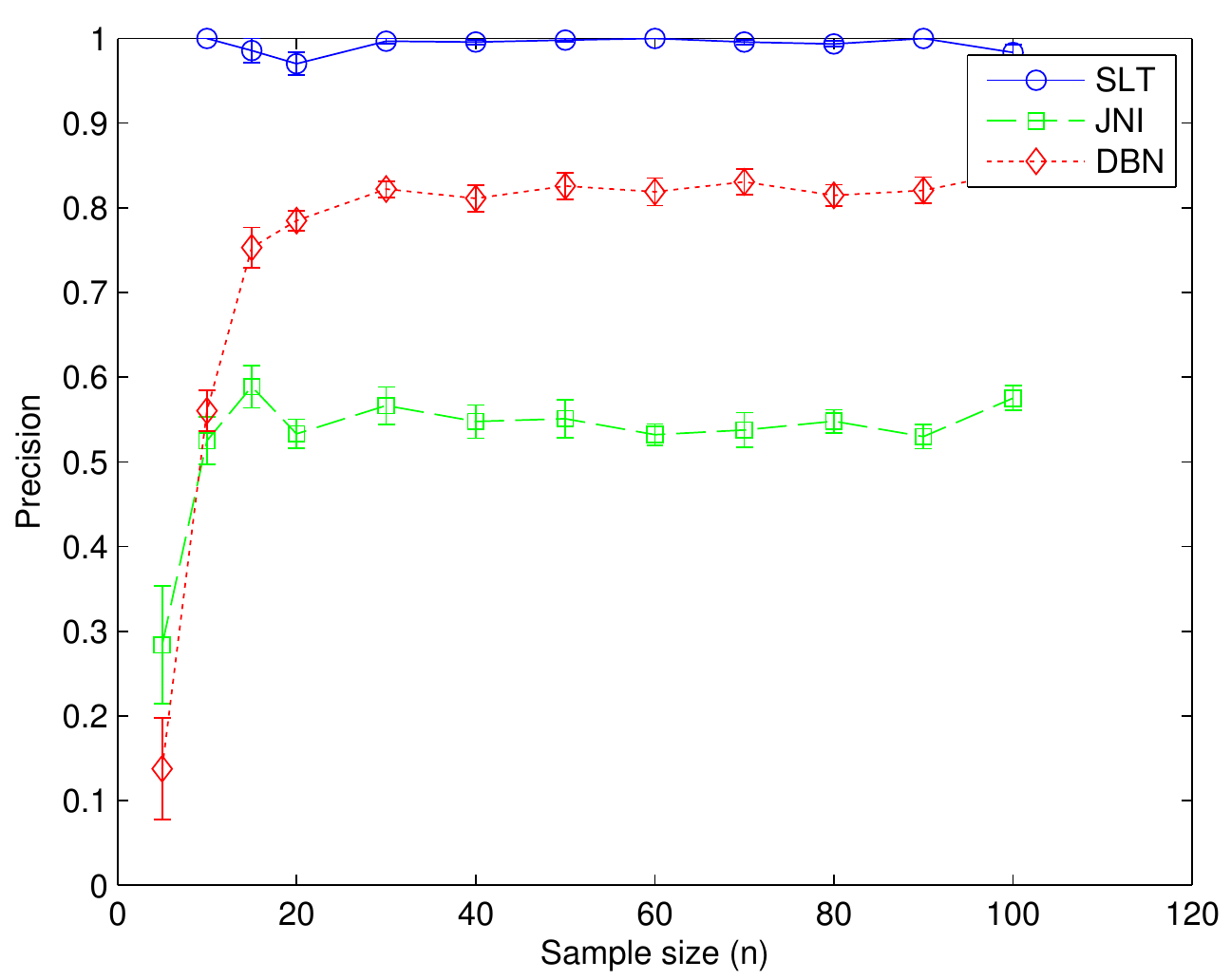}

\includegraphics[width = 0.24\textwidth]{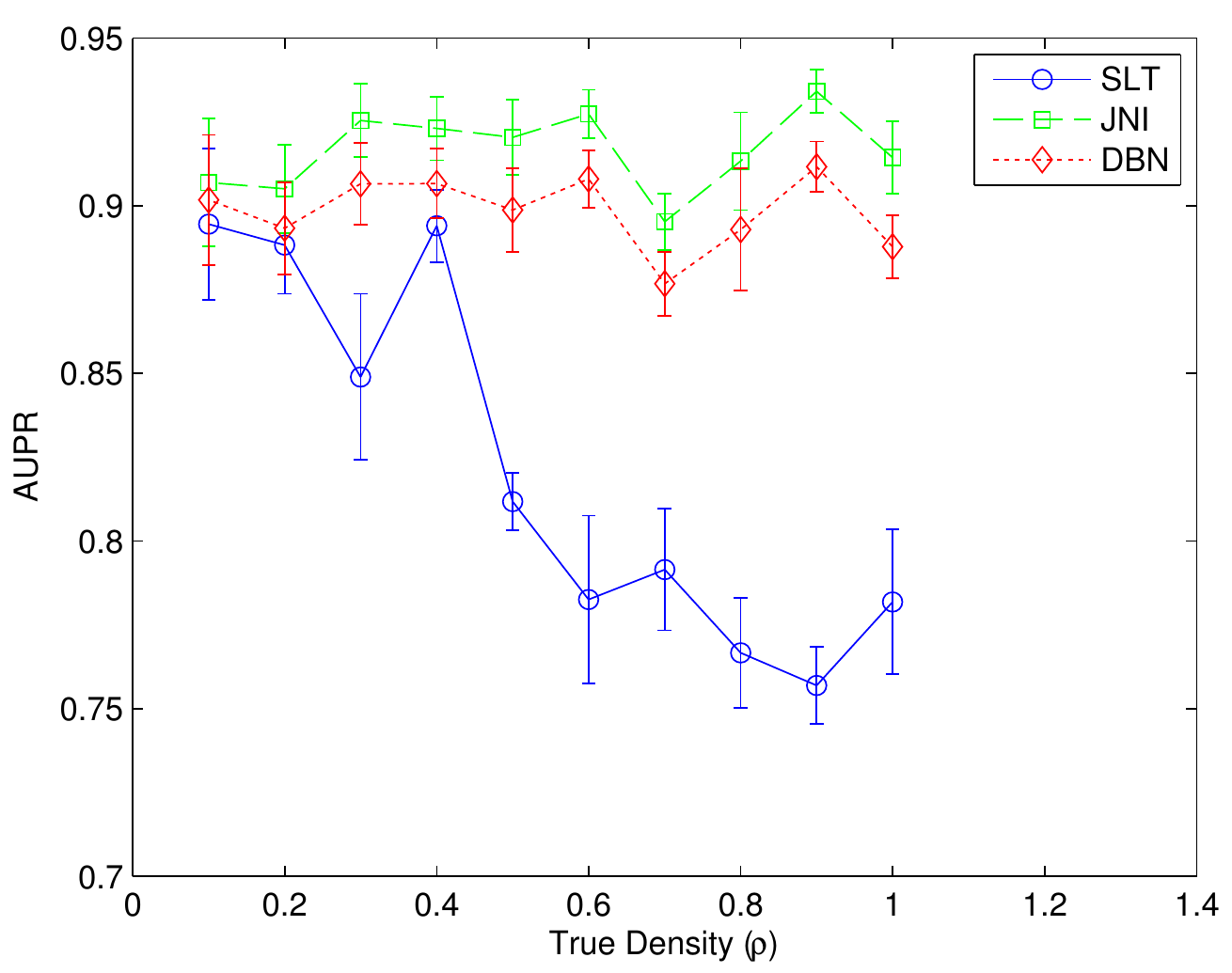}
\includegraphics[width = 0.24\textwidth]{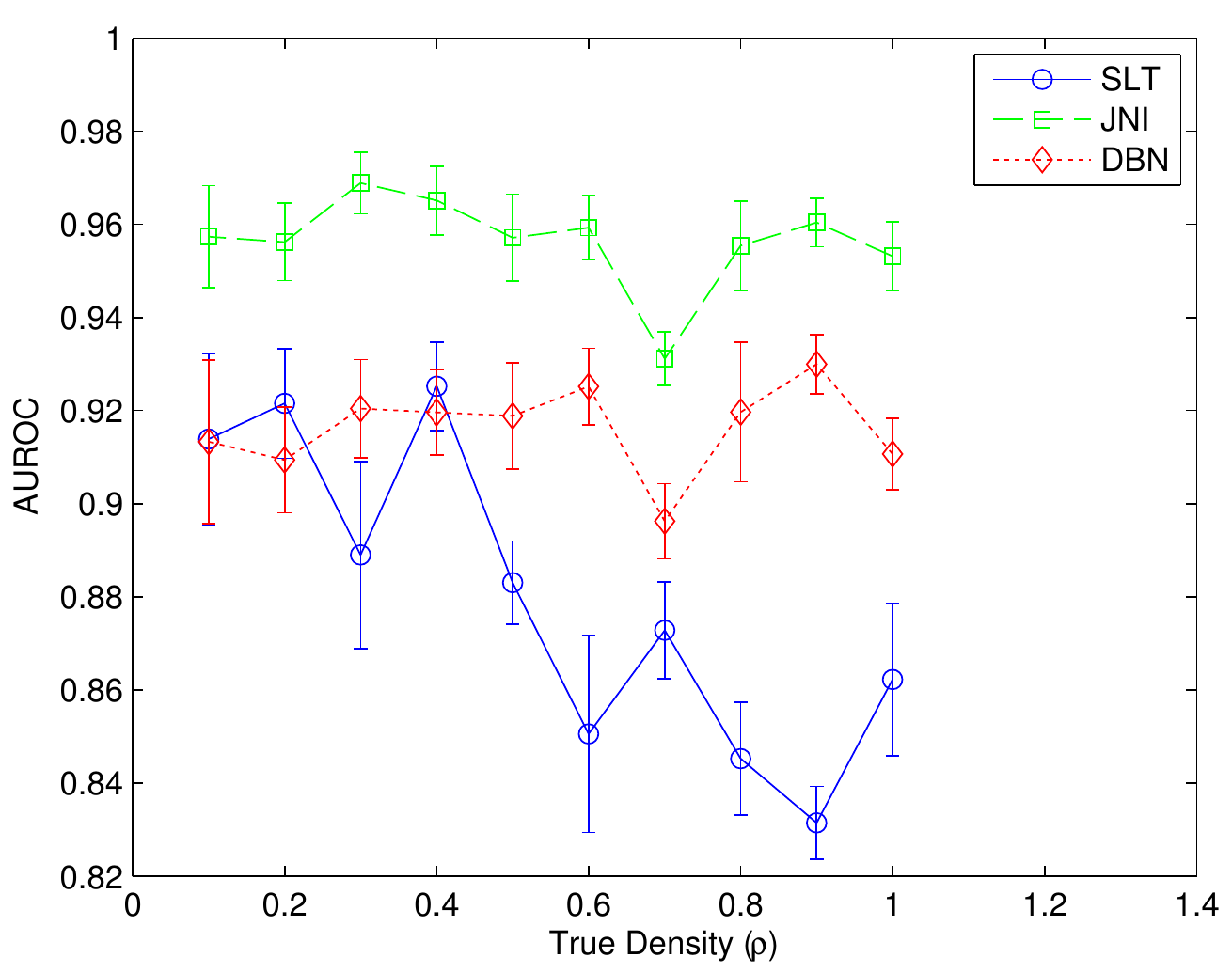}
\includegraphics[width = 0.24\textwidth]{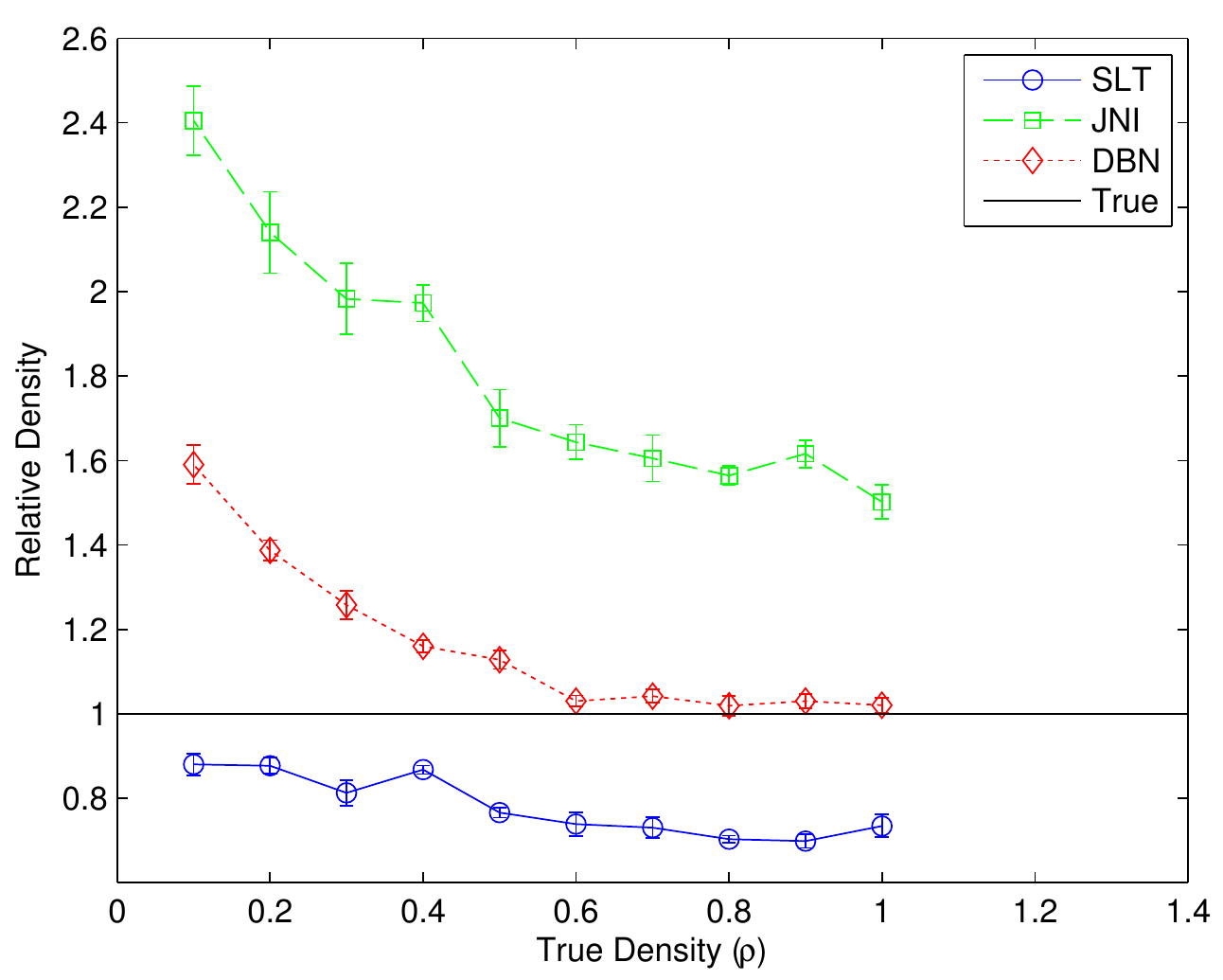}
\includegraphics[width = 0.24\textwidth]{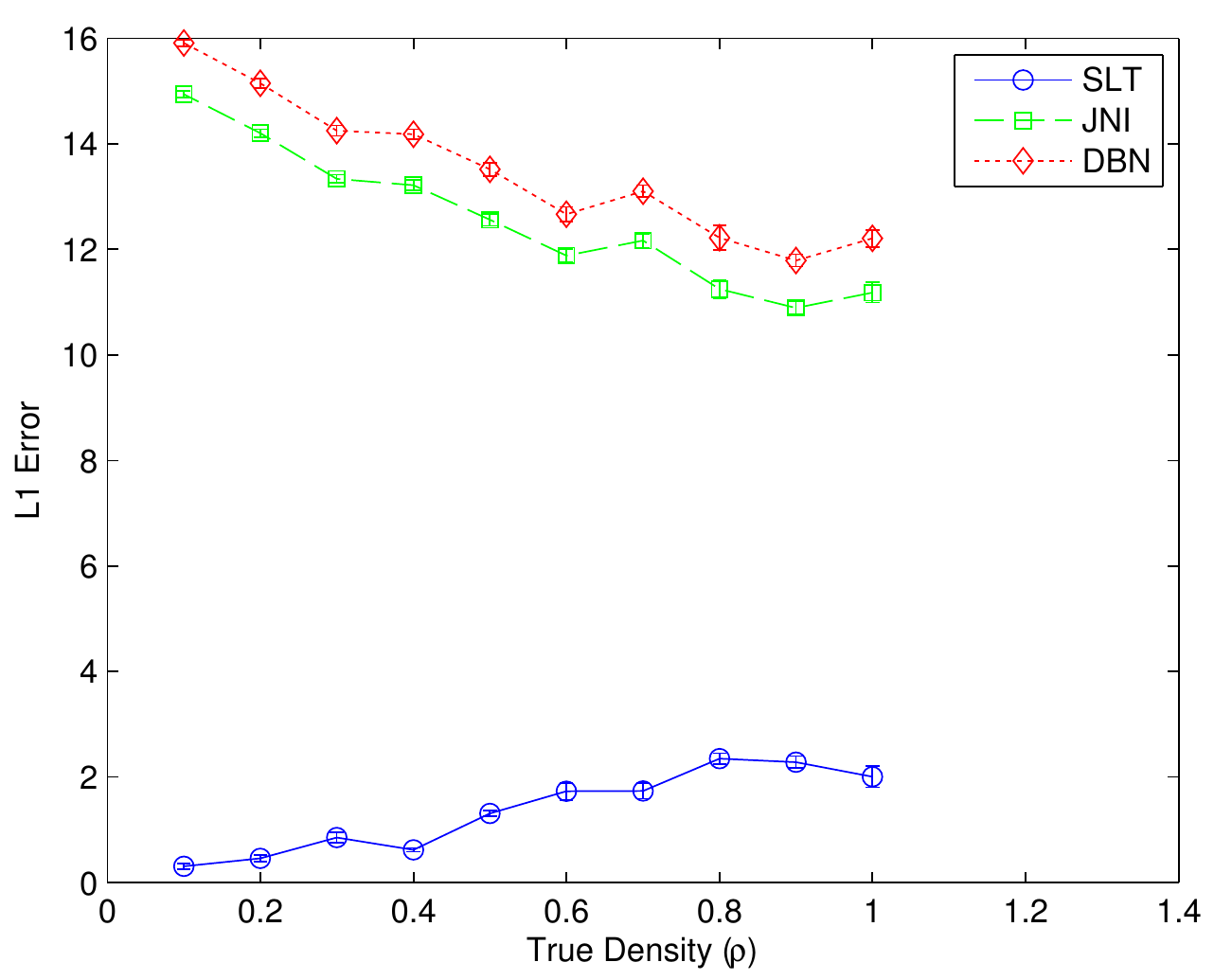}

\includegraphics[width = 0.24\textwidth]{Figures/vary_sparsity_5_Matthews_Correlation_Coefficient.pdf}
\includegraphics[width = 0.24\textwidth]{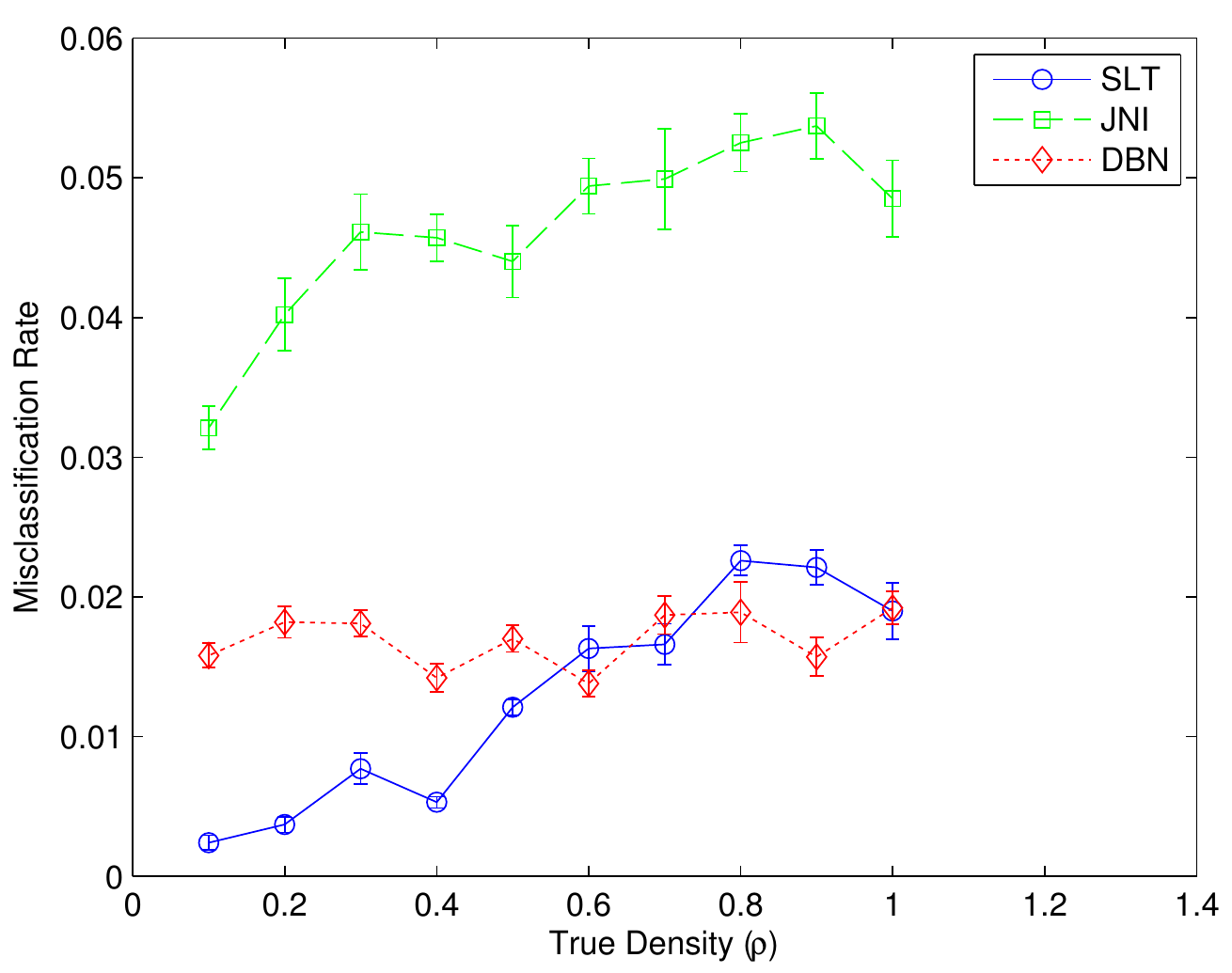}
\includegraphics[width = 0.24\textwidth]{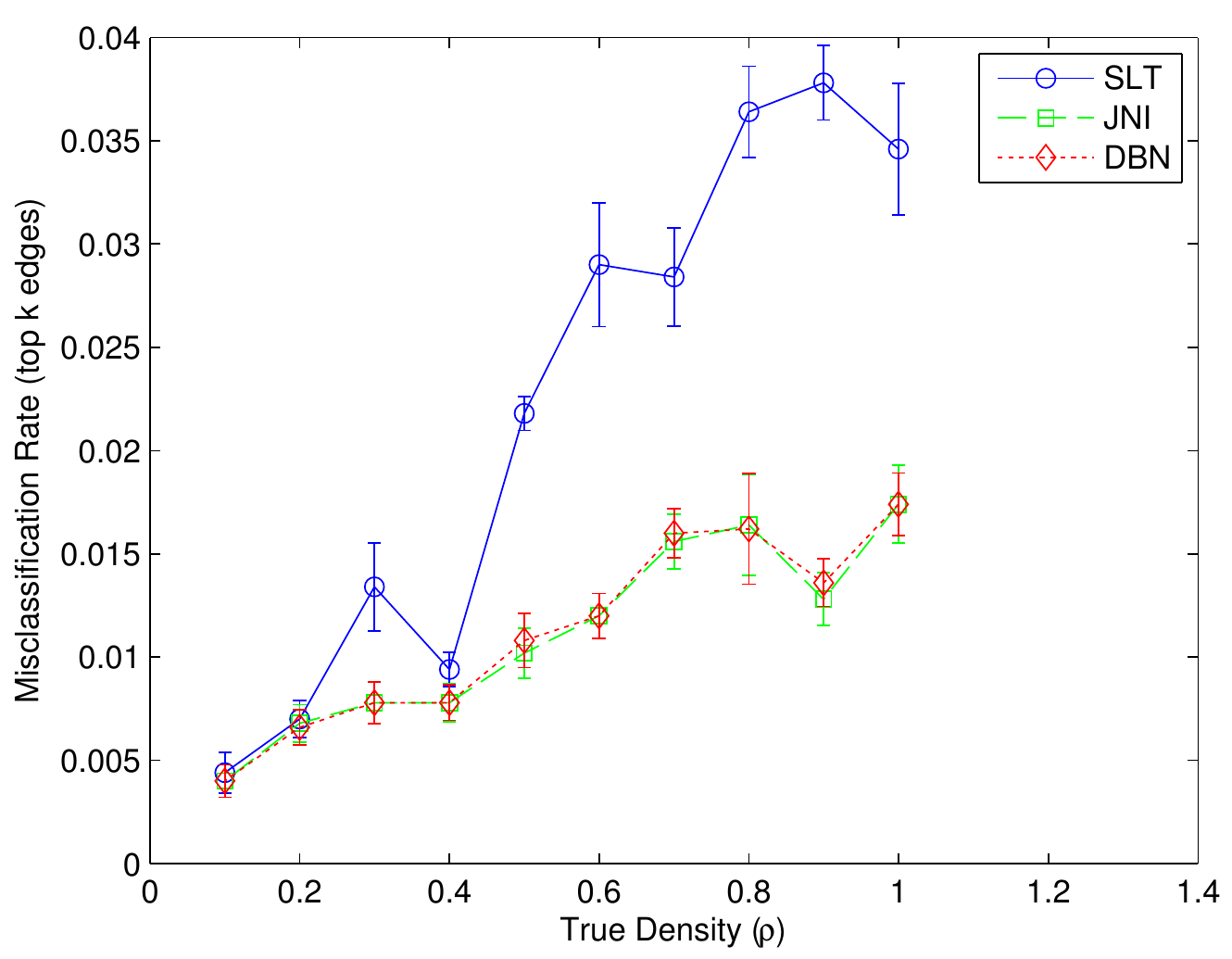}
\includegraphics[width = 0.24\textwidth]{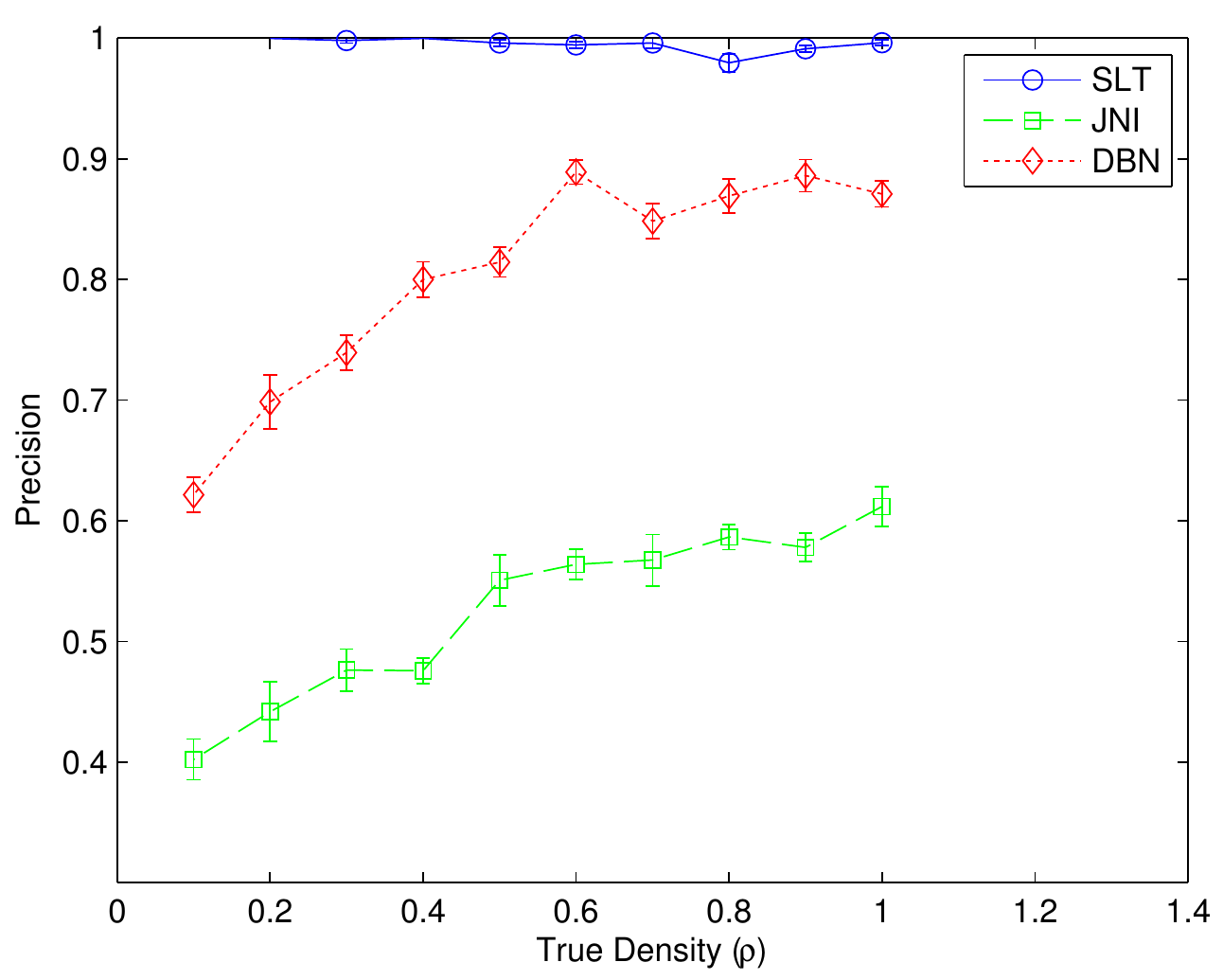}

\caption{Results on simulated data generated from a 2-tier SLT with structure which violates the subset assumption. [Network estimators: ``SLT'' = structure learning trees; ``JNI'' = joint network inference; ``DBN'' = inference for each tier-3 network independently. For each estimator we considered both thresholded and un-thresholded adjacency matrices. Performance scores: ``AUROC'' = area under the receiver operating characteristic curve; ``AUPR'' = area under the precision-recall curve; ``L1 Error'' = $\ell_1$ distance from the true adjacency matrices to the inferred weighted adjacency matrices; ``top k edges'' = the $\rho P$ most probable edges. Performance scores were averaged over all 10 data-generating networks and all 10 datasets; error bars denote standard errors of mean performance over datasets. We considered both varying $n$ for fixed $\rho = 0.5$ and varying $\rho$ for fixed $n=60$.]}
\label{sim res5}
\end{figure*}

In order to probe robustness of SLT to prior mis-specification we considered two scenarios in which assumptions encoded in the joint structural prior are  violated.
Firstly, we investigated whether the performance of SLT deteriorates when the tree $T$ itself is misspecified (in fact chosen randomly; regime (4), SFig. \ref{ex4}).
These results showed that SLT remains superior to JNI and DBN terms of MCC, and remains competitive in terms of AUPR and AUROC (SFig. \ref{sim res4}).
Secondly, we considered strongly violating the subset inclusions (regime (5); SFig. \ref{ex5}). 
The MCC performance of SLT in this regime was competitive with JNI and DBN (Fig. \ref{reg3}).
However SLT performed worse than JNI and DBN in terms of AUPR, AUROC and misclassification rate (SFig. \ref{sim res5}).
Robustness of SLTs is therefore dependent upon which aspects of performance are being considered.

The above experiments were performed at constant edge density $\rho = 1/2$, however SLT tends to produce sparser networks {\it a priori}. We therefore repeated the above experiments whilst varying the true density $\rho$ and holding the number of samples constant at $n = 60$. 
Results (Fig. \ref{reg}, SFigs. \ref{sim res1}-\ref{sim res5}) showed that, in all regimes, performance of SLT improves in sparse settings whilst the performance of both JNI and DBN deteriorate.
Examining the density of estimated networks relative to the data-generating networks (SFigs. \ref{sim res1}-\ref{sim res5}) we found that JNI and DBN dramatically over-estimate density in sparse regimes; in contrast SLT automatically adjusts to the density of the data-generating networks.
This appealing property results from our novel subset prior of Eqn. \ref{model eq}.

\subsection{Biological data} \label{results 2}

This work was motivated by the problem of inference for protein signalling networks (PSNs) over a diverse panel of breast cancer cell lines. 
The cell lines under study are expected to differ with respect to PSN structure but can be grouped into sub-types based on underlying biology, as described below. 
Here independent estimation is likely to be inefficient, since the cell lines have a common lineage and share much of their biology. 
On there other hand, since sub-types may be quite different from one another, exchangeability within sub-type is arguably a more reasonable assumption  than exchangeability between sub-type.

\begin{figure*}[t!]
\centering
\includegraphics[clip,trim = 5cm 14cm 4cm 4.5cm,width = 0.98\textwidth]{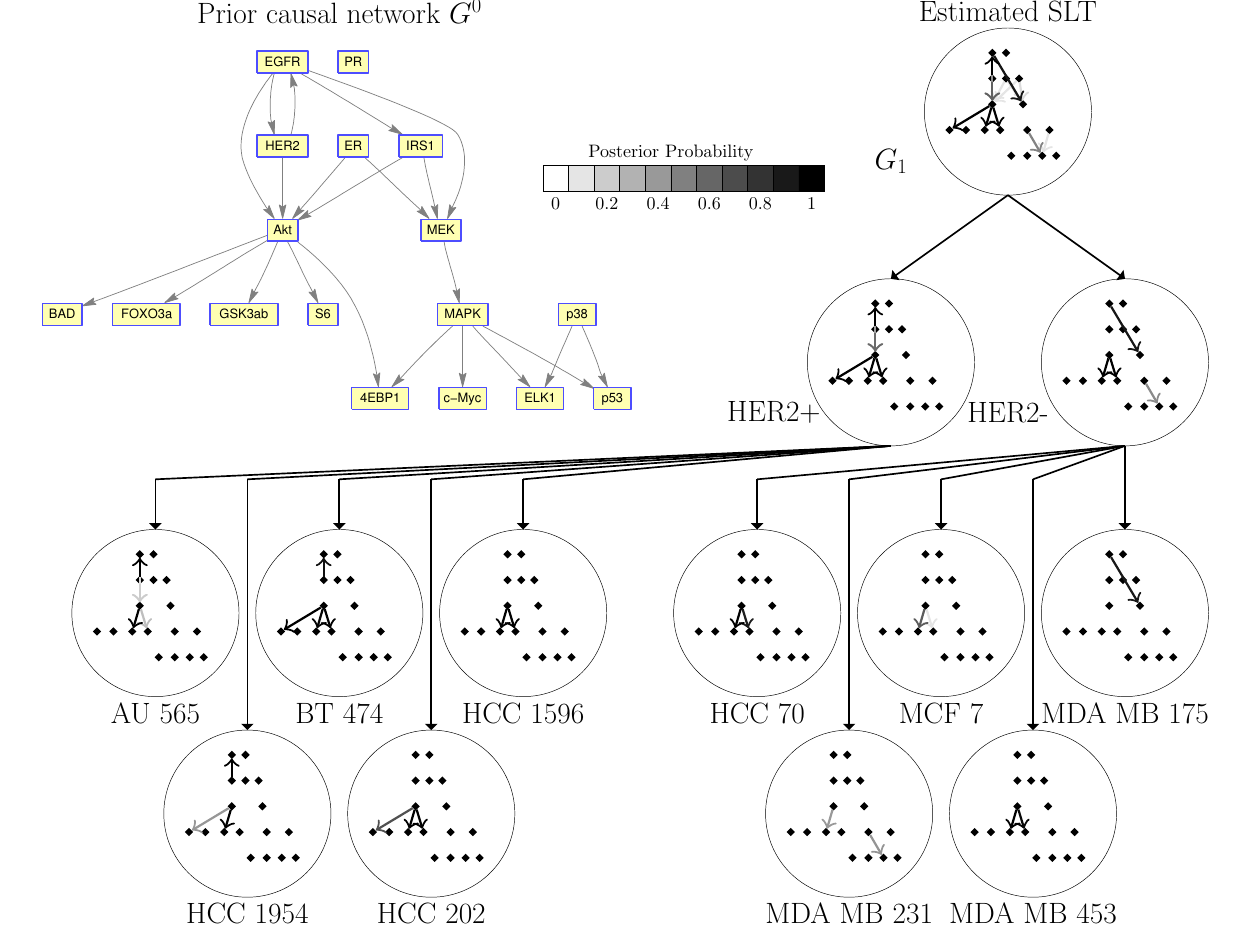}
\caption{Results, experimental data. Sub-type and cell-line-specific protein signalling networks were inferred from proteomic data obtained from a panel of breast cancer cell lines. [The prior network $G^0$ (top left) may be used as a key for the vertex labels on smaller networks. Edge shading indicates posterior marginal inclusion probabilities as shown in the legend.]}
\label{panel}
\end{figure*}

Amplification of the HER2 gene  (denoted as ``HER2+") is a key biomarker used to stratify breast cancer samples and cell lines.
HER2 codes for a receptor that is a member of the EGFR family of receptors and it is believed that signalling related to these receptors 
may differ between these two sub-types.
However, it is challenging to study signalling at the group level {\it per se}, since  within
each sub-type there remains considerable genetic diversity. 
We therefore applied SLT to learn both cell-line-specific and group-level PSNs, whilst controlling for confounding due to both HER2 status and line-specific genomic characteristics.
Specifically, we constructed a tree $T$ such that the doubly latent networks $\overline{G_{1i}}$ define HER2+/- sub-types respectively and the data-generating networks $\overline{G_{1ij}}$ correspond to PSNs in cell lines $j$ of sub-type $i$.
We used an informative prior network $G^0$ derived from the signalling literature (Fig. \ref{panel}, top left).

Reverse phase protein array data \citep{Hennessy} were obtained over a panel of 10 breast cancer cell lines \citep{Neve} of which half were HER2+ and half HER2-. 
Data consisted of $P = 17$ protein expression levels, observed at 0.5,1,2,4,8,24,48,72 hours following ligand stimulation.
A total of 4 time series were obtained, under treatment with DMSO, a EGFR/HER2 inhibitor (Lapatinib), an AKT inhibitor (AKTi) and Lapatinib + AKTi in combination, giving a total sample size of $n = 4 \times 8 = 32$.
From a modelling perspective, the drugs Lapatinib and Akti are perturbations in the causal sense of intervening upon a node in the network. We assumed perfect interventions, corresponding to 100\% removal of the target's activity with 100\% specificity.
Full experimental protocol is provided in the Supp. Sec. \ref{protocol}.
Fig. \ref{panel} displays the inferred root network $\overline{G_1}$, the sub-type and cell line networks. 
It is noticeable that HER2 signalling plays a more prominent role in the HER2+ sub-type in line with biological intuition. 
Interestingly we infer regulation of BAD by HER2 (via AKT); dephosphorylation of BAD initiates apoptosis and this may help to explain a differential efficacy of HER2 inhibitors observed between HER2+/- sub-types.
These results illustrate application of SLTs in a topical applied problem; however, inference of network structure from biological data remains extremely challenging \citep{Oates7} and experimental validation of inferred topology is necessary.

\section{Discussion} \label{discussion}

In this paper we introduced a novel methodology, SLT, which generalises joint estimation of multiple networks to the non-exchangeable setting. 
Our empirical results support the notion that SLTs can offer improved estimation relative to existing estimators based on exchangeability.
We illustrated the use of the SLT framework using FFDBNs for which joint estimation could be carried out in a computationally efficient manner. 
However the general SLT approach is applicable in principle to any probabilistic network model for which marginal likelihoods are available.
Thus, in principle SLT formulations could be developed for Bayesian networks, GGMs, or more sophisticated local likelihoods, for example based on differential equations \citep{Nelander}.

In empirical studies we considered FFDBNs of dimension $P =$ 10 and 17; in this setting, exact inference using SLTs was massively faster than (exchangeable) alternatives based on MCMC \citep{Penfold,Werhli}. 
The (serial) computational complexity of our approach applied to a tree $T$ is at worst $\mathcal{O}(h_1 h_2 \dots h_t c(P))$, where $h_i$ is the number of networks that are tree distance $i$ from the root network $G_1$ and $t$ is the number of tiers in $T$.
Thus in our cancer example, inclusion of more cancer sub-types or cell lines is computationally cheap (linear in both $h_1$ and $h_2$).
For FFDBNs, $c(P) = P^{1+2d_{\max}}$ so that SLT has the same computational complexity as a fully exchangeable formulation (JNI), but requires $\mathcal{O}(P^{d_{\max}})$ more computation than the classical non-joint approach.

Extensions of theoretical interest include: (i) The case where $T$ itself is unknown; here the challenge is to jointly learn both individual-specific networks and tree structure. In principle this could be accomplished using the SLT model described here, but further work would be needed to render this tractable for non-trivial applications.  (ii) The case of arbitrarily-structured populations, where $T$ need not be a tree, or where data may be associated with multiple networks; here MCMC methods similar to \cite{Dondelinger_12} or approximate inference algorithms such as loopy belief propagation may prove effective.

\subsubsection*{Acknowledgments} 

The authors wish to thank Frank Dondelinger, Steven Hill, Dan Woodcock and Chris Penfold.
Data courtesy James Korkola and Joe W. Gray, Oregon Health and Science University.
CJO supported by UK EPSRC EP/E501311/1.
SM supported by NCI U54 CA112970 and the Cancer Systems Center grant from the Netherlands Organisation for Scientific Research.

\FloatBarrier

\clearpage

\onecolumn

\section{Supplement for ``Joint Structure Learning of Multiple Non-Exchangeable Networks'', Chris. J. Oates and Sach Mukherjee, AISTATS 2014.}

\subsection{Belief propagation for SLTs} \label{prop alg}

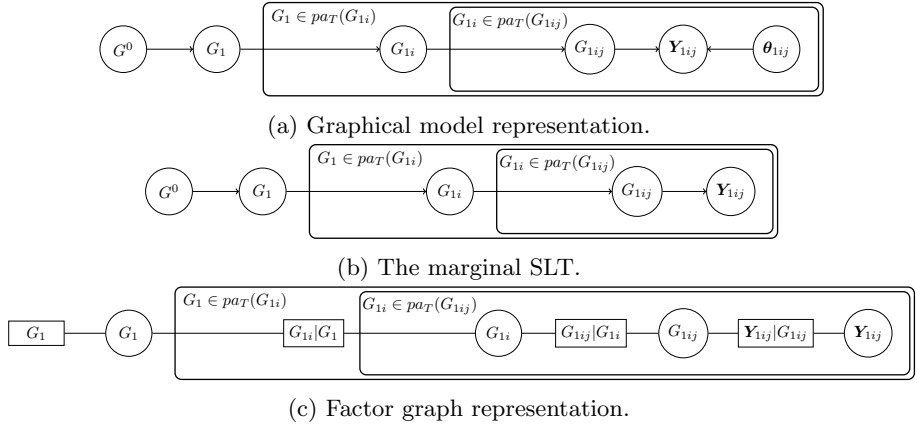
\begin{figure}[t!]
\centering

\begin{subfigure}{0.8\textwidth}
\resizebox{\textwidth}{!}{
\begin{tikzpicture}
\tikzstyle{var}=[draw,circle,fill = black!0,minimum width=1cm];
\tikzstyle{arrow}=[->];

\node[var] at (0,0) (G0) {$G^0$};
\node[var] at (2,0) (G) {$G_1$};
\node[var] at (6,0) (Gi) {$G_{1i}$};
\node[var] at (10,0) (Gij) {$G_{1ij}$};
\node[var] at (12,0) (Yij) {$\bm{Y}_{1ij}$};
\node[var] at (14,0) (tij) {$\bm{\theta}_{1ij}$};
\node at (4.3,0.7) (i) {$G_1 \in pa_T(G_{1i})$};
\node at (8.3,0.6) (j) {$G_{1i} \in pa_T(G_{1ij})$};

\path[arrow] (G0) edge (G);
\path[arrow] (G) edge (Gi);
\path[arrow] (Gi) edge (Gij);
\path[arrow] (Gij) edge (Yij);
\path[arrow] (tij) edge (Yij);

\begin{pgfonlayer}{background}
\tikzstyle{compartment} = [rectangle,draw=black, top color=white, bottom color=black!0,thick,rounded corners, inner sep=1cm, minimum size=1cm]
\filldraw [compartment]
(3,1)  rectangle (15,-1)
(7,0.9)  rectangle (14.9,-0.9);;
\end{pgfonlayer}
\end{tikzpicture}}
\caption{Graphical model representation.}
\label{model}
\end{subfigure}

\begin{subfigure}{0.7\textwidth}
\resizebox{\textwidth}{!}{
\begin{tikzpicture}
\tikzstyle{var}=[draw,circle,fill = black!0,minimum width=1cm];
\tikzstyle{arrow}=[->];

\node[var] at (0,0) (G0) {$G^0$};
\node[var] at (2,0) (G) {$G_1$};
\node[var] at (6,0) (Gi) {$G_{1i}$};
\node[var] at (10,0) (Gij) {$G_{1ij}$};
\node[var] at (12,0) (Yij) {$\bm{Y}_{1ij}$};
\node at (4.3,0.7) (i) {$G_1 \in pa_T(G_{1i})$};
\node at (8.3,0.6) (j) {$G_{1i} \in pa_T(G_{1ij})$};

\path[arrow] (G0) edge (G);
\path[arrow] (G) edge (Gi);
\path[arrow] (Gi) edge (Gij);
\path[arrow] (Gij) edge (Yij);

\begin{pgfonlayer}{background}
\tikzstyle{compartment} = [rectangle,draw=black, top color=white, bottom color=black!0,thick,rounded corners, inner sep=1cm, minimum size=1cm]
\filldraw [compartment]
(3,1)  rectangle (13,-1)
(7,0.9)  rectangle (12.9,-0.9);;
\end{pgfonlayer}
\end{tikzpicture}}
\caption{The marginal SLT.}
\label{marginal model}
\end{subfigure}

\begin{subfigure}{\textwidth}
\resizebox{\textwidth}{!}{
\begin{tikzpicture}
\tikzstyle{var}=[draw,circle,fill = black!0,minimum width=1cm];
\tikzstyle{factor}=[draw,rectangle,fill = black!0,minimum width=1.2cm];
\tikzstyle{arrow}=[];

\node[var] at (0,0) (G) {$G_1$};
\node[factor] at (-2,0) (fG) {$G_1$};
\node[factor] at (4,0) (fGiG) {$G_{1i}|G_1$};
\node[var] at (8,0) (Gi) {$G_{1i}$};
\node[factor] at (10,0) (fGijGi) {$G_{1ij}|G_{1i}$};
\node[var] at (12,0) (Gij) {$G_{1ij}$};
\node[factor] at (14,0) (fYijGij) {$\bm{Y}_{1ij}|G_{1ij}$};
\node[var] at (16,0) (Yij) {$\bm{Y}_{1ij}$};
\node at (2.3,0.7) (i) {$G_1 \in pa_T(G_{1i})$};
\node at (6.3,0.6) (j) {$G_{1i} \in pa_T(G_{1ij})$};

\path[arrow] (G) edge (fG);
\path[arrow] (G) edge (fGiG);
\path[arrow] (fGiG) edge (Gi);
\path[arrow] (Gi) edge (fGijGi);
\path[arrow] (fGijGi) edge (Gij);
\path[arrow] (Gij) edge (fYijGij);
\path[arrow] (fYijGij) edge (Yij);

\begin{pgfonlayer}{background}
\tikzstyle{compartment} = [rectangle,draw=black, top color=white, bottom color=black!0,thick,rounded corners, inner sep=1cm, minimum size=1cm]
\filldraw [compartment]
(1,1)  rectangle (17,-1)
(5,0.9)  rectangle (16.9,-0.9);;
\end{pgfonlayer}
\end{tikzpicture}}
\caption{Factor graph representation.}
\label{factor}
\end{subfigure}
\caption{Structure Learning Trees (SLT); 2-tier example. (a) Graphical model representation. [$G^0 =$ prior network, $G_1=$ root network, $G_{1i}=$ tier-1 networks, $G_{1ij} =$ tier-2 networks, $\bm{Y}_{1ij} =$ data available on network $G_{1ij}$, $\bm{\theta}_{1ij} =$ parameters describing the distribution of the data $\bm{Y}_{1ij}$. Bounding boxes are used to denote multiplicity of variables.] (b) The marginal SLT is obtained from (a) by integrating out continuous parameters $\bm{\theta}_{1ij}$. (c) Factor graph representation of the marginal SLT (b). [Circled nodes are random variables, rectangular nodes are factors. Dependence on the prior network is suppressed.]}
\label{various graphs}
\end{figure}

Following marginalisation of continuous parameters $\bm{\theta}$, inference for SLTs reduces to inference for a discrete Bayesian network whose nodes are themselves graphical models (SFig. \ref{various graphs}).
In this Section we describe the use of belief propagation (BP; \cite{Pearl3}) for inference in this setting and provide pseudocode for the 2-tier SLT model.

Denote by $\bm{X}$ a vector of random variables whose density factorizes according to
\begin{eqnarray}
p_{\bm{X}}(\bm{x}) = \prod_{f \in \mathcal{F}} f(\bm{x}_f) 
\label{factorise}
\end{eqnarray}
where $\bm{x}_f$ denotes the components of vector $\bm{x}$ upon which the factor $f$ depends. 
The  factor graph corresponding to the 2-tier SLT model is shown in SFig. \ref{factor}.
We use $\mu_{v \rightarrow f}$ to denote a message passed from a variable $v$ to a factor $f$, whereas $\overline{\mu}_{f \rightarrow v}$ will be used to denote a message passed from a factor $f$ to a variable $v$.
The message from a variable $v$ to a factor $f$ takes the following form:
\begin{eqnarray}
\mu_{v \rightarrow f} (x_v) = \prod_{f^* \in N(v) \setminus \{f\}} \overline{\mu}_{f^* \rightarrow v} (x_v)
\end{eqnarray}
where $N(v)$ denotes the neighbours of variable $v$ according to the factor graph.
Similarly the message from a factor $f$ to a variable $v$ takes the form
\begin{eqnarray}
\overline{\mu}_{f \rightarrow v} (x_v) = \sum_{\bm{x}' : x'_v = x_v} f(\bm{x}'_f) \prod_{v^* \in N(f) \setminus \{v\}} \mu_{v^* \rightarrow f} (x'_{v^*})
\end{eqnarray}
where $N(f)$ denotes the neighbours of factor $f$ according to the factor graph.

To simplify notation, we describe our algorithm using subscript notation as in the Main Text; e.g. $\overline{G_{1ij}}$ denotes the network that is the $j$th child of the $i$th child of the root network $\overline{G_1}$ in $T$.
BP nominates one node in the factor graph as a ``root''; of the remaining nodes, those with degree one are known as ``leaves''.
For BP applied to SLT we nominate the network $\overline{G_1}$ as the root node.
Messages are initiated at the leaves of the factor graph; specifically, in our 2-tier example, each variable node $\bm{Y}_{1ij}$ is initialised with an atomic distribution $\delta\{\bm{Y}_{1ij} = \bm{y}_{1ij}\}$ centered on the observed data $\bm{y}_{1ij}$.
Messages are passed through to the root node before being returned to the leaves.

Once the message passing has been completed, it is possible to extract marginals of interest by taking products of messages from factors neighboring the random variable of interest:
\begin{eqnarray}
p_{X_v}(x_v) \propto \prod_{f \in N(v)} \overline{\mu}_{f \rightarrow v} (x_v)
\end{eqnarray}
Alg. \ref{MP} contains pseudocode for the BP algorithm in the context of 2-tier SLTs.

\begin{algorithm}[h!]
\caption{Belief propagation (BP) for the 2-tier SLT model. Here we list all steps of the BP algorithm in order; at each stage messages are passed for all relevant networks indexed by $i$ and $j$, but we leave this implicit for clarity.}
\begin{spacing}{2}
\begin{algorithmic}[1]
\State $\mu_{\bm{Y}_{1ij} \rightarrow \bm{Y}_{1ij}|G_{1ij}}(\bm{Y}_{1ij}) = \delta\{\bm{Y}_{1ij} = \bm{y}_{1ij}\}$
\State $\overline{\mu}_{\bm{Y}_{1ij}|G_{1ij} \rightarrow G_{1ij}}(G_{1ij}) = \int_{\bm{Y}_{1ij}} p(\bm{Y}_{1ij}|G_{1ij}) \mu_{\bm{Y}_{1ij} \rightarrow \bm{Y}_{1ij}|G_{1ij}}(\bm{Y}_{1ij}) d\bm{Y}_{1ij}$ 
\State $\mu_{G_{1ij} \rightarrow G_{1ij}|G_{1i}}(G_{1ij}) = \overline{\mu}_{\bm{Y}_{1ij}|G_{1ij} \rightarrow G_{1ij}}(G_{1ij}) $ 
\State $\overline{\mu}_{G_{1ij}|G_{1i} \rightarrow G_{1i}}(G_{1i}) = \sum_{G_{1ij}} p(G_{1ij}|G_{1i}) \mu_{G_{1ij} \rightarrow G_{1ij}|G_{1i}}(G_{1ij}) $
\State $\mu_{G_{1i} \rightarrow G_{1i}|G_1}(G_{1i}) = \prod_j \overline{\mu}_{G_{1ij}|G_{1i} \rightarrow G_{1i}}(G_{1i}) $ 
\State $\overline{\mu}_{G_{1i}|G_1 \rightarrow G_1}(G_1) = \sum_{G_{1i}} p(G_{1i}|G_1) \mu_{G_{1i} \rightarrow G_{1i}|G_1}(G_{1i}) $ 
\State $\overline{\mu}_{G_1 \rightarrow G_1}(G_1) = p(G_1)$
\State $\mu_{G_1 \rightarrow G_{1i}|G_1}(G_1) = \overline{\mu}_{G_1 \rightarrow G_1}(G_1) \prod_{i' \neq i} \overline{\mu}_{G_{1i'}|G_1 \rightarrow G_1}(G_1) $ 
\State $\overline{\mu}_{G_{1i}|G_1 \rightarrow G_{1i}}(G_{1i}) = \sum_{G_1} p(G_{1i}|G_1) \mu_{G_1 \rightarrow G_{1i}|G_1}(G_1) $ 
\State $\mu_{G_{1i} \rightarrow G_{1ij}|G_{1i}}(G_{1i}) = \overline{\mu}_{G_{1i}|G_1 \rightarrow G_{1i}}(G_{1i}) $ 
\State $\overline{\mu}_{G_{1ij}|G_{1i} \rightarrow G_{1ij}}(G_{1ij}) = \sum_{G_{1i}} p(G_{1ij}|G_{1i}) \mu_{G_{1i} \rightarrow G_{1ij}|G_{1i}}(G_{1i}) $ 
\State $p(G_1|\bm{y}) = \overline{\mu}_{G_1 \rightarrow G_1}(G_1) \prod_i \overline{\mu}_{G_{1i}|G_1 \rightarrow G_1}(G_1)$
\State $p(G_{1i}|\bm{\bm{y}}) = \overline{\mu}_{G_{1i}|G_1 \rightarrow G_{1i}}(G_{1i}) \prod_j \overline{\mu}_{G_{1ij}|G_{1i} \rightarrow G_{1i}}(G_{1i})$
\State $p(G_{1ij}|\bm{\bm{y}}) = \overline{\mu}_{G_{1ij}|G_{1i} \rightarrow G_{1ij}}(G_{1ij}) \overline{\mu}_{\bm{Y}_{1ij}|G_{1ij} \rightarrow G_{1ij}}(G_{1ij})$
\end{algorithmic}
\end{spacing}
\label{MP}
\end{algorithm}

\subsection{Simulation study}

\subsubsection{Data generation} \label{sim set up}

From each network $\overline{G_{1ij}}$ we generated time series data $\bm{y}_{1ij}$, each containing $n$ time points, according to a linear VAR$(1)$ process. For each time series one variable was selected uniformly at random to be the target of a perfect intervention \citep{Spencer}.
Dynamical parameters were assigned such that for each edge $(i,j)$ we select a data-generating coefficient $\beta \in \{-1,+1\}$ uniformly at random.
For all experiments we used a noise magnitude $\sigma = 1$.
In each regime we generated data of varying sample size $n$ and edge density $\rho$.
Specifically, we considered both varying $n$ for fixed $\rho = 0.5$ and varying $\rho$ for fixed $n=60$.

\subsubsection{Performance measures} \label{metrics}

Denote the true data-generating (binary) adjacency matrix by $\bm{A}^0$.
In this work we considered the performance of two kinds of estimator; (i) the weighted adjacency matrices $\bm{A}$ produced by collecting together posterior marginal inclusion probabilities, and (ii) the binary adjacency matrices $\bm{A}(\tau)$ with $(i,j)$th entry $\mathbb{I}(A_{ij}>\tau)$, i.e. including edges if and only if the corresponding posterior marginal inclusion probabilities exceed a threshold $\tau$.  Write TP$(\tau)$, FP$(\tau)$, TN$(\tau)$, FN$(\tau)$ for, respectively, the true positive, false positive, true negative and false negative counts obtained by comparing $\bm{A}(\tau)$ to $\bm{A}^0$. Further write TPR$(\tau)$ = TP$(\tau)$ / (TP$(\tau)$ + FN$(\tau)$), FPR$(\tau)$ = FP$(\tau)$ / (TN$(\tau)$ + FP$(\tau)$), PPV$(\tau)$ = TP$(\tau)$ / (TP$(\tau)$ + FP$(\tau)$).

For (i) we considered the following performance measures:
\begin{enumerate}[(1)]
\item L1 Error $=\sum_{ij} |A_{ij} - A_{ij}^0|$ \\
\item Relative Density $=\sum_{ij} |A_{ij}| / \sum_{ij} |A_{ij}^0|$ \\
\item AUROC $=\int \text{TPR}(\tau) d\text{FPR}(\tau)$ \\
\item AUPR $=\int \text{PPV}(\tau) d\text{TPR}(\tau)$ \\
\end{enumerate}
For (ii) special attention is afforded to the ``median'' estimator with $\tau = 0.5$. Specifically we considered the performance measures
\begin{enumerate}[(1)]
\item Matthews Correlation Coefficient $= (\text{TP} \times \text{TN} - \text{FP} \times \text{FN}) / \sqrt{}((\text{TP} + \text{FP})(\text{TP} + \text{FN})(\text{TN} + \text{FP})(\text{TN} + \text{FN}))$ \\
\item Misclassification Rate $=$ (FP + FN) / $P^2$ \\
\item Misclassification Rate (top $k$ edges) $=$ As for the misclassification rate, but with $\tau$ chosen such that $\bm{A}(\tau)$ contains exactly $k$ non-zero entries, where $k$ is the number of edges in the true data-generating network. \\
\item Precision $=$ TP / (TP + FP).
\end{enumerate}

\subsubsection{Additional results}

\begin{figure*}
\centering
\begin{subfigure}{\textwidth}
\includegraphics[width = \textwidth,clip,trim = 2.5cm 0cm 0cm 1.5cm]{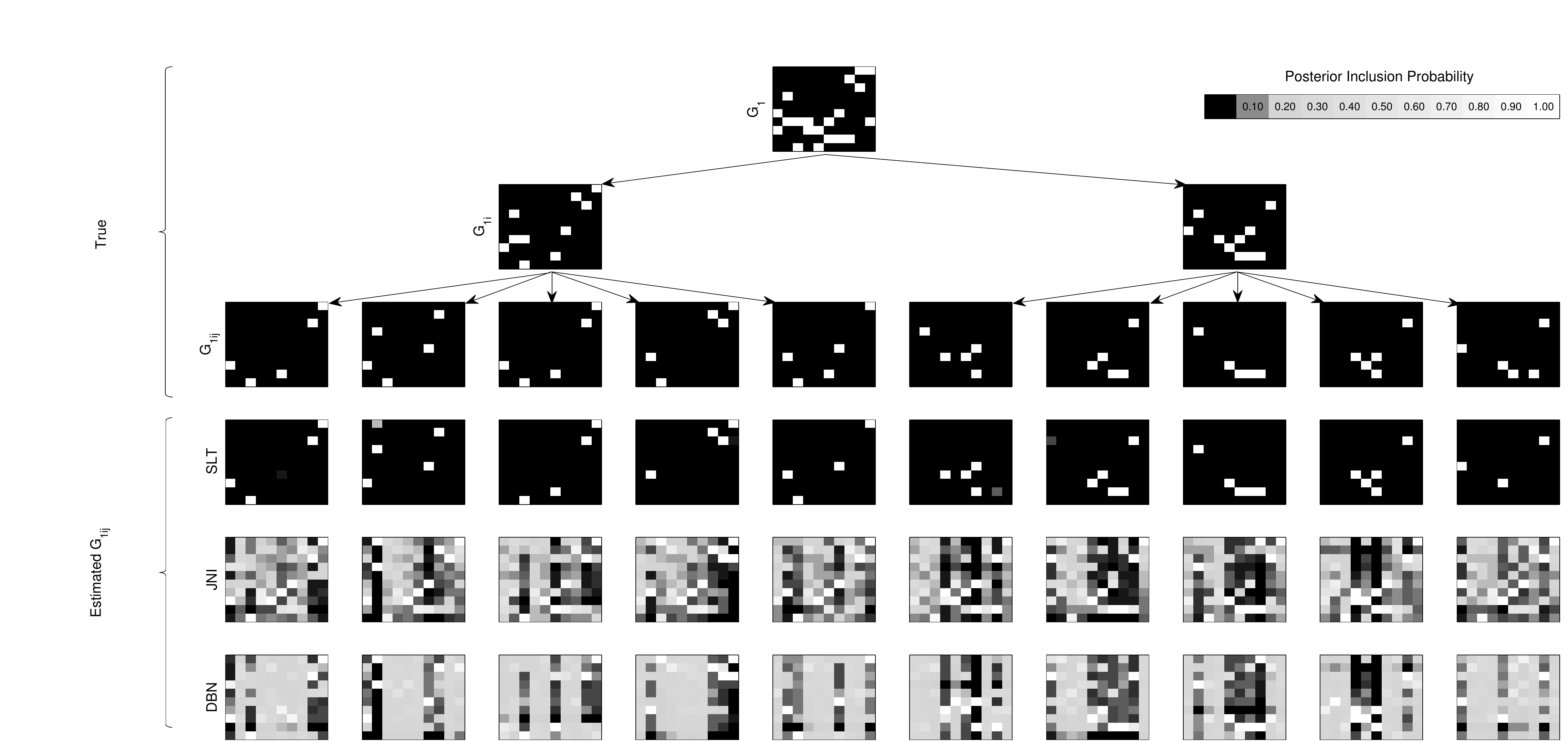}
\caption{Weakly exchangeable}
\label{ex2}
\end{subfigure}
\begin{subfigure}{\textwidth}
\includegraphics[width = \textwidth,clip,trim = 2.5cm 0cm 0cm 1.5cm]{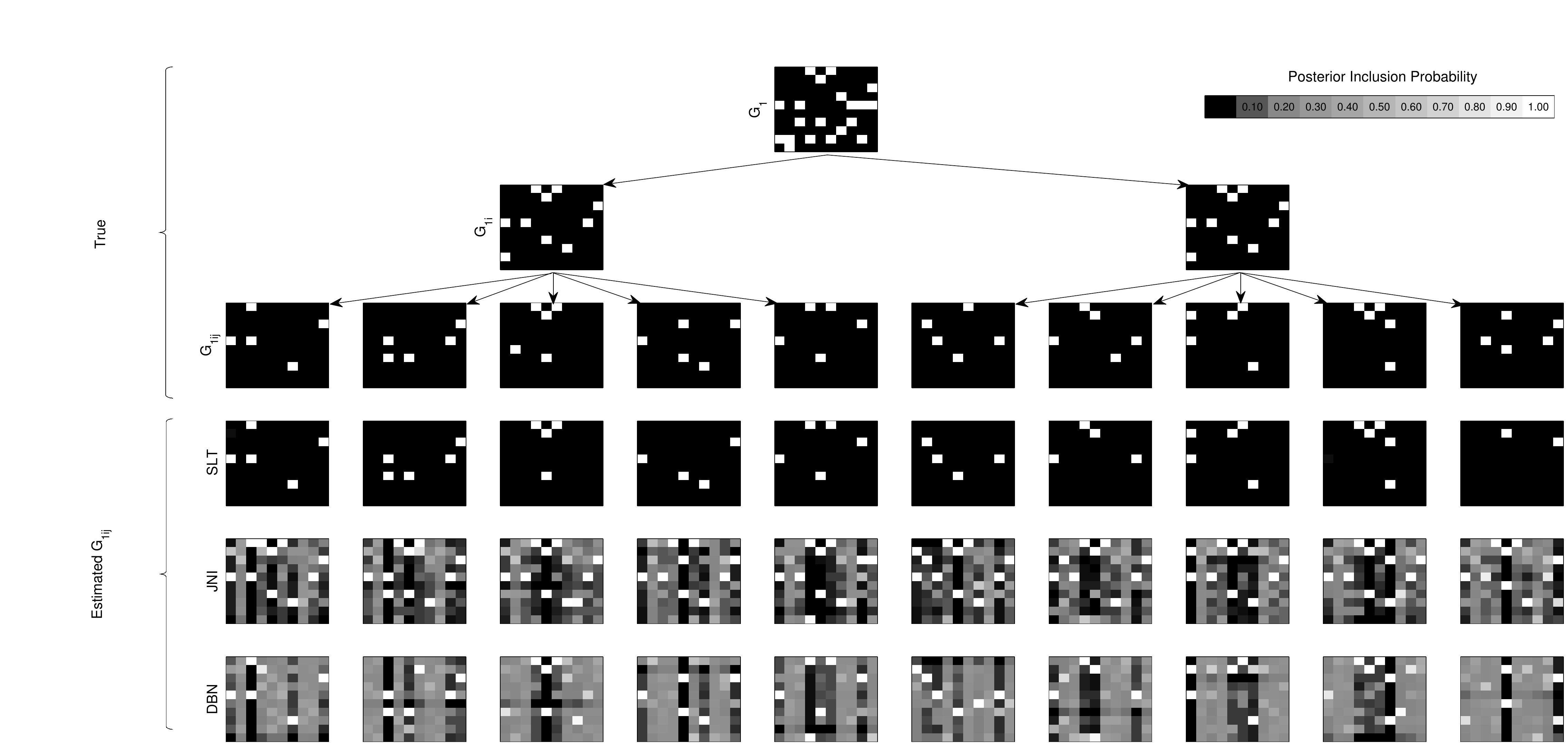}
\caption{Fully exchangeable}
\label{ex3}
\end{subfigure}
\caption{Results on simulated data generated from 2-tier SLTs; (a) a weakly exchangeable population, where $\overline{G_{11}}$, $\overline{G_{12}}$ are likely to share edges, and (b) a fully exchangeable population. [Inference methods: ``SLT'' = structure learning trees, ``JNI'' = joint network inference \citep{Oates4}, ``DBN'' = independent network inference (this corresponds to structure learning under the same local likelihood as SLT and JNI but applied separately to the data-generating networks located at the leaves of the tree).]}
\end{figure*}

\begin{figure*}
\centering
\begin{subfigure}{\textwidth}
\includegraphics[width = \textwidth,clip,trim = 2.5cm 0cm 0cm 1.5cm]{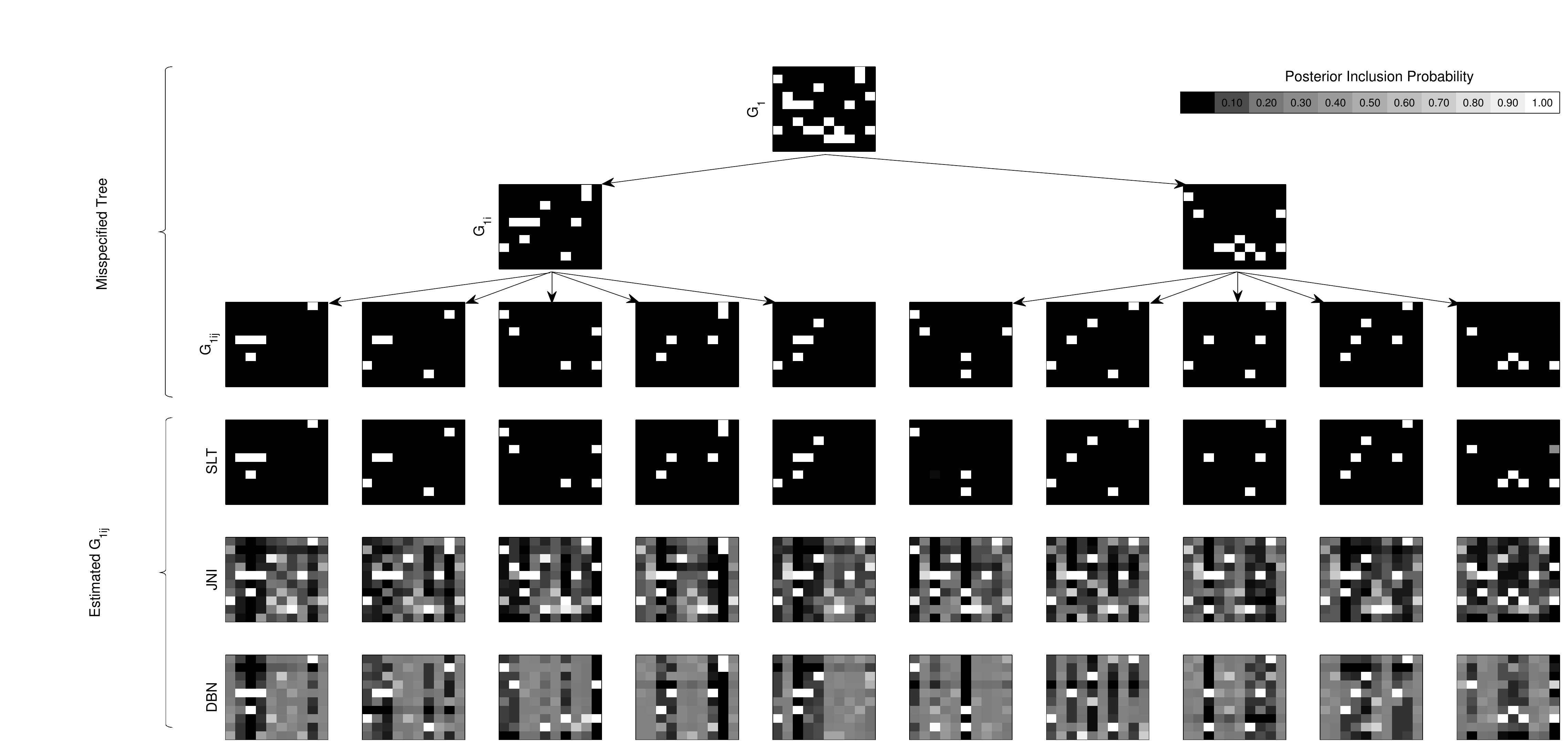}
\caption{Misspecified Tree}
\label{ex4}
\end{subfigure}
\begin{subfigure}{\textwidth}
\includegraphics[width = \textwidth,clip,trim = 2.5cm 0cm 0cm 1.5cm]{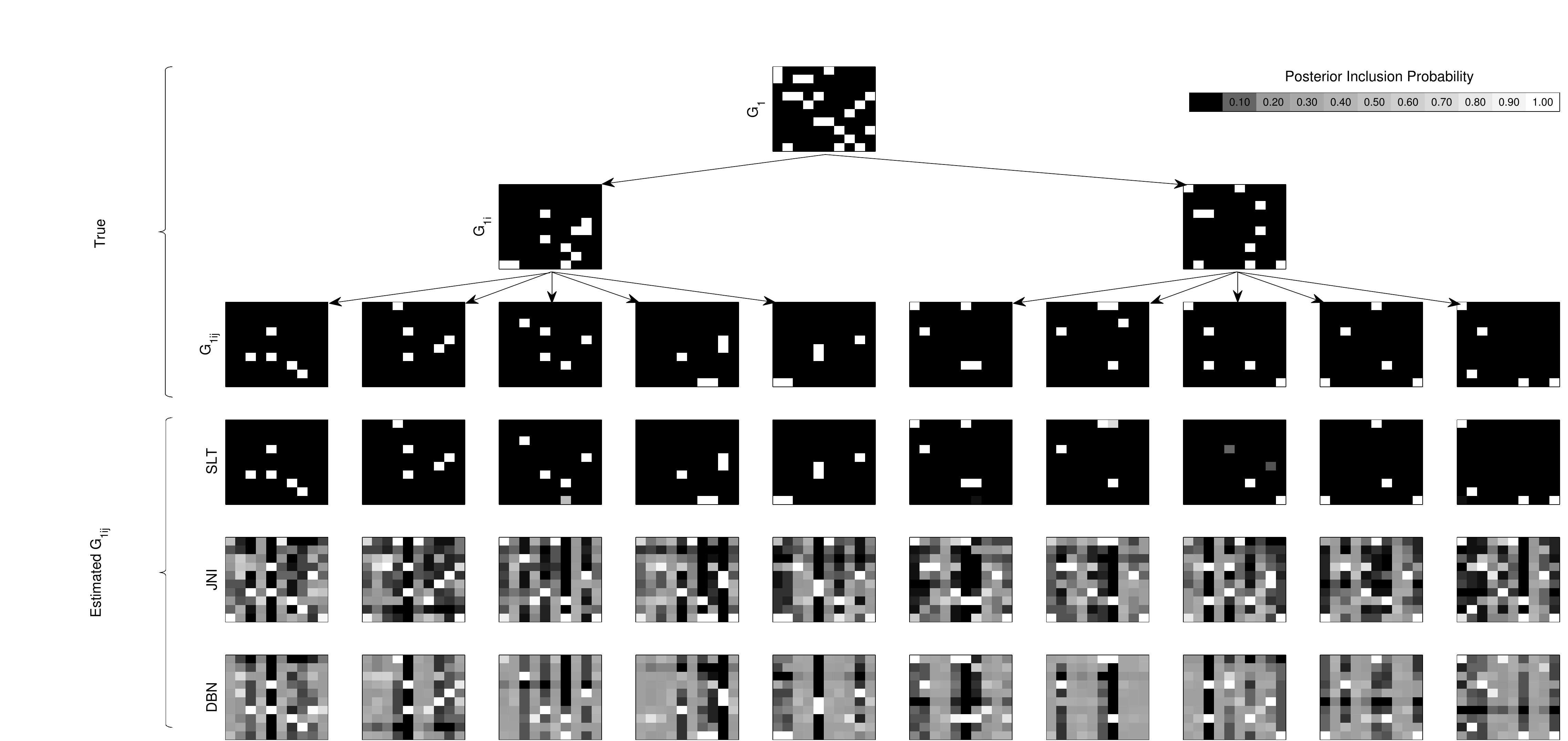}
\caption{Subset violation}
\label{ex5}
\end{subfigure}
\caption{Results on simulated data generated from 2-tier SLTs; (a) a misspecified tree structure $T$, and (b) a weakly exchangeable population which violates the subset assumptions encoded in the joint structural prior used by SLT. [Inference methods: ``SLT'' = structure learning trees, ``JNI'' = joint network inference \citep{Oates4}, ``DBN'' = independent network inference (this corresponds to structure learning under the same local likelihood as SLT and JNI but applied separately to the data-generating networks located at the leaves of the tree).]}
\end{figure*}

SFigs. \ref{ex2}, \ref{ex3}, \ref{ex4}, \ref{ex5} display typical simulation examples for regimes 2-5 respectively, and SFigs. \ref{sim res1}-\ref{sim res5} display full simulation results for each of the the 5 regimes described in the Main Text.

\subsection{Experimental protocol} \label{protocol}

Cells were plated into 10 cm$^2$ dishes at a density of $1-2 \times 10^6$ cells.  After 24 hours, cells were treated with 250 $nM$ lapatinib or 250 $nM$ AKTi (GSK690693). 
DMSO served as a control.  Cells were grown in 10\% FBS and harvested in RPPA lysis buffer at 30 min, 1h, 2h, 4h, 8h, 24h, 48h, and 72h post-treatment.  Cell lysates were quantitated, diluted, arrayed, and probed following \citet{Tibes}.  Imaging and quantitation of signal intensity was done following \citet{Tibes}. 
The particular protein species analysed were 4EBP1(pT37), AKT(pS473), BAD(pS112), c-Myc(pT58), EGFR(pY1173), ELK1(pS383), ER, FOXO3a(pS318), GSK3ab(pS21), HER2, IRS1(pS307), MAPK(pT202), MEK1/2(pS217), p38(pT180), p53, PR and S6(pS240).

\FloatBarrier

\end{document}